\documentclass[3p,12pt]{elsarticle}
\usepackage{threeparttable,booktabs,tabularx}
\usepackage[fleqn]{amsmath}
\usepackage{cases}
\usepackage{multirow}
\usepackage{xcolor}
\usepackage[normalem]{ulem}
\usepackage{tabularx}
\usepackage{siunitx}
\usepackage{comment}
\usepackage{algorithm}
\usepackage{algpseudocode}
\usepackage{algorithmicx}
\usepackage{grffile}
\usepackage{rotating}
\usepackage{array}
\usepackage{lineno}
\usepackage{hyperref}
\usepackage{makecell}
\usepackage{graphicx,enumerate}
\usepackage{amssymb}
\usepackage{bm,amsmath}
\allowdisplaybreaks  % Allow equations to break across pages
\usepackage{caption}
\usepackage{subcaption}
\usepackage{threeparttable}
\usepackage{subeqnarray}
\usepackage{multirow}
\usepackage{lscape}
\usepackage{rotating}
\usepackage{graphics}
\floatname{algorithm}{Algorithm}
\usepackage{epstopdf}
\journal{}

\biboptions{numbers,sort&compress} % 压制引用，[1,2,3]--> [1-3]

\usepackage{natbib}
\AtBeginEnvironment{thebibliography}{%
  \footnotesize
  \interlinepenalty=10000
}

\begin{document}
	
\begin{frontmatter}

\title{A general synthetic iterative solver for axisymmetric rarefied gas and electrostatic charged-particle flows}

\author[]{Yifan Wen}
\author[]{Lei Wu\corref{Boss}}

\cortext[Boss]{Corresponding author: wul@sustech.edu.cn}

\address{Department of Mechanics and Aerospace Engineering, Southern University of Science and Technology, 518055 Shenzhen, China}

\begin{abstract}
An axisymmetric general synthetic iterative scheme (AxiGSIS) is proposed to simulate rarefied gas flows and charged-particle transport under prescribed electrostatic fields. This solver adopts a finite-volume discrete velocity method defined over the two-dimensional axisymmetric meridian plane paired with a three-dimensional molecular velocity space.
Under the GSIS framework, the kinetic solver computes non-equilibrium stress and heat flux, which are subsequently imported as corrective source terms into the macroscopic synthetic system. Fast iterative updates of low-order flow primitive variables are performed on this macroscopic system, whose corrected flow fields are then fed back to the kinetic solver. This bidirectional coupling enables rapid propagation of macroscopic information and substantially accelerates steady-state convergence, particularly in near-continuum flow regimes. Four benchmark flows are examined: the Taylor–Couette flow, neutral nozzle expansion flow, charged-particle flow past an electrostatic sphere, and electrostatically accelerated charged-particle nozzle flow. Results show that AxiGSIS reproduces the reference kinetic solutions and accurately captures axisymmetric flow physics and charged-particle responses to prescribed electrostatic fields. Utilizing fewer spatial cells and iteration steps, AxiGSIS substantially cuts computational overhead relative to conventional kinetic iterations, particularly for low- and moderate-Knudsen-number flows.
\end{abstract}

\begin{keyword}
rarefied gas dynamics, axisymmetry, fast convergence, charged-particle transport
\end{keyword}

\end{frontmatter}

%\linenumbers

\section{Introduction}\label{sec:1}

Axisymmetric kinetic flows are ubiquitous in a wide range of engineering systems, including atmosphere-breathing electric propulsion, Hall thrusters, plasma accelerators, particle beam devices, and Z-pinches \cite{kaganovich2007kinetic,coche2014two,davidson2015implementation,bruhwiler2001particle,pointon2001particle,arber1996hybrid,schmidt2012fully}. The macroscopic flow fields of these flows are essentially 2D in physical space, whereas the velocity distribution function (VDF) in the Boltzmann equation remains fully 3D. Therefore, direct kinetic simulations involving both 3D physical space and 3D velocity space usually entail high computational cost. An axisymmetric kinetic formulation, which adopts a 2D physical space while retaining the 3D molecular velocity space, can significantly reduce the computational cost, without compromising the key kinetic effects induced by the circumferential velocity component.

% A number of canonical and application-oriented configurations have been investigated for axisymmetric rarefied gas flows. Sone and Aoki established asymptotic and hydrodynamic descriptions for steady rarefied gas flows past bodies at small Knudsen numbers~\cite{sone1987steady}, while Takata, Sone and Aoki performed a direct numerical analysis of uniform rarefied gas flow past a sphere using the Boltzmann equation~\cite{takata1993sphere}. Cylindrical Couette flow, another fundamental axisymmetric configuration with azimuthal velocity, has also been extensively studied. For example, Aoki and co-workers analyzed the inverted velocity profile in rarefied cylindrical Couette flow~\cite{aoki2003inverted}, and Sharipov and Kremer investigated linear and non-isothermal Couette flows of rarefied gases between two rotating cylinders~\cite{sharipov1996linear,sharipov1999nonisothermal}. In the deterministic numerical framework, Mieussens developed discrete-velocity models and numerical schemes for the Boltzmann--BGK and BGK--ES equations in plane and axisymmetric geometries~\cite{mieussens2000dvm}. These studies demonstrate the importance of axisymmetric kinetic formulations for reducing the physical-space dimensionality while retaining the 3D molecular velocity effects.

Numerical investigations into axisymmetric kinetic flows have been performed via both stochastic and deterministic numerical approaches. Within rarefied gas dynamics, the stochastic direct simulation Monte Carlo (DSMC) method has established itself as a longstanding mainstream technique \cite{bird1994molecular}, whereas particle-in-cell serves as the canonical tool for tracking charged particles evolving under electromagnetic fields in plasma simulations \cite{birdsall2018plasma,verboncoeur2005particle,tskhakaya2007particle}.
Nevertheless, their inherent stochasticity generates statistical fluctuations and imposes substantial computational overhead, particularly for low-speed, near-continuum flow regimes.
The deterministic discrete velocity method, which solves the VDF directly on a phase-space grid and thereby eliminates statistical noise, offers an alternative framework \cite{aristov2001direct,mieussens2000dvm,vogman2018conservative}. Sone and collaborators conducted kinetic-theory analyses of steady neutral rarefied gas flows over bluff bodies and cylindrical Couette configurations \cite{sone1987steady,takata1993sphere,aoki2003inverted}, while Sharipov and Kremer explored linear and non-isothermal rarefied Couette flows confined between rotating cylinders \cite{sharipov1996linear,sharipov1999nonisothermal}. Mieussens constructed conservative and entropy-stable discretizations tailored to axisymmetric domains, and underscored the necessity of consistent discretization for cylindrical-coordinate velocity derivative terms \cite{mieussens2000dvm}. Subsequent research yielded multiple quasi-2D axisymmetric kinetic schemes built upon kinetic equation under relaxation-time approximation and gas-kinetic flux formulations \cite{mieussens2000dvm,li2018ugksas}.
For charged-particle flows, Vlasov--Poisson solvers have been devised via semi-Lagrangian, finite-volume, spectral, and gas-kinetic frameworks \cite{cheng1976integration,sonnendrucker1999semi,filbet2001conservative,valentini2005numerical,vogman2018conservative,wang2022gas,ni2022fourier}, and applied to charge-dominated particle beams, plasma-edge charge separation, magnetized dynamics, and electrostatic Z-pinch geometries \cite{filbet2002direct,shoucri2004study,valentini2005numerical,vogman2018conservative,wang2022gas,ni2022fourier}.

However, the reduction of physical-space dimensionality does not remove the main computational bottleneck of kinetic simulations, especially for near-continuum steady flows. For stochastic methods, the cell size and time step are usually constrained by the molecular mean free path and collision time, and a large number of particles or samples are required to reduce statistical noise. As a result, low-speed or near-continuum flow simulations are often computationally expensive~\cite{bird1994molecular,mieussens2000dvm}. Conventional discrete velocity method also suffers from slow convergence in the small-Knudsen-number regime~\cite{wang2018comparative}. In this regime, the flow approaches the hydrodynamic limit, and the evolution of the VDF is strongly constrained by slowly varying macroscopic modes. As a result, purely kinetic iterations propagate flow information inefficiently over the computational domain and require a large number of iterations before reaching the steady state. Hence, despite the dimensional reduction afforded by axisymmetric frameworks, high-efficiency steady-state computations for near-continuum kinetic flows still pose a prominent challenge.

Over the past decades, significant advances have been achieved in deterministic methods for multiscale gas kinetic simulations, which are developed based on either the Boltzmann equation or simplified kinetic models.
Asymptotic preserving and fast convergence  are two essential properties that guarantee high-efficiency flow simulations. Specifically, asymptotic preserving  allows simulations to proceed without resolving fine kinetic scales, whereas fast convergence greatly accelerates the computation of steady-state solutions. The unified gas-kinetic scheme serves as a typical asymptotic preserving method. By coupling particle transport and collision in the numerical flux construction, it permits the use of spatial cell sizes considerably larger than the molecular mean free path~\cite{xu2010ugks,guo2013dugks}. For axisymmetric rarefied gas flows, it has been proposed to accurately capture multiscale flow physics across a wide range of Knudsen number regimes~\cite{li2018ugksas}.
The general synthetic iterative scheme (GSIS) represents another pivotal development, as it simultaneously inherits both asymptotic preserving and fast convergence characteristics~\cite{Su2020GSIS,Su2020SIAM,Zhang2024CaF}. High-order nonequilibrium moments, such as stress tensors and heat fluxes, are extracted from kinetic solutions and then integrated into the synthetic macroscopic equations. These macroscopic synthetic equations admit far more efficient numerical solution than the underlying kinetic equation, enabling rapid propagation and transfer of macroscopic flow quantities. Accordingly, when the solution to these macroscopic synthetic equations is employed to steer the evolution of low-order moments within the collision operator of the kinetic equation, GSIS drastically cuts the total number of kinetic iterations, while accurately retaining the correct asymptotic physical characteristics for near-continuum flow regimes.

To our knowledge, GSIS has never been extended to axisymmetric kinetic simulations or charged-particle transport, and both extensions entail numerical difficulties. For axisymmetric flows, the scheme must preserve complete 3D velocity-space physics on the 2D meridian plane while maintaining consistent cylindrical geometric source terms in both kinetic and synthetic macroscopic equations. For charged-particle flows, electrostatic acceleration terms are introduced into the kinetic equation, with associated momentum and energy source terms embedded into the macroscopic synthetic system. These mandatory numerical treatments underpin the construction of the present axisymmetric GSIS solver, named AxiGSIS.

The rest of the paper is organized as follows. Section~\ref{sec:Gov_eq} presents the governing equations for axisymmetric kinetic flows, including the kinetic model equation, the prescribed electrostatic acceleration, and the corresponding macroscopic moment equations. Section~\ref{sec:3} describes the numerical method, including the finite-volume discrete velocity formulation, the treatment of axisymmetric geometric and electrostatic source terms, the velocity-space integration compensation, and the GSIS iteration procedure. Section \ref{sec:4} evaluates the accuracy, convergence characteristics, and computational efficiency of the proposed scheme across four distinct flow configurations. Concluding remarks are provided in Section \ref{sec:conclusion}.

% \leir{
% \section{charged particle flow}

% In addition to neutral rarefied gases, axisymmetric transport of a single charged-particle species under electrostatic fields is also of interest in kinetic simulations. For such problems, the particle distribution function evolves under the acceleration induced by the electrostatic field. In cylindrical coordinates, the governing equation includes not only the physical-space transport and collision terms, but also velocity-space acceleration terms induced by the electric field and by cylindrical geometry. In the present work, the electrostatic field is prescribed or obtained from an electrostatic potential problem independent of the simulated charged-particle density; the self-generated electric field of the charged particles is not considered. Therefore, the charged-particle module focuses on the kinetic response of particles to a given electrostatic field in an axisymmetric geometry.

\section{Governing equations}\label{sec:Gov_eq}

For axisymmetric kinetic flows, the physical space is described by the axial coordinate \(x\) and the radial coordinate \(r\). The molecular velocity space $\boldsymbol{v}=(v_x,v_r,v_\theta)$ remains 3D, where \(v_x\), \(v_r\), and \(v_\theta\) are the axial, radial, and circumferential velocity components, respectively. The transverse molecular velocity can be further expressed in polar coordinates as $v_r=\zeta\cos\omega$ and $v_\theta=\zeta\sin\omega$, where \(\zeta\ge 0\) is the magnitude and \(\omega\) is the polar angle. Therefore, the VDF can be written as $f=f(t,x,r,v_x,\zeta,\omega)$. 

Macroscopic quantities, such as the number density $n$, flow velocity $\boldsymbol{u}$, temperature $T$, pressure tensor $P_{ij}$, and heat flux $\boldsymbol{q}$, are obtained by taking moments of the VDF:
\begin{equation}\label{eq:basic_moments}
\begin{aligned}
\left[n,n\boldsymbol{u},\frac{3}{2}n k_B T,  P_{ij}, \boldsymbol{q}\right]=\int \left[1,\boldsymbol{v}, \frac{1}{2}m|\boldsymbol{c}|^2, mc_i c_j, \frac{m}{2}|\boldsymbol{c}|^2\boldsymbol{c}\right]f\,d\boldsymbol{v}, 
% \\
%     n=&\int f\,d\boldsymbol{v},
%     \qquad
%     n\boldsymbol{u}=\int \boldsymbol{v}f\,d\boldsymbol{v},
%     \qquad
%     \frac{3}{2}n k_B T
%     =
%     \int
%     \frac{1}{2}m|\boldsymbol{c}|^2 f\,d\boldsymbol{v}, \\
%     P_{ij}=&m\int c_i c_j f d\boldsymbol{v}, 
%    \qquad
%   \boldsymbol{q}
% =
% \frac{m}{2}
% \int
% |\boldsymbol{c}|^2\boldsymbol{c}f d\boldsymbol{v},  
\end{aligned}
\end{equation}
where $i,j\in{x,r,\theta}$, $\boldsymbol{c}=\boldsymbol{v}-\boldsymbol{u}=(v_x-u_x, v_r-u_r, v_\theta-u_\theta)$ is the peculiar velocity, \(m\) is the molecular mass, $k_B$ is the Boltzmann constant, and $d\boldsymbol{v}= \zeta\,dv_x\,d\zeta\,d\omega$, with the integration carried out over the full domain of cylindrical coordinates in velocity space.  The scalar pressure follows the ideal-gas law:
$
p=(P_{xx}+P_{rr}+P_{\theta\theta})/3=n k_B T$.
The non-equilibrium stress tensor is written as
\begin{equation}
\Pi_{ij}=P_{ij}-p\delta_{ij}
=
m\int
\left(
c_i c_j-\frac{1}{3}|\boldsymbol{c}|^2\delta_{ij}
\right)
f d\boldsymbol{v}.
\label{eq:stress_tensor}
\end{equation}

% Likewise, the pressure tensor is defined as
% \begin{equation}
% P_{ij}=m\int c_i c_j f d\boldsymbol{v},
% \qquad
% i,j\in{x,r,\theta},
% \label{eq:pressure_tensor}
% \end{equation}
% where $c_x=v_x-u_x$, $c_r=v_r-u_r$, and $c_\theta=v_\theta-u_\theta$ are the peculiar velocity. 
% For a general axisymmetric flow with possible swirl, the retained stress components are
% \begin{equation}
% \begin{aligned}
% \Pi_{xx}
% &=
% m\int
% \left(
% c_x^2-\frac{1}{3}|\boldsymbol{c}|^2
% \right)f\,d\boldsymbol{v},
% \\
% \Pi_{rr}
% &=
% m\int
% \left(
% c_r^2-\frac{1}{3}|\boldsymbol{c}|^2
% \right)f\,d\boldsymbol{v},
% \\
% \Pi_{\theta\theta}
% &=
% m\int
% \left(
% c_\theta^2-\frac{1}{3}|\boldsymbol{c}|^2
% \right)f\,d\boldsymbol{v},
% \\
% \Pi_{xr}
% &=
% m\int c_x c_r f\,d\boldsymbol{v},
% \\
% \Pi_{x\theta}
% &=
% m\int c_x c_\theta f\,d\boldsymbol{v},
% \\
% \Pi_{r\theta}
% &=
% m\int c_r c_\theta f\,d\boldsymbol{v}.
% \end{aligned}
% \label{eq:stress_components}
% \end{equation}

% The heat flux is defined as
% \begin{equation}
% \boldsymbol{q}
% =
% \frac{m}{2}
% \int
% |\boldsymbol{c}|^2\boldsymbol{c}f d\boldsymbol{v},
% \label{eq:heat_flux}
% \end{equation}
% with components
% \begin{equation}
% \begin{aligned}
% q_x
% &=
% \frac{m}{2}
% \int |\boldsymbol{c}|^2 c_x f d\boldsymbol{v},
% \qquad
% q_r
% &=
% \frac{m}{2}
% \int |\boldsymbol{c}|^2 c_r f d\boldsymbol{v},
% \qquad
% q_\theta
% &=
% \frac{m}{2}
% \int |\boldsymbol{c}|^2 c_\theta f d\boldsymbol{v}.
% \end{aligned}
% \label{eq:heat_flux_components}
% \end{equation}

Two types of axisymmetric flows are considered in the present work. For non-swirling axisymmetric flows, the macroscopic velocity has no circumferential component, i.e., $\boldsymbol{u}=(u_x,u_r,0)$. In this case, the symmetry of the VDF with respect to the circumferential molecular velocity leads to $q_\theta=\Pi_{x\theta}=\Pi_{r\theta}=0$. However, the circumferential molecular velocity $v_\theta$ and the normal stress component $\Pi_{\theta\theta}$ must still be retained, because they contribute to the radial momentum balance and represent the 3D velocity-space effect.
For swirling axisymmetric flows, the circumferential velocity is non-zero. In a purely rotational configuration, such as cylindrical Couette flow between two coaxial cylinders, the macroscopic velocity is  written as $\boldsymbol{u}=(0,0,u_\theta)$, and the shear stress $\Pi_{r\theta}$ plays an important role in the radial transport of circumferential momentum. 

%More generally, in axisymmetric flows with both meridian motion and swirl, the velocity may take the form $\boldsymbol{u}=(u_x,u_r,u_\theta)$, and the circumferential heat flux and shear stresses, such as $q_\theta$, $\Pi_{x\theta}$, and $\Pi_{r\theta}$, may be non-zero. These quantities are therefore retained in the present formulation.

\subsection{Kinetic model equation}

Within the axisymmetric formulation, temporal evolution of  the VDF arises from the free transport $\mathcal{T}^{x,r}$, geometrical acceleration $\mathcal{T}^{geo}$, external acceleration $\mathcal{T}^{E}$, and collisional effects. The  kinetic model equation can be written in the conservative
phase-space form as
\begin{equation}
\begin{split}
\frac{\partial (r f)}{\partial t}
&+
\underbrace{\frac{\partial (r v_x f)}{\partial x}
+
\frac{\partial (r v_r f)}{\partial r}}_\text{free transport: $\mathcal{T}^{x,r}$}
\underbrace{-\zeta
\frac{\partial(\sin\omega f)}{\partial\omega}}_\text{geometrical acceleration: $\mathcal{T}^{geo}$}
\\
&+
r\underbrace{\left[
\frac{\partial(a_x f)}{\partial v_x}
+
\frac{1}{\zeta}
\frac{\partial(\zeta a_\zeta f)}{\partial \zeta}
+
\frac{1}{\zeta}
\frac{\partial(a_\omega f)}{\partial \omega}
\right]}_\text{external acceleration: $\mathcal{T}^{E}$}
=
r\underbrace{\frac{g^S-f}{\tau}}_\text{collision},
\end{split}
\label{eq:axis_kinetic}
\end{equation}
where $\tau=\mu/p$ is the relaxation time, with $\mu$ being the dynamic viscosity of the gas; $g^S$ is the Shakhov reference VDF \cite{shakhov1968approximate}: 
\begin{equation}\label{eq:shakhov}
g^S
=
g^M(n,\boldsymbol{u},T)
\left[
1
+
(1-\Pr)
\frac{m\boldsymbol{c}\cdot\boldsymbol{q}}
{5pk_BT}
\left(
\frac{m|\boldsymbol{c}|^2}{k_BT}
-
5
\right)
\right],
\end{equation}
with $\Pr=2/3$ being the Prandtl number for monatomic gases, and  $g^M$  the Maxwellian equilibrium VDF:
\begin{equation}
g^M(n,\boldsymbol{u},T)
=
n
\left(
\frac{m}{2\pi k_B T}
\right)^{3/2}
\exp
\left[
-\frac{m|\boldsymbol{v}-\boldsymbol{u}|^2}{2k_B T}
\right].
\label{eq:maxwellian}
\end{equation}

For charged particles under an electrostatic field, the external acceleration is $\boldsymbol{a} =q_p\boldsymbol{E}/m$, where \(q_p\) is the particle charge and \(\boldsymbol{E}\) is the external electric field. In the axisymmetric meridian plane,
\[
    \boldsymbol{E}
    =
    (E_x,E_r)
    =
    -\nabla_{x,r}\phi
    =
    \left(
    -\frac{\partial \phi}{\partial x},
    -\frac{\partial \phi}{\partial r}
    \right),
    \label{eq:electric_field}
\]
where \(\phi\) is the electrostatic potential satisfying the axisymmetric
Laplace equation
\begin{equation}
    \frac{\partial^2\phi}{\partial x^2}
    +
    \frac{1}{r}
    \frac{\partial}{\partial r}
    \left(
    r\frac{\partial \phi}{\partial r}
    \right)
    =
    0 .
    \label{eq:laplace}
\end{equation}
In the present work, the electrostatic field is determined only by the prescribed electrostatic boundary conditions $\phi_D$ and $g_N$:
\begin{equation}
    \phi=\phi_D
    \quad \text{on } \Gamma_D,
    \qquad
    \nabla_{x,r}\phi\cdot\boldsymbol{n}=g_N
    \quad \text{on } \Gamma_N,
    \label{eq:laplace_bc}
\end{equation}
where \(\Gamma_D\) and \(\Gamma_N\) denote the Dirichlet and Neumann boundary
parts, respectively.
The particle density computed from the VDF is not used to update the potential or the electric field. Therefore, the electric field acts as an externally imposed
field in the kinetic equation, rather than a self-consistent field coupled with
the particle distribution.

In the cylindrical velocity coordinates
\((v_x,\zeta,\omega)\), the acceleration components are
\[
    a_x=\frac{q_p}{m}E_x,\qquad
    a_\zeta=\frac{q_p}{m}E_r\cos\omega,\qquad
    a_\omega=-\frac{q_p}{m}E_r\sin\omega .
    \label{eq:acceleration_components}
\]
For neutral gases, \(q_p=0\), and hence
$a_x=a_\zeta=a_\omega=0$.

The geometric acceleration term in Eq.~\eqref{eq:axis_kinetic} arises from the cylindrical coordinate transformation. Absent in its Cartesian counterpart, this term is critical to retaining full 3D velocity-space dynamics within the 2D axisymmetric physical domain.

\subsection{Kinetic boundary conditions}\label{sec:bc}
The kinetic boundary conditions are needed to fully determine a rarefied gas flow. First, for a solid boundary, let \(\boldsymbol{n}=(n_x,n_r)\) be the outward unit normal vector of the gas domain, pointing from the gas to the solid wall. The normal molecular velocity is
$
v_n
=
\boldsymbol{v}\cdot\boldsymbol{n}
=
v_x n_x+v_r n_r
=
v_x n_x+\zeta\cos\omega n_r .
$
The boundary condition is prescribed for molecules entering the computational domain, i.e., $v_n<0$. At a solid wall, the diffuse-reflection condition is adopted:
\begin{equation}
f(\boldsymbol{x}_w,\boldsymbol{v})
=
g^M(n_w,\boldsymbol{u}_w,T_w),
\qquad v_n<0,
\label{eq:diffuse_wall_bc}
\end{equation}
where $T_w$ and $\boldsymbol{u}_w$ are the wall temperature and velocity, respectively. For a stationary wall, $\boldsymbol{u}_w=\boldsymbol{0}$, while for a rotating cylindrical wall, $\boldsymbol{u}_w=(0,0,\Omega r)$. The wall number density $n_w$ is determined from the impermeable condition: $\int_{v_n<0} v_n g^M(n_w,\boldsymbol{u}_w,T_w) d\boldsymbol{v}
+
\int_{v_n>0} v_n f d\boldsymbol{v}
=
0$.
% \begin{equation}
% \int_{v_n<0} v_n g_w^M d\boldsymbol{v}
% +
% \int_{v_n>0} v_n f d\boldsymbol{v}
% =
% 0 .
% \label{eq:zero_wall_flux}
% \end{equation}

Second, at an inflow or far-field boundary, the incoming VDF is specified by the prescribed macroscopic state:
\begin{equation}
f(\boldsymbol{x}_b,\boldsymbol{v})
=
g^M(n_{\rm in},\boldsymbol{u}_{\rm in},T_{\rm in}),
\qquad v_n<0 .
\label{eq:inflow_bc}
\end{equation}
At an outflow boundary, the VDF is extrapolated from the interior cells. 

Third, on the symmetry axis $r=0$, the regularity condition is imposed through specular symmetry in the radial molecular velocity: $f(t,x,0,v_x,v_r,v_\theta)
=
f(t,x,0,v_x,-v_r,v_\theta)$,
or equivalently, in $(v_x,\zeta,\omega)$ coordinates,
\begin{equation}
f(t,x,0,v_x,\zeta,\omega)
=
f(t,x,0,v_x,\zeta,\pi-\omega).
\label{eq:axis_bc_cylindrical_velocity}
\end{equation}

\subsection{Macroscopic equations}
\label{subsec:macroscopic_equations}

Macroscopic equations for conservative variables 
\begin{equation}
    \boldsymbol{W}
    =
    \left(
        \rho,\,
        \rho u_x,\,
        \rho u_r,\,
        \rho u_{\theta},\,
        \rho E
    \right)^T,
\label{eq:conservative_variables}
\end{equation}
where $\rho = mn$  is the mass density and  
$ E = \frac{1}{2}
    \left(
        u_x^2+u_r^2+u_{\theta}^2
    \right)
    +
    \frac{3}{2}
    \frac{k_B T}{m}$ is the specific energy, can be established by
taking the conservative moments of Eq.~\eqref{eq:axis_kinetic}. For instance, the mass conservation equation is obtained by multiplying Eq.~\eqref{eq:axis_kinetic} by $m$ and then integrating over the whole molecular velocity space:
\begin{equation}
    \frac{\partial \rho}{\partial t}
    +
    \frac{\partial}{\partial x}
    \left(
        \rho u_x
    \right)
    +
    \frac{1}{r}
    \frac{\partial}{\partial r}
    \left(
        r\rho u_r
    \right)
    =
    0.
\label{eq:mass_equation}
\end{equation}
The axial, radial, and azimuthal momentum equations are obtained by taking the $m v_x$, $mv_r$, and $mv_\theta$ moment of Eq.~\eqref{eq:axis_kinetic}:
\begin{equation}
\begin{split}
    \frac{\partial (\rho u_x)}{\partial t}
    &+
    \frac{\partial}{\partial x}
    \left(
        \rho u_x^2+p+\Pi_{xx}
    \right)
    +
    \frac{1}{r}
    \frac{\partial}{\partial r}
    \left[
        r
        \left(
            \rho u_xu_r+\Pi_{xr}
        \right)
    \right]
    =
    S_x, \\
    \frac{\partial (\rho u_r)}{\partial t}
    &+
    \frac{\partial}{\partial x}
    \left(
        \rho u_xu_r+\Pi_{xr}
    \right)
    +
    \frac{1}{r}
    \frac{\partial}{\partial r}
    \left[
        r
        \left(
            \rho u_r^2+p+\Pi_{rr}
        \right)
    \right]
    -
    \frac{
        \rho u_\theta^2
        +
        p+\Pi_{\theta\theta}
    }{r}
    =
    S_r, \\
   \frac{\partial (\rho u_\theta)}{\partial t}
    &+
    \frac{\partial}{\partial x}
    \left(
        \rho u_xu_\theta+\Pi_{x\theta}
    \right)
    +
    \frac{1}{r^2}
    \frac{\partial}{\partial r}
    \left[
        r^2
        \left(
            \rho u_ru_\theta+\Pi_{r\theta}
        \right)
    \right]
    =
    S_\theta . 
\end{split}
\label{eq:axial_momentum_equation}
\end{equation}
The total energy equation is obtained by taking the $\frac{1}{2}m|\boldsymbol{v}|^2$ moment of Eq.~\eqref{eq:axis_kinetic}:
\begin{equation}
\begin{split}
    \frac{\partial (\rho E)}{\partial t}
    &+
    \frac{\partial}{\partial x}
    \left[
        (\rho E+p)u_x
        +
        \Pi_{xx}u_x
        +
        \Pi_{xr}u_r
        +
        \Pi_{x\theta}u_\theta
        +
        q_x
    \right]
    \\
    &+
    \frac{1}{r}
    \frac{\partial}{\partial r}
    \left\{
        r
        \left[
            (\rho E+p)u_r
            +
            \Pi_{xr}u_x
            +
            \Pi_{rr}u_r
            +
            \Pi_{r\theta}u_\theta
            +
            q_r
        \right]
    \right\}
    =
    S_E .
\end{split}
\label{eq:energy_equation}
\end{equation}

% The radial momentum equation is obtained by taking the $m v_r$ moment of Eq.~\eqref{eq:axis_kinetic}:
% \begin{equation}
% \begin{split}
%     \frac{\partial (\rho u_r)}{\partial t}
%     &+
%     \frac{\partial}{\partial x}
%     \left(
%         \rho u_xu_r+\Pi_{xr}
%     \right)
%     +
%     \frac{1}{r}
%     \frac{\partial}{\partial r}
%     \left[
%         r
%         \left(
%             \rho u_r^2+p+\Pi_{rr}
%         \right)
%     \right]
%     -
%     \frac{
%         \rho u_\theta^2
%         +
%         p+\Pi_{\theta\theta}
%     }{r}
%     =
%     S_r .
% \end{split}
% \label{eq:radial_momentum_equation}
% \end{equation}
% The azimuthal momentum equation is obtained by taking the $m v_\theta$ moment of Eq.~\eqref{eq:axis_kinetic}:
% \begin{equation}
% \begin{split}
%     \frac{\partial (\rho u_\theta)}{\partial t}
%     &+
%     \frac{\partial}{\partial x}
%     \left(
%         \rho u_xu_\theta+\Pi_{x\theta}
%     \right)
%     +
%     \frac{1}{r^2}
%     \frac{\partial}{\partial r}
%     \left[
%         r^2
%         \left(
%             \rho u_ru_\theta+\Pi_{r\theta}
%         \right)
%     \right]
%     =
%     S_\theta .
% \end{split}
% \label{eq:azimuthal_momentum_equation}
% \end{equation}

For charged particles under a prescribed electrostatic field, the force and work terms are given below: 
\begin{equation}
    S_x
    =
    q_pnE_x,
   \quad
    S_r
    =
    q_pnE_r,
    \quad
    S_\theta
    =    
    q_pnE_\theta,
   \quad
    S_E
    =
    q_pn
    \left(
        u_xE_x+u_rE_r+u_\theta E_\theta
    \right),
\label{eq:charged_sources}
\end{equation}
while for neutral gases, these source terms vanish.

These macroscopic equations are not closed because the stress tensor
$\boldsymbol{\Pi}$ and heat flux $\boldsymbol{q}$ are not expressed by the conservative variables. In the continuum limit, the
Chapman--Enskog expansion gives the Navier-Stokes-Fourier (NSF) constitutive relations \cite{CE},
\begin{equation}\label{eq:nsf_constitutive_relations}
\begin{aligned}
    \Pi_{xx}^{NSF}
    &=
    -2\mu
    \left(
        \frac{\partial u_x}{\partial x}
        -
        \frac{1}{3}
        \nabla\cdot\boldsymbol{u}
    \right),
\quad
    \Pi_{rr}^{NSF}
    =
    -2\mu
    \left(
        \frac{\partial u_r}{\partial r}
        -
        \frac{1}{3}
        \nabla\cdot\boldsymbol{u}
    \right),
    \\
    \Pi_{\theta\theta}^{NSF}
    &=
    -2\mu
    \left(
        \frac{u_r}{r}
        -
        \frac{1}{3}
        \nabla\cdot\boldsymbol{u}
    \right),
\quad
    \Pi_{xr}^{NSF}
   =
    -\mu
    \left(
        \frac{\partial u_x}{\partial r}
        +
        \frac{\partial u_r}{\partial x}
    \right),
    \\
    \Pi_{x\theta}^{NSF}
    &=
    -\mu
    \frac{\partial u_\theta}{\partial x},
\quad
    \Pi_{r\theta}^{NSF}
    =
    -\mu
    \left(
        \frac{\partial u_\theta}{\partial r}
        -
        \frac{u_\theta}{r}
    \right),
    \\
     q_x^{NSF}
    &=
    -\kappa
    \frac{\partial T}{\partial x},
   \quad
    q_r^{NSF}
    =
    -\kappa
    \frac{\partial T}{\partial r},
\end{aligned}
\end{equation}
where $\nabla\cdot\boldsymbol{u}
    = \frac{\partial u_x}{\partial x}
    +
    \frac{\partial u_r}{\partial r}
    +
    \frac{u_r}{r}$, 
$\mu$ is the viscosity, and $\kappa$ is the thermal conductivity.
In this paper, the power-law viscosity is considered,
\begin{equation}
    \mu(T)
    =
    \mu_0
    \left(
        \frac{T}{T_0}
    \right)^{\omega_{\mu}},
\label{eq:viscosity_power_law}
\end{equation}
where $\mu_0$ is the reference viscosity at the reference temperature $T_0$, and $\omega_{\mu}$ is the viscosity index. The thermal conductivity is written as
\begin{equation}
    \kappa
    =
    \frac{5}{2}
    \frac{k_B}{m}
    \frac{\mu}{Pr}.
\label{eq:thermal_conductivity}
\end{equation}

The NSF constitutive relation is valid for small Knudsen numbers. The Knudsen number is defined as the ratio of the molecular mean free path $\lambda$ to the characteristic flow length $L$, as formulated below:
\begin{equation}
    Kn
    =
    \frac{\lambda}{L},
   \quad
    \lambda
    =
    \frac{\mu(T_0)}{n_0}
    \sqrt{
        \frac{\pi }{2mk_BT_0}
    },
\label{eq:knudsen_number}
\end{equation}
When the Knudsen number becomes appreciable, high-order corrective terms are required, which, in general flow conditions, can only be retrieved from the numerical solution of the kinetic equation \eqref{eq:axis_kinetic}.
Details will be given in Section \ref{subsec:The general synthetic iterative scheme} below.

\section{The numerical methods}\label{sec:3}

This section details the numerical framework implemented in AxiGSIS. The finite-volume solver for electric potential is briefly outlined in \ref{subsec:poisson_solver}.
The kinetic solver adopts a finite-volume discrete velocity formulation for 2D axisymmetric physical domains and 3D molecular velocity space. First, a velocity-space integration correction scheme is presented in \ref{subsec:velocity_space_integration_compensation}, which mitigates quadrature inconsistencies arising from finite, truncated velocity grids in the discrete velocity method. Second, to expedite steady-state convergence, the kinetic solver is coupled to axisymmetric macroscopic synthetic equations via the GSIS framework \cite{Su2020GSIS,Su2020SIAM}, whose details are elaborated in the subsequent text.

\subsection{Conventional iterative scheme for the kinetic equation}
\label{subsec:Conventional iterative scheme}

The conventional iterative scheme (CIS) is first introduced for solving the axisymmetric kinetic equation \eqref{eq:axis_kinetic}. In this scheme, the VDF is advanced directly by the finite-volume discrete velocity method, while the macroscopic quantities in the collision term are evaluated from the VDF at the previous iteration step. The CIS serves as the baseline kinetic solver and also provides the kinetic prediction step in the GSIS.

Let \(\Omega_i\) be the \(i\)-th physical control volume, with area \(|\Omega_i|\) and cell-center radius \(r_i\). The interface between \(\Omega_i\) and its neighboring cell \(\Omega_j\) is denoted by \(\Gamma_{ij}\), with length \(|\Gamma_{ij}|\), interface radius \(r_{ij}\), and outward unit normal vector \(\boldsymbol{n}_{ij}=(n_x,n_r)^T\) pointing from cell \(i\) to cell \(j\). The discrete velocity point is indexed by \(\alpha\), with cylindrical velocity-space coordinates \((v_{x,\alpha},\zeta_\alpha,\omega_\alpha)\). The corresponding radial and azimuthal molecular velocity components are
$v_{r,\alpha}=\zeta_\alpha\cos\omega_\alpha$
and $v_{\theta,\alpha}=\zeta_\alpha\sin\omega_\alpha.$
The normal molecular velocity at the edge \(\Gamma_{ij}\) is
\begin{equation}
v_{n,\alpha}
=
v_{x,\alpha}n_x
+
v_{r,\alpha}n_r
=
v_{x,\alpha}n_x
+
\zeta_\alpha\cos\omega_\alpha n_r .
\label{eq:normal_velocity}
\end{equation}

\subsubsection{Discretization of the free transport term}

% the conservative axisymmetric kinetic equation is
% \begin{equation}
% \frac{\partial(rv_x f)}{\partial x}
% +
% \frac{\partial(rv_r f)}{\partial r}.
% \end{equation}

After integrating over the control volume $\Omega_i$, the finite-volume discretization of the physical-space free streaming term in Eq.~\eqref{eq:axis_kinetic} is written as
\begin{equation}
\mathcal{T}^{x,r}_{i,\alpha}(f)
=
\frac{1}{|\Omega_i|}
\sum_{j\in N(i)}
r_{ij}
|\Gamma_{ij}|
v_{n,\alpha}
f_{ij,\alpha},
\label{eq:physical_transport}
\end{equation}
where $N(i)$ is the set of neighboring cells of $\Omega_i$, and $f_{ij,\alpha}$ is the VDF reconstructed at the interface $\Gamma_{ij}$. The factor $r_{ij}$ is introduced by the axisymmetric conservative formulation and represents the geometrical weight of the interface flux.

The physical-space interface value is evaluated by a second-order upwind reconstruction. The left and right states at $\Gamma_{ij}$, denoted by superscripts L and R respectively, are given by
\begin{equation}
\begin{aligned}
f_{ij,\alpha}^{L}
=
f_{i,\alpha}
+
\varphi_{i,\alpha}
\nabla_{\boldsymbol{x}} f_{i,\alpha}
\cdot
\left(
\boldsymbol{x}_{ij}
-
\boldsymbol{x}_{i}
\right),\\
f_{ij,\alpha}^{R}
=
f_{j,\alpha}
+
\varphi_{j,\alpha}
\nabla_{\boldsymbol{x}} f_{j,\alpha}
\cdot
\left(
\boldsymbol{x}_{ij}
-
\boldsymbol{x}_{j}
\right),
\end{aligned}
\end{equation}
where $\boldsymbol{x}_{ij}$ is the face center, $\nabla_{\boldsymbol{x}} f$ is obtained by the least-squares reconstruction, and $\varphi$ is the Venkatakrishnan limiter \cite{venkatakrishnan1993accuracy}. The second-order upwind interface value is then
\begin{equation}
f_{ij,\alpha}^{(2)}
=
\begin{cases}
f_{ij,\alpha}^{L},
& v_{n,\alpha}\ge 0,\\[3pt]
f_{ij,\alpha}^{R},
& v_{n,\alpha}<0 .
\end{cases}
\label{eq:physical_second_order_reconstruction}
\end{equation}
Consequently, the second-order physical-space transport operator is discretized as 
\begin{equation}
\mathcal{T}^{x,r,(2)}_{i,\alpha}(f)
=
\frac{1}{|\Omega_i|}
\sum_{j\in N(i)}
r_{ij}
|\Gamma_{ij}|
v_{n,\alpha}
f_{ij,\alpha}^{(2)} .
\label{eq:physical_second_order_operator}
\end{equation}
At boundary faces, the same upwind flux formulation is used, while the incoming VDF is determined by the corresponding kinetic boundary condition in Section \ref{sec:bc}. Specifically, inflow or far-field boundaries are prescribed by the Maxwellian associated with the given macroscopic state, outflow boundaries are treated by extrapolation from the interior cell, solid walls are described by the gas-surface interaction model, and the symmetry axis is treated by the specular symmetry with respect to the radial molecular velocity.

\subsubsection{Discretization of the acceleration terms}

%velocity-space acceleration term in the $(v_x,\zeta,\omega)$ coordinates reads
% \begin{equation}
% r
% \left[
% \frac{\partial(a_x f)}{\partial v_x}
% +
% \frac{1}{\zeta}
% \frac{\partial(\zeta a_\zeta f)}{\partial \zeta}
% +
% \frac{1}{\zeta}
% \frac{\partial(a_\omega f)}{\partial \omega}
% \right],
% \end{equation}

% For charged-particle flows under a prescribed electrostatic field, the external acceleration term is also discretized by the finite-volume method. Let $C_\alpha$ be the velocity-space control volume associated with the discrete velocity point $\alpha$, with volume $|C_\alpha|$. The interface between velocity cells $C_\alpha$ and $C_\beta$ is denoted by $\Sigma_{\alpha\beta}$, with area $|\Sigma_{\alpha\beta}|$ and outward unit normal vector $\boldsymbol{n}^{v}_{\alpha\beta}$ in velocity space. The normal acceleration at this velocity-space interface is
% \begin{equation}
% a_{n,\alpha\beta}
% =
% \boldsymbol{a}_{\alpha\beta}
% \cdot
% \boldsymbol{n}^{v}_{\alpha\beta}.
% \end{equation}

For charged-particle flows under a prescribed electrostatic field, the external acceleration term is also discretized by the finite-volume method in velocity space. Let \(C_\alpha\) be the velocity-space control volume associated with the discrete velocity point \(\alpha\), with volume \(|C_\alpha|\). The interface between velocity cells \(C_\alpha\) and \(C_\beta\) has area \(S_{\alpha\beta}^{v}\), and its outward unit normal vector in velocity space is denoted by \(\boldsymbol{n}_{\alpha\beta}^{v}\), pointing from \(C_\alpha\) to \(C_\beta\). The normal acceleration at this velocity-space interface is
\[
a_{n,\alpha\beta}
=
\boldsymbol{a}_{\alpha\beta}
\cdot
\boldsymbol{n}_{\alpha\beta}^{v}.
\]

% The external electric acceleration term in Eq.~\eqref{eq:axis_kinetic} is evaluated with a second-order upwind finite-volume discretization in velocity space:
% \begin{equation}
% \mathcal{T}_{i,\alpha}^{E,(2)}(f)
% =
% \frac{r_i}{|C_\alpha|}
% \sum_{\beta\in N_v(\alpha)}
% S_{\alpha\beta}^{v}
% a_{n,\alpha\beta}
% f_{i,\alpha\beta}^{v,(2)},
% \end{equation}
% where \(N_v(\alpha)\) is the set of neighboring velocity cells of \(C_\alpha\).
% The left and right reconstructed values at the velocity-space face are defined as
The external electric acceleration term in Eq.~\eqref{eq:axis_kinetic} is evaluated with a second-order upwind finite-volume discretization in velocity space. 
For each velocity-space control volume \(C_\alpha\), let \(N_v(\alpha)\) be the set of its neighboring velocity cells. 
The reconstructed values on the two sides of the velocity-space face between \(C_\alpha\) and \(C_\beta\) are defined as
\begin{equation}
\begin{aligned}
f_{i,\alpha\beta}^{L}
&=
f_{i,\alpha}
+
\varphi_{i,\alpha}^{v}
\nabla_v f_{i,\alpha}
\cdot
\left(
\boldsymbol{v}_{\alpha\beta}
-
\boldsymbol{v}_{\alpha}
\right),\\
f_{i,\alpha\beta}^{R}
&=
f_{i,\beta}
+
\varphi_{i,\beta}^{v}
\nabla_v f_{i,\beta}
\cdot
\left(
\boldsymbol{v}_{\alpha\beta}
-
\boldsymbol{v}_{\beta}
\right).
\end{aligned}
\end{equation}
Here, \(L\) and \(R\) denote the two sides of the velocity-space face, corresponding to \(C_\alpha\) and \(C_\beta\), respectively.
The upwind interface value is then
\begin{equation}
f_{i,\alpha\beta}^{v,(2)}
=
\begin{cases}
f_{i,\alpha\beta}^{L},
& a_{n,\alpha\beta}\ge 0,\\
f_{i,\alpha\beta}^{R},
& a_{n,\alpha\beta}<0 .
\end{cases}
\end{equation}
Consequently, the second-order electric acceleration operator in velocity space is discretized as
\begin{equation}
\mathcal{T}_{i,\alpha}^{E,(2)}(f)
=
\frac{r_i}{|C_\alpha|}
\sum_{\beta\in N_v(\alpha)}
S_{\alpha\beta}^{v}
a_{n,\alpha\beta}
f_{i,\alpha\beta}^{v,(2)} .
\end{equation}

In addition to the electric acceleration term, the axisymmetric kinetic equation~\eqref{eq:axis_kinetic} incorporates a geometric velocity-space transport term. This term originates from the cylindrical coordinate transformation of transverse molecular velocity components and accounts for angular transport within the $(v_r,v_\theta)$ velocity plane. To preserve the positivity, conservation properties, and uniform-flow consistency of the axisymmetric transport operator, this term is discretized by the trigonometric upwind conservative (T-UCE) scheme~\cite{mieussens2000dvm}. That is, at physical cell $i$ and velocity point $(v_{x,k},\zeta_l,\omega_\gamma)$, the axisymmetric geometrical term is discretized as
\begin{equation}\label{eq:geo_tuce_operator}
\begin{aligned}
\mathcal{T}^{geo}_{i,k,l,\gamma}(f)
=&
-
\zeta_l
D_{\omega}^{T\text{-}UCE}
\left(
\sin\omega f
\right)_{i,k,l,\gamma}\\
=&
-
\frac{\zeta_l}{2\sin(\Delta\omega/2)}
\left(
H_{i,k,l,\gamma+1/2}
-
H_{i,k,l,\gamma-1/2}
\right),
\end{aligned}
\end{equation}
where
\begin{equation}
\begin{aligned}
H_{i,k,l,\gamma+1/2}
=
\left(
\sin\omega_{\gamma+1/2}
\right)^+
f_{i,k,l,\gamma+1}
+
\left(
\sin\omega_{\gamma+1/2}
\right)^-
f_{i,k,l,\gamma},\\
H_{i,k,l,\gamma-1/2}
=
\left(
\sin\omega_{\gamma-1/2}
\right)^+
f_{i,k,l,\gamma}
+
\left(
\sin\omega_{\gamma-1/2}
\right)^-
f_{i,k,l,\gamma-1}. 
\end{aligned}
\end{equation}
Note that in the above equation, the angular velocity grid is discretized uniformly into $N_\omega$ points:
\begin{equation}
   \omega_\gamma=
\gamma\Delta\omega, \quad \text{with} \quad 
\gamma=0,1,\cdots,N_\omega-1.  
\end{equation}
The angular interfaces are
$
\omega_{\gamma+1/2}
=
\omega_\gamma
+
\frac{\Delta\omega}{2}$ and $
\omega_{\gamma-1/2}
=
\omega_\gamma
-
\frac{\Delta\omega}{2}$.
Here, the superscripts $+$ and $-$ denote the positive and negative
parts, respectively:
\[
\left(\sin\omega_{\gamma+\delta}\right)^{\pm}
=
\frac{1}{2}
\left[
\sin\omega_{\gamma+\delta}
\pm
\left|\sin\omega_{\gamma+\delta}\right|
\right],
\qquad
\delta=\pm\frac{1}{2}.
\]

% Also, we define $a^+
% =
% \frac{1}{2}
% \left(
% a+|a|
% \right)$ and $
% a^-
% =
% \frac{1}{2}
% \left(
% a-|a|
% \right)$ for a scalar $a$.

The essential feature of the T-UCE scheme is that the angular spacing $\Delta\omega$ is replaced by the trigonometric correction factor $2\sin(\Delta\omega/2)$. This correction gives
$
D_{\omega}^{T\text{-}UCE}
\left(
\sin\omega
\right)_\gamma
=
\cos\omega_\gamma$,
which is required to preserve uniform flows in the discrete axisymmetric transport equation. In addition, the upwind construction based on the sign of $\sin\omega_{\gamma+1/2}$ preserves the positivity of the VDF. The T-UCE discretization also satisfies the discrete conservation of mass, axial momentum and energy, and provides entropy dissipation for the conservative-form axisymmetric transport operator.

The angular velocity direction is periodic, so that $f_{i,k,l,-1}
=
f_{i,k,l,N_\omega-1}$ and  $
f_{i,k,l,N_\omega}
=
f_{i,k,l,0}$.

\subsubsection{Pseudo-time stepping}

Combining the above discretizations, the CIS over one pseudo-time step from iteration $k$ to $k+1$ is written in the delta form:
\begin{equation}
r_i
\frac{
\Delta f_{i,\alpha}
}{
\Delta t
}
+
\frac{r_i}{\tau_i^k}
\Delta f_{i,\alpha}
+
\mathcal{T}^{x,r,(1)}_{i,\alpha}
\left(
\Delta f
\right)
+
\mathcal{T}^{E,(1)}_{i,\alpha}
\left(
\Delta f
\right)
+
\mathcal{T}^{geo,(1)}_{i,\alpha}
\left(
\Delta f
\right)
=
R_{i,\alpha}^{k},
\label{eq:cis_delta_form}
\end{equation}
where $\Delta f_{i,\alpha}=f_{i,\alpha}^{k+1}-f_{i,\alpha}^{k}$, and the right-hand-side residual is evaluated from the VDF at the current iteration step:
\begin{equation}
R_{i,\alpha}^{k}
=
r_i
\frac{
g_{i,\alpha}^{S,k}
-
f_{i,\alpha}^{k}
}{
\tau_i^k
}
-
\mathcal{T}^{x,r,(2)}_{i,\alpha}
\left(
f^k
\right)
-
\mathcal{T}^{E,(2)}_{i,\alpha}
\left(
f^k
\right)
-
\mathcal{T}^{geo}_{i,\alpha}
\left(
f^k
\right).
\label{eq:cis_residual}
\end{equation}
Here $g_{i,\alpha}^{S,k}$ is the Shakhov target distribution constructed from the macroscopic variables at the $k$-th iteration step, and $\tau_i^k$ is the corresponding relaxation time. 

% It should be emphasized that the residual $R_{i,\alpha}^{k}$ determines the discretization accuracy of the final steady-state solution. Therefore, the physical-space transport term and the electric velocity-space transport term are evaluated by second-order upwind reconstructions in the residual. In contrast, the left-hand side of Eq.~\eqref{eq:cis_delta_form} is used only to accelerate the pseudo-time convergence. The increment fluxes in $\mathcal{T}^{x,r,(1)}$, $\mathcal{T}^{E,(1)}$, and $\mathcal{T}^{geo,(1)}$ are approximated by first-order upwind schemes, 
% \leir{you did not introduce the first-order scheme}. Since $\Delta f_{i,\alpha}\rightarrow 0$ after convergence, the first-order approximation in the implicit operator does not affect the accuracy of the final converged solution.

It should be emphasized that the residual \(R_{i,\alpha}^k\) determines the discretization accuracy of the final steady-state solution. Therefore, the physical-space transport term and the electric velocity-space transport term are evaluated by second-order upwind reconstructions in the residual.

In contrast, the left-hand side of Eq~\eqref{eq:cis_delta_form} is used only to accelerate the pseudo-time convergence. 
The transport terms in the implicit operator are approximated by first-order upwind schemes for the increment \(\Delta f\).  Since \(\Delta f_{i,\alpha}\rightarrow 0\) after convergence, the first-order approximation in the implicit operator does not affect the accuracy of the final converged solution.
Specifically, the first-order physical-space transport operator is written as
\[
\mathcal{T}_{i,\alpha}^{x,r,(1)}(\Delta f)
=
\frac{1}{|\Omega_i|}
\sum_{j\in N(i)}
r_{ij}|\Gamma_{ij}|
v_{n,\alpha}
\Delta f_{ij,\alpha}^{(1)},
\]
where the face value of the increment is taken from the upwind cell:
\[
\Delta f_{ij,\alpha}^{(1)}
=
\begin{cases}
\Delta f_{i,\alpha}, & v_{n,\alpha}\ge 0,\\
\Delta f_{j,\alpha}, & v_{n,\alpha}<0 .
\end{cases}
\]
Similarly, the first-order electric acceleration operator in velocity space is written as
\[
\mathcal{T}_{i,\alpha}^{E,(1)}(\Delta f)
=
\frac{r_i}{|C_\alpha|}
\sum_{\beta\in N_v(\alpha)}
S_{\alpha\beta}^{v}
a_{n,\alpha\beta}
\Delta f_{i,\alpha\beta}^{v,(1)},
\]
where
\[
\Delta f_{i,\alpha\beta}^{v,(1)}
=
\begin{cases}
\Delta f_{i,\alpha}, & a_{n,\alpha\beta}\ge 0,\\
\Delta f_{i,\beta}, & a_{n,\alpha\beta}<0 .
\end{cases}
\]
%Here \(S_{\alpha\beta}^{v}\) is the area of the velocity-space face between \(C_\alpha\) and \(C_\beta\), and \(a_{n,\alpha\beta}\) determines the upwind direction in velocity space.
The cylindrical geometrical term in the implicit operator is treated by applying the same T-UCE discretization introduced above to the increment \(\Delta f\), namely,
$
\mathcal{T}_{i,\alpha}^{\mathrm{geo},(1)}(\Delta f)
=
\mathcal{T}_{i,\alpha}^{\mathrm{geo}}(\Delta f)$.
%Therefore, the implicit operator uses first-order upwind increments only to enhance numerical stability and convergence.

After solving the increment equation, the VDF is updated by $f_{i,\alpha}^{k+1}=f_{i,\alpha}^{k}+\Delta f_{i,\alpha}.$
The macroscopic variables are then recomputed from the updated VDF and used to construct the  Shakhov distribution \eqref{eq:shakhov} for the next iteration. This procedure is repeated until the relative change of the macroscopic quantities between two consecutive iterations satisfies the prescribed convergence criterion.

The CIS is usually efficient in transition and free-molecular flow regimes. However, in the near-continuum regime with small Knudsen numbers, the evolution of the VDF is strongly constrained by slowly varying macroscopic modes \cite{wang2018comparative}. Since the gain term of the collision operator in the CIS is constructed from the macroscopic variables at the previous iteration step, macroscopic information is propagated mainly through kinetic transport and collision relaxation. This leads to slow convergence in the near-continuum regime and motivates the introduction of the GSIS described in the following subsection.

\subsection{The general synthetic iterative scheme}
\label{subsec:The general synthetic iterative scheme}

To accelerate the convergence toward steady state, the GSIS is introduced. The key idea is to use the kinetic equation to provide high-order non-equilibrium information, while the low-order macroscopic quantities are updated by solving the macroscopic synthetic equations. In this way, macroscopic flow information can be propagated much faster throughout the computational domain, thereby substantially accelerating convergence toward the steady state.
In our prior studies, macroscopic synthetic equations are implemented via two distinct strategies. For deterministic solvers, these equations are typically invoked upon completion of each kinetic solve cycle \cite{Su2020GSIS,Zhang2024CaF}. For stochastic DSMC solvers, by contrast, the synthetic correction is enforced every 100 DSMC evolution steps \cite{luo2024directIntermittentDIG}, a practice necessitated by the inherent statistical noise of DSMC simulations, which requires averaging to suppress fluctuations. In the current work, macroscopic synthetic equations are applied every three successive CIS iterations. This configuration mitigates numerical instabilities that may arise from direct coupling with macroscopic synthetic equations when geometric or external acceleration terms are present.

Let $f^{k,0}=f^k$. For $s=0,1,2$, the macroscopic variables are first evaluated from $f^{k,s}$, and the corresponding reference VDF and relaxation time are constructed. The CIS increment equation is then solved to update the VDF as
\begin{equation}
f_{i,\alpha}^{k,s+1}
=
f_{i,\alpha}^{k,s}
+
\Delta f_{i,\alpha}^{k,s}.
\end{equation}
After these three CIS iterations, the predicted VDF used for the synthetic correction is defined as
\begin{equation}
f_{i,\alpha}^{k+1/2}
=
f_{i,\alpha}^{k,3}.
\end{equation}
The superscript $k+1/2$ indicates that this VDF has been advanced by the kinetic solver but has not yet been corrected by the macroscopic synthetic equations.

The conservative variables predicted from the three-step CIS solution $f^{k+1/2}$ are denoted by
$\boldsymbol{W}^{k+1/2}=(\rho,\rho u_x,\rho u_r,\rho u_\theta,\rho E)^T$. Meanwhile, the stress and heat flux are directly computed from the kinetic solution. The high-order terms (HoTs) are defined as the difference between the kinetic non-equilibrium moments and their NSF approximations:
\begin{equation}
\boldsymbol{\Pi}^{\mathrm{HoT},k+1/2}
=
\boldsymbol{\Pi}^{k+1/2}
-
\boldsymbol{\Pi}^{\mathrm{NSF},k+1/2},
\qquad
\boldsymbol{q}^{\mathrm{HoT},k+1/2}
=
\boldsymbol{q}^{k+1/2}
-
\boldsymbol{q}^{\mathrm{NSF},k+1/2}.
\label{eq:hot_terms}
\end{equation}
Equivalently, the kinetic parts in Eq.~\eqref{eq:hot_terms} are evaluated as
\begin{equation}
\boldsymbol{\Pi}^{k+1/2}
=
m
\int
\left(
\boldsymbol{c}\boldsymbol{c}
-
\frac{1}{3}
|\boldsymbol{c}|^2
\boldsymbol{I}
\right)
f^{k+1/2}
d\boldsymbol{v},
\qquad
\boldsymbol{q}^{k+1/2}
=
\frac{m}{2}
\int
|\boldsymbol{c}|^2
\boldsymbol{c}
f^{k+1/2}
d\boldsymbol{v}.
\end{equation}
The NSF stress and heat flux are evaluated from $\boldsymbol{W}^{k+1/2}$ using the constitutive relations given in Eq.~\eqref{eq:nsf_constitutive_relations}. During the subsequent macroscopic iteration, the HoTs are fixed and used as source corrections to the NSF equations.

The macroscopic synthetic equations are constructed by replacing the stress tensor and heat flux in the exact macroscopic equations with their NSF parts plus the HoTs. Eventually, Eqs.~\eqref{eq:mass_equation} to \eqref{eq:energy_equation} can be written in the following compact form:
\begin{equation}
\frac{\partial \boldsymbol{W}}{\partial t}
+
\frac{\partial \boldsymbol{F}_{x}^{\mathrm{NSF}}}{\partial x}
+
\frac{1}{r}
\frac{\partial}{\partial r}
\left(
r\boldsymbol{F}_{r}^{\mathrm{NSF}}
\right)
=
\boldsymbol{S}^{E}
+
\boldsymbol{S}^{\mathrm{axi,NSF}}
+
\boldsymbol{R}^{\mathrm{HoT}},
\label{eq:synthetic_equations}
\end{equation}
where the NSF fluxes are
\begin{equation}
\boldsymbol{F}_{x}^{\mathrm{NSF}}
=
\left(
\begin{array}{c}
\rho u_x \\
\rho u_x^2+p+\Pi_{xx}^{\mathrm{NSF}} \\
\rho u_xu_r+\Pi_{xr}^{\mathrm{NSF}} \\
\rho u_xu_\theta+\Pi_{x\theta}^{\mathrm{NSF}} \\
(\rho E+p)u_x+\Pi_{xx}^{\mathrm{NSF}}u_x+\Pi_{xr}^{\mathrm{NSF}}u_r+\Pi_{x\theta}^{\mathrm{NSF}}u_\theta+q_x^{\mathrm{NSF}}
\end{array}
\right),
\end{equation}
and
\begin{equation}
\boldsymbol{F}_{r}^{\mathrm{NSF}}
=
\left(
\begin{array}{c}
\rho u_r \\
\rho u_xu_r+\Pi_{xr}^{\mathrm{NSF}} \\
\rho u_r^2+p+\Pi_{rr}^{\mathrm{NSF}} \\
\rho u_ru_\theta+\Pi_{r\theta}^{\mathrm{NSF}} \\
(\rho E+p)u_r+\Pi_{xr}^{\mathrm{NSF}}u_x+\Pi_{rr}^{\mathrm{NSF}}u_r+\Pi_{r\theta}^{\mathrm{NSF}}u_\theta+q_r^{\mathrm{NSF}}
\end{array}
\right).
\end{equation}
The electrostatic force and work terms are included through
\begin{equation}
\boldsymbol{S}^{E}
=
\left(
0,\,
q_pnE_x,\,
q_pnE_r,\,
q_pnE_\theta,\,
q_pn(u_xE_x+u_rE_r+u_\theta E_\theta)
\right)^T .
\end{equation}
For the axisymmetric electrostatic field considered in this work, $E_\theta=0$. For neutral gases, $q_p=0$, and hence $\boldsymbol{S}^{E}=0$.

The geometrical source term in Eq.~\eqref{eq:synthetic_equations} is
\begin{equation}
\boldsymbol{S}^{\mathrm{axi,NSF}}
=
\left(
0,\,
0,\,
\frac{\rho u_\theta^2+p+\Pi_{\theta\theta}^{\mathrm{NSF}}}{r},\,
-\frac{\rho u_ru_\theta+\Pi_{r\theta}^{\mathrm{NSF}}}{r},\,
0
\right)^T .
\end{equation}
This term comes from the cylindrical-coordinate form of the radial and azimuthal momentum equations. It is essential for swirling axisymmetric flows, such as the Taylor--Couette flow, and also retains the contribution of the circumferential normal stress in non-swirling flows.

The HoT source term in Eq.~\eqref{eq:synthetic_equations} is given by
\begin{equation}
\boldsymbol{R}^{\mathrm{HoT}}
=
-
\frac{\partial \boldsymbol{F}_{x}^{\mathrm{HoT}}}{\partial x}
-
\frac{1}{r}
\frac{\partial}{\partial r}
\left(
r\boldsymbol{F}_{r}^{\mathrm{HoT}}
\right)
+
\boldsymbol{S}^{\mathrm{axi,HoT}},
\end{equation}
where
\begin{equation}
\boldsymbol{F}_{x}^{\mathrm{HoT}}
=
\left(
\begin{array}{c}
0 \\
\Pi_{xx}^{\mathrm{HoT}} \\
\Pi_{xr}^{\mathrm{HoT}} \\
\Pi_{x\theta}^{\mathrm{HoT}} \\
\Pi_{xx}^{\mathrm{HoT}}u_x+\Pi_{xr}^{\mathrm{HoT}}u_r+\Pi_{x\theta}^{\mathrm{HoT}}u_\theta+q_x^{\mathrm{HoT}}
\end{array}
\right),
\end{equation}
and
\begin{equation}
\boldsymbol{F}_{r}^{\mathrm{HoT}}
=
\left(
\begin{array}{c}
0 \\
\Pi_{xr}^{\mathrm{HoT}} \\
\Pi_{rr}^{\mathrm{HoT}} \\
\Pi_{r\theta}^{\mathrm{HoT}} \\
\Pi_{xr}^{\mathrm{HoT}}u_x+\Pi_{rr}^{\mathrm{HoT}}u_r+\Pi_{r\theta}^{\mathrm{HoT}}u_\theta+q_r^{\mathrm{HoT}}
\end{array}
\right).
\end{equation}
The corresponding axisymmetric HoT source is
\begin{equation}
\boldsymbol{S}^{\mathrm{axi,HoT}}
=
\left(
0,\,
0,\,
\frac{\Pi_{\theta\theta}^{\mathrm{HoT}}}{r},\,
-\frac{\Pi_{r\theta}^{\mathrm{HoT}}}{r},\,
0
\right)^T .
\end{equation}
When the iteration converges, the sum of the NSF part and the HoT part recovers the stress tensor and heat flux directly obtained from the kinetic equation. Therefore, the macroscopic synthetic equations are consistent with the kinetic moment equations at steady state.

The synthetic equations are solved on the same finite-volume mesh as the kinetic equation. Let $\boldsymbol{W}_i^\ell$ be the macroscopic conservative variables at the $\ell$-th inner iteration, with $\boldsymbol{W}_i^0=\boldsymbol{W}_i^{k+1/2}$. The finite-volume residual of Eq.~\eqref{eq:synthetic_equations} is evaluated as
\begin{equation}
\boldsymbol{R}_{i}^{M,\ell}
=
-
\sum_{j\in N(i)}
r_{ij}
|\Gamma_{ij}|
\boldsymbol{F}_{n,ij}^{\mathrm{NSF},\ell}
+
r_i|\Omega_i|
\left(
\boldsymbol{S}_{i}^{E,\ell}
+
\boldsymbol{S}_{i}^{\mathrm{axi,NSF},\ell}
+
\boldsymbol{R}_{i}^{\mathrm{HoT}}
\right),
\end{equation}
where $\boldsymbol{F}_{n,ij}^{\mathrm{NSF}}=\boldsymbol{F}_{x,ij}^{\mathrm{NSF}}n_x+\boldsymbol{F}_{r,ij}^{\mathrm{NSF}}n_r$ is the normal macroscopic flux at the interface $\Gamma_{ij}$. The flux is computed by the Rusanov scheme:
\begin{equation}
\boldsymbol{F}_{n,ij}^{\mathrm{NSF}}
=
\frac{1}{2}
\left[
\boldsymbol{F}_{n}^{\mathrm{NSF}}(\boldsymbol{W}_{ij}^{L})
+
\boldsymbol{F}_{n}^{\mathrm{NSF}}(\boldsymbol{W}_{ij}^{R})
\right]
-
\frac{1}{2}
\Lambda_{ij}
\left(
\boldsymbol{W}_{ij}^{R}
-
\boldsymbol{W}_{ij}^{L}
\right),
\end{equation}
where the left and right states are obtained by the same second-order reconstruction strategy as used in the kinetic solver. The numerical flux is evaluated in a Rusanov-type form~\cite{RUSANOV1962304}, with the dissipation coefficient chosen as
\begin{equation}
\Lambda_{ij}
=
|u_{n,ij}|
+
c_{s,ij}
+
\frac{2\mu_{ij}}{\rho_{ij}d_{ij}},
\end{equation}
where $u_{n,ij}$ is the normal macroscopic velocity, $c_{s,ij}$ is the local sound speed, and $d_{ij}$ is the distance between the two cell centers in the normal direction. The last term accounts for the viscous contribution to the spectral radius.

A point-implicit iteration is used to update the macroscopic conservative variables. The increment $\Delta\boldsymbol{W}_i^\ell=\boldsymbol{W}_i^{\ell+1}-\boldsymbol{W}_i^\ell$ is solved from
\begin{equation}
\left(
\frac{r_i|\Omega_i|}{\Delta t_M}
+
D_i^\ell
\right)
\Delta\boldsymbol{W}_i^\ell
=
\boldsymbol{R}_{i}^{M,\ell},
\label{eq:macro_implicit_update}
\end{equation}
where $\Delta t_M$ is the pseudo-time step of the macroscopic solver, and $D_i^\ell$ is the local diagonal coefficient assembled from the convective, viscous, and source-term contributions. After each inner iteration, the macroscopic variables are updated by
\begin{equation}
\boldsymbol{W}_i^{\ell+1}
=
\boldsymbol{W}_i^\ell
+
\Delta\boldsymbol{W}_i^\ell .
\end{equation}
The macroscopic inner iteration is terminated when the prescribed residual tolerance is reached or when the maximum number of 1000 inner iterations is attained. The resulting macroscopic state is denoted by $\boldsymbol{W}^{k+1}$.

Note that conventional NSF equations enforce no-slip velocity and zero temperature-jump boundary conditions, whereas the synthetic governing equation necessitates velocity-slip and temperature-jump boundary conditions aligned with the kinetic equation. This requirement is fulfilled by implementing the consistent boundary formulations presented in \ref{app:gsis_boundary_treatment}.

Finally, the VDF is corrected using the macroscopic solution. The correction keeps the non-equilibrium part of $f^{k+1/2}$ and replaces only the equilibrium part associated with the updated macroscopic variables:
\begin{equation}
f_{i,\alpha}^{k+1}
=
f_{i,\alpha}^{k+1/2}
+
\eta_f
\left[
g_{\alpha}^{M}(\boldsymbol{W}_{i}^{k+1})
-
g_{\alpha}^{M}(\boldsymbol{W}_{i}^{k+1/2})
\right],
\label{eq:gsis_vdf_correction}
\end{equation}
where $g^{M}_{\alpha}$ denotes the Maxwellian distribution evaluated at discrete velocity node $\alpha$, and $\eta_f$ is a relaxation coefficient fixed at 0.2 across all numerical cases considered herein. This correction feeds rapidly updated macroscopic information back into the kinetic solution whilst retaining the high-order nonequilibrium structure predicted by the kinetic equation. The corrected macroscopic variables are then used to construct the reference VDF \eqref{eq:shakhov} in the next kinetic iteration.

The convergence of the GSIS is monitored by the relative change of the macroscopic quantities between two consecutive GSIS outer iterations. Here, the index $k$ denotes the outer GSIS iteration after the three CIS prediction steps and the macroscopic synthetic correction. In this work, the residual is defined as
\begin{equation}
\epsilon^k
=
\max_{\chi\in\{\rho,u_x,u_r,u_\theta,T\}}
\left[
\frac{
\sum_i
r_i|\Omega_i|
\left(
\chi_i^k-\chi_i^{k-1}
\right)^2
}{
\sum_i
r_i|\Omega_i|
\left(
\chi_i^{k-1}
\right)^2
}
\right]^{1/2}.
\end{equation}
The iteration is terminated when $\epsilon^k$ is smaller than the prescribed tolerance. Since the macroscopic synthetic equations provide fast propagation of the low-order moments, the GSIS significantly accelerates the steady-state convergence. At convergence, $\Delta f\rightarrow 0$ and the HoT-corrected synthetic equations are consistent with the original kinetic moment equations, so that the converged solution remains the same as that of the kinetic model.

\subsection{Overview of the GSIS}
\label{subsec:Overview of the GSIS}

The overall AxiGSIS procedure is summarized as follows:
\begin{enumerate}
\item Generate the physical mesh and the discrete velocity grid.
\item Initialize the macroscopic variables and construct the initial VDF \(f^0\).
\item For charged-particle flows, solve the electrostatic potential equation and compute the prescribed electric field.
\item At the \(k\)-th GSIS outer iteration, set \(f^{k,0}=f^k\).
\item Perform three CIS iterations to obtain the predicted VDF \(f^{k+1/2}=f^{k,3}\).
\item Compute the predicted macroscopic variables, stress tensor, and heat flux from \(f^{k+1/2}\).
\item Evaluate the HoT corrections and solve the macroscopic synthetic equations.
\item Correct the VDF using the updated macroscopic variables to obtain \(f^{k+1}\).
\item Check the convergence criterion. If it is not satisfied, repeat the GSIS outer iteration with the updated VDF; otherwise, terminate the iteration.
\end{enumerate}

\subsection{Parallel computing strategy}

All simulations are performed on a high-performance computing cluster, where the computational nodes are equipped with Intel(R) Xeon(R) Gold 6140 CPUs @ 2.30 GHz. The parallel implementation is based on MPI. The unstructured physical mesh in the 2D meridian plane is partitioned into several subdomains using METIS, and each MPI process is assigned a subset of physical cells. The partition is constructed to balance the number of cells among processes while reducing the number of inter-process interfaces. 

Both the kinetic equation and macroscopic synthetic equations are solved on the same domain decomposition. Each MPI process stores the conservative variables, macroscopic variables, and the VDFs for its owned cells. To evaluate fluxes across subdomain interfaces, ghost cells are introduced for neighboring subdomains. During the CIS iterations, the VDFs in ghost cells are exchanged between neighboring MPI processes. Similarly, in the macroscopic synthetic solver, the macroscopic variables, gradients, high-order terms, and increments are exchanged through the same ghost-cell communication framework. 

The discrete velocity space is stored locally for each owned physical cell. Unlike the parallel strategy adopted in Ref.~\cite{Zhang2024CaF}, we avoid distributing the velocity space across multiple computational processes for two reasons: first, parallelizing acceleration terms severely degrades parallel efficiency; second, the total memory overhead remains moderate under axisymmetric configurations, making velocity-space domain decomposition unnecessary. Hence, the velocity-space quadrature, collision term, geometrical transport term, and electrostatic acceleration term are evaluated independently within each subdomain. The main communication cost comes from the exchange of ghost-cell data across physical subdomain interfaces. For charged-particle cases, the electrostatic potential equation is also solved on the same distributed physical mesh, and the computed electric field is stored locally for the subsequent kinetic and macroscopic updates.

%The convergence errors are evaluated from global residuals obtained by MPI reductions over all subdomains. The parallel output is gathered according to the global cell numbering for post-processing. 
%In the cost comparisons reported in Sec.~4, the computational cost is measured by core-hours, defined as the product of the wall-clock time and the number of CPU cores. This metric allows a fair comparison between AxiGSIS and CIS when different numbers of cores are used.

%\linenumbers

\section{Numerical results}\label{sec:4}

Numerical simulations are carried out to evaluate the accuracy, convergence characteristics, and computational efficiency of the proposed AxiGSIS solver.

Throughout this work, argon is used as the working species, being treated as neutral argon in the neutral-flow cases and as singly charged argon ions in the charged-particle cases. The viscosity index is set to $\omega_\mu=0.81$. Unless otherwise stated, the global convergence tolerance for both CIS and AxiGSIS is set to $\epsilon_{\mathrm{K}}=10^{-5}$. In AxiGSIS, the macroscopic inner iteration is terminated when its relative residual falls below $\epsilon_{\mathrm{M}}=10^{-5}$ or when the maximum number of 1000
inner iterations is reached. For the charged-particle cases, the electrostatic potential equation is solved until the relative change in the potential falls below $\epsilon_{\varphi}=10^{-10}$.

Four representative benchmark cases are investigated as follows.
First, the Taylor–Couette flow is adopted to validate the method’s ability to capture rarefied gas dynamics with substantial azimuthal velocity components.
Second, neutral rarefied gas flow through an axisymmetric nozzle is simulated to verify the implemented axisymmetric kinetic framework and benchmark the computational efficiency of AxiGSIS against its full 3D GSIS counterpart.
Third, charged-particle flow around an electrostatic sphere is considered to test the solver’s capacity to handle particle transport driven by imposed electrostatic fields.
Fourth, axisymmetric nozzle flow of charged particles is computed to further demonstrate the scheme’s applicability to internal charged-particle flows with a pressure inlet, a vacuum outlet, and prescribed electrostatic forcing.

In the following convergence comparisons, the iteration number of AxiGSIS denotes the number of GSIS outer iterations. At the beginning of each GSIS outer iteration, three CIS iterations are first performed to obtain a kinetic prediction. The macroscopic synthetic equations are then solved using the high-order non-equilibrium information extracted from this predicted distribution function, and the VDF is subsequently corrected by the updated macroscopic variables. For CIS, the iteration number denotes the number of conventional kinetic iterations.

% \leir{At the beginning of the $k$-th GSIS outer iteration, several CISs are first performed to obtain a kinetic prediction. ....}

\subsection{Taylor--Couette flow}

\begin{figure}[h]
    \centering
    \subfloat[Computational setup]{
        \includegraphics[height=0.23\textheight]{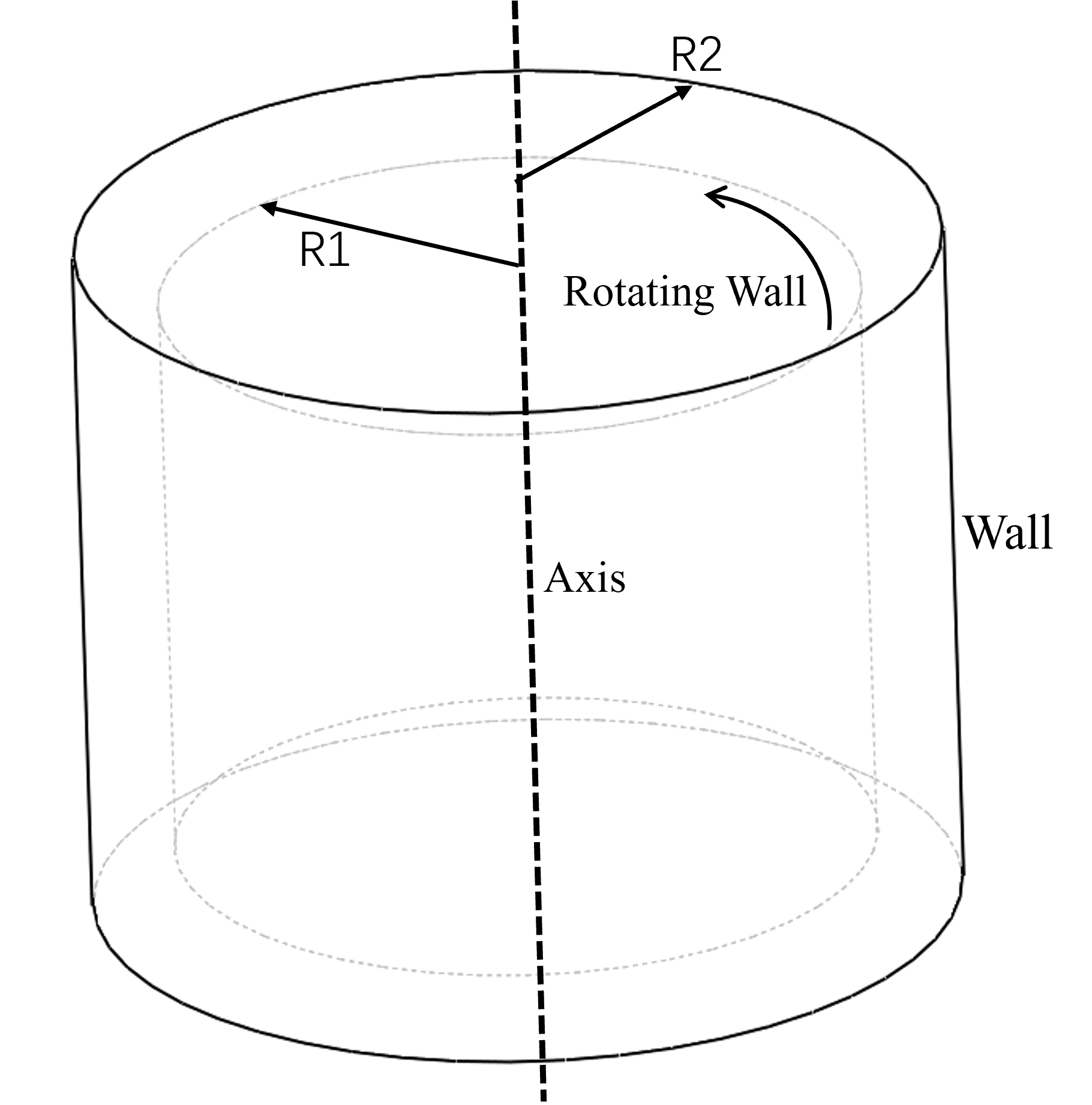}
    }
    \hfill
    \subfloat[Computational mesh]{
        \includegraphics[height=0.23\textheight]{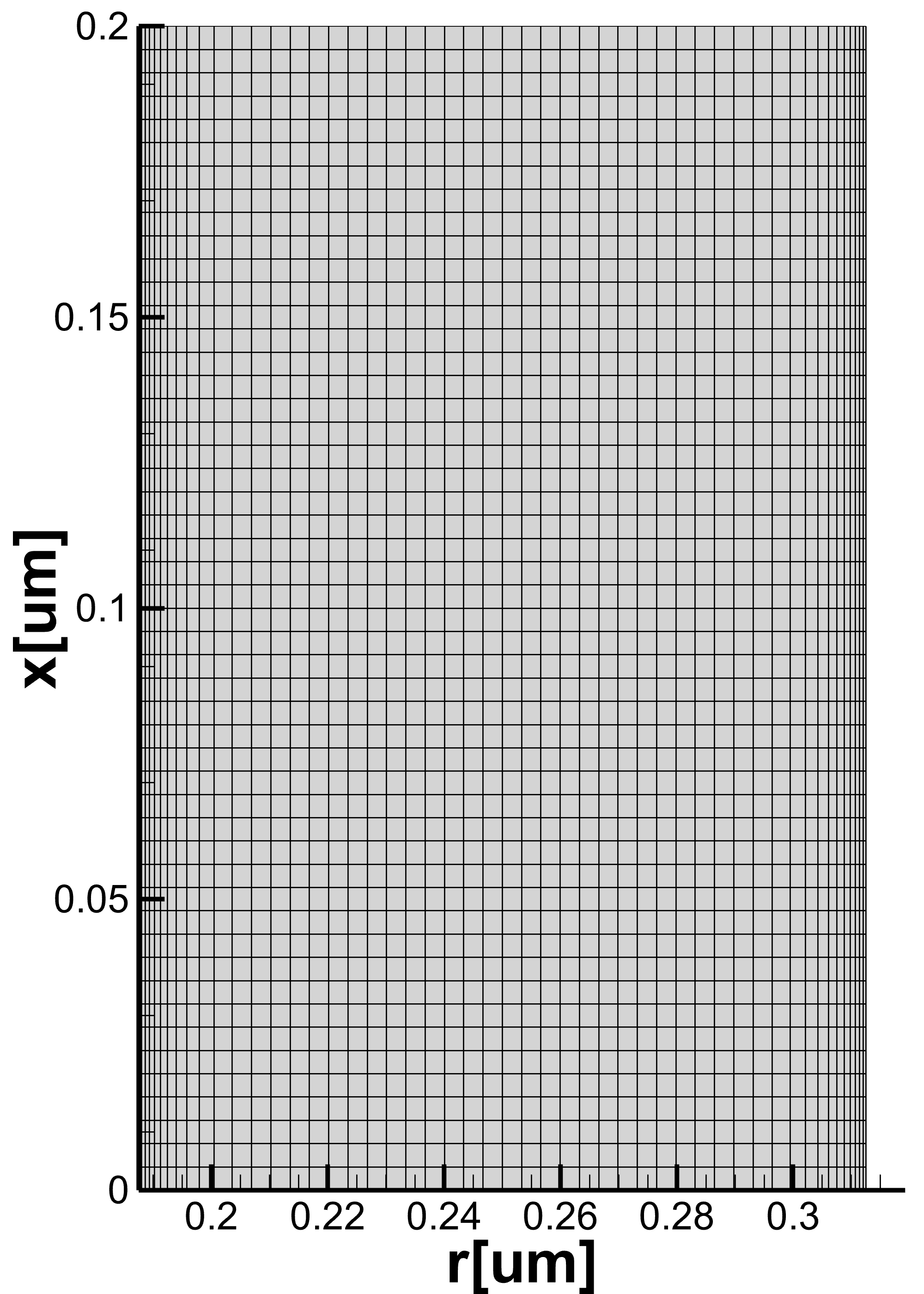}
    }
    \hfill
    \subfloat[Velocity-profile comparison]{
        \includegraphics[height=0.23\textheight]{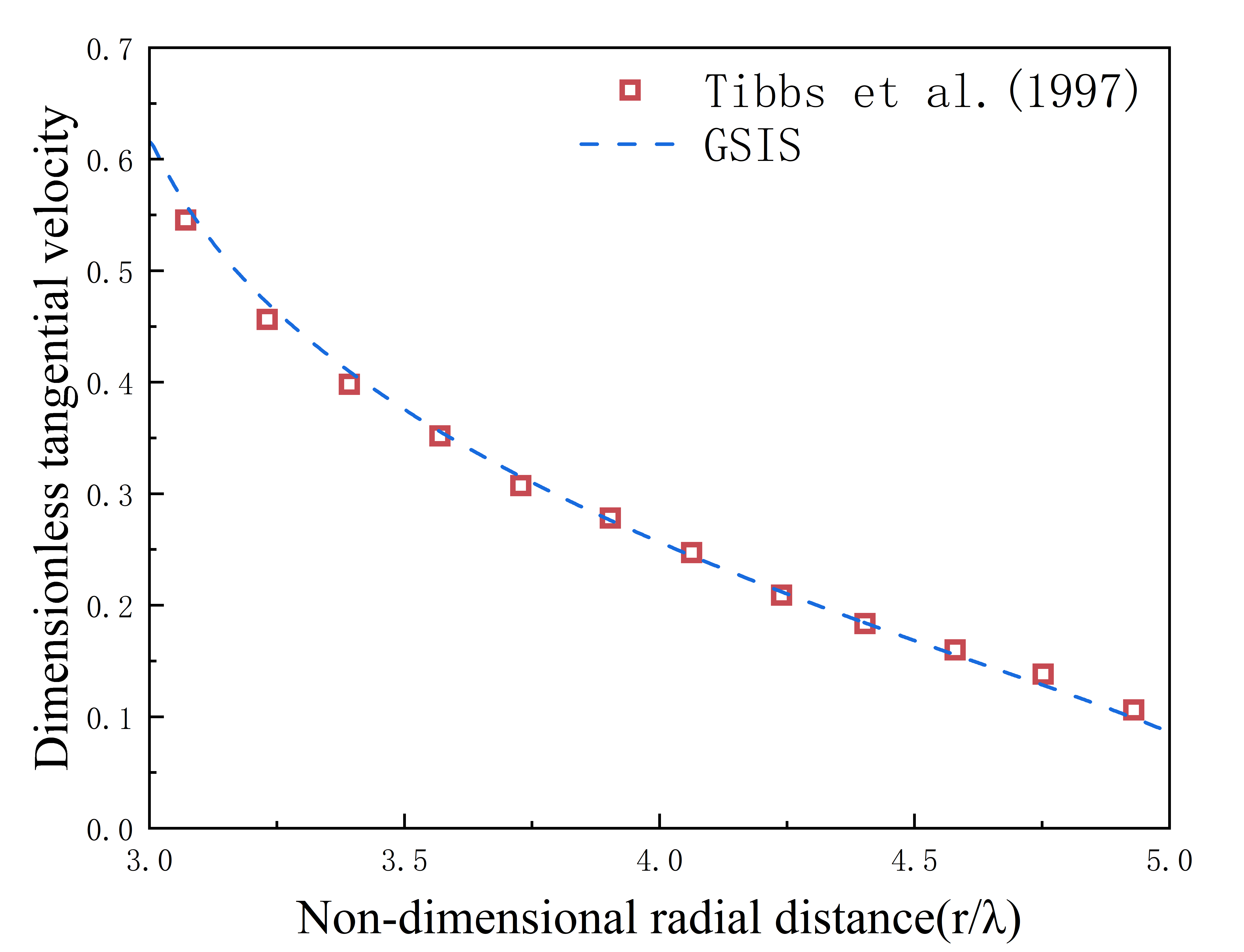}
    }

    \caption{Computational setup, mesh, and validation results for the Taylor--Couette flow.
    (a) Schematic of the cylindrical configuration.
    (b) Computational mesh in the \(r\)-\(x\) meridian plane.
    (c) Comparison of the dimensionless tangential velocity profile with the reference data of Tibbs \textit{et al.}~\cite{tibbs1997}, where the radial distance is normalized by mean free path, and the velocity is normalized by rotating speed.
    % \leir{fonts too small}
    }
    \label{fig:taylor_couette_setup}
\end{figure}

We first consider the Taylor--Couette flow~\cite{tibbs1997}. The computational domain is the annular region between two coaxial cylinders, as shown in Fig.~\ref{fig:taylor_couette_setup}(a). The radius of the inner cylinder is \(R_1=0.1875~\mu\mathrm{m}\), and that of the outer cylinder is \(R_2=0.3125~\mu\mathrm{m}\). A diffuse reflection kinetic boundary condition at wall temperature  $T_0=273.15$~K is enforced along the cylindrical surfaces. The Knudsen number is defined as
\(\mathrm{Kn}=\lambda/(R_2-R_1)\). With the  mean free path \(\lambda=0.0633~\mu\mathrm{m}\), the corresponding Knudsen number is \(\mathrm{Kn}=0.506\). The inner wall is prescribed as a rotating wall (the rotating speed is \(96.94~\mathrm{m/s}\), corresponding approximately to \(Ma=0.3\)), while the outer wall is kept stationary. 
Figure~\ref{fig:taylor_couette_setup}(b) illustrates the computational grid defined on the meridian plane, which comprises 2500 discrete cells.
Periodic boundary conditions are imposed at the upper and lower boundaries in the axial direction. 
The truncated molecular velocity domain is set as
$v_x\in[-6,6]\sqrt{k_BT_0/m}$, 
$ \zeta\in[0,6]\sqrt{k_BT_0/m}$, and $\omega\in[0,2\pi]$,
which is uniformly discretized using \(30\times30\times30\) points. 

The dimensionless tangential velocity profile obtained by the AxiGSIS is compared with the reference data of Tibbs \textit{et al.}~\cite{tibbs1997} in Fig.~\ref{fig:taylor_couette_setup}(c). 
Good agreement is observed between the AxiGSIS results and the reference data, indicating that the present method can accurately capture the velocity-slip characteristics and the radial variation of tangential velocity in rarefied Taylor--Couette flow.

\subsection{Neutral nozzle flow}

\begin{figure}[p]
    \centering
    \subfloat[Computational mesh]{
        \includegraphics[width=0.48\textwidth, trim={10 10 10 10}, clip=true]{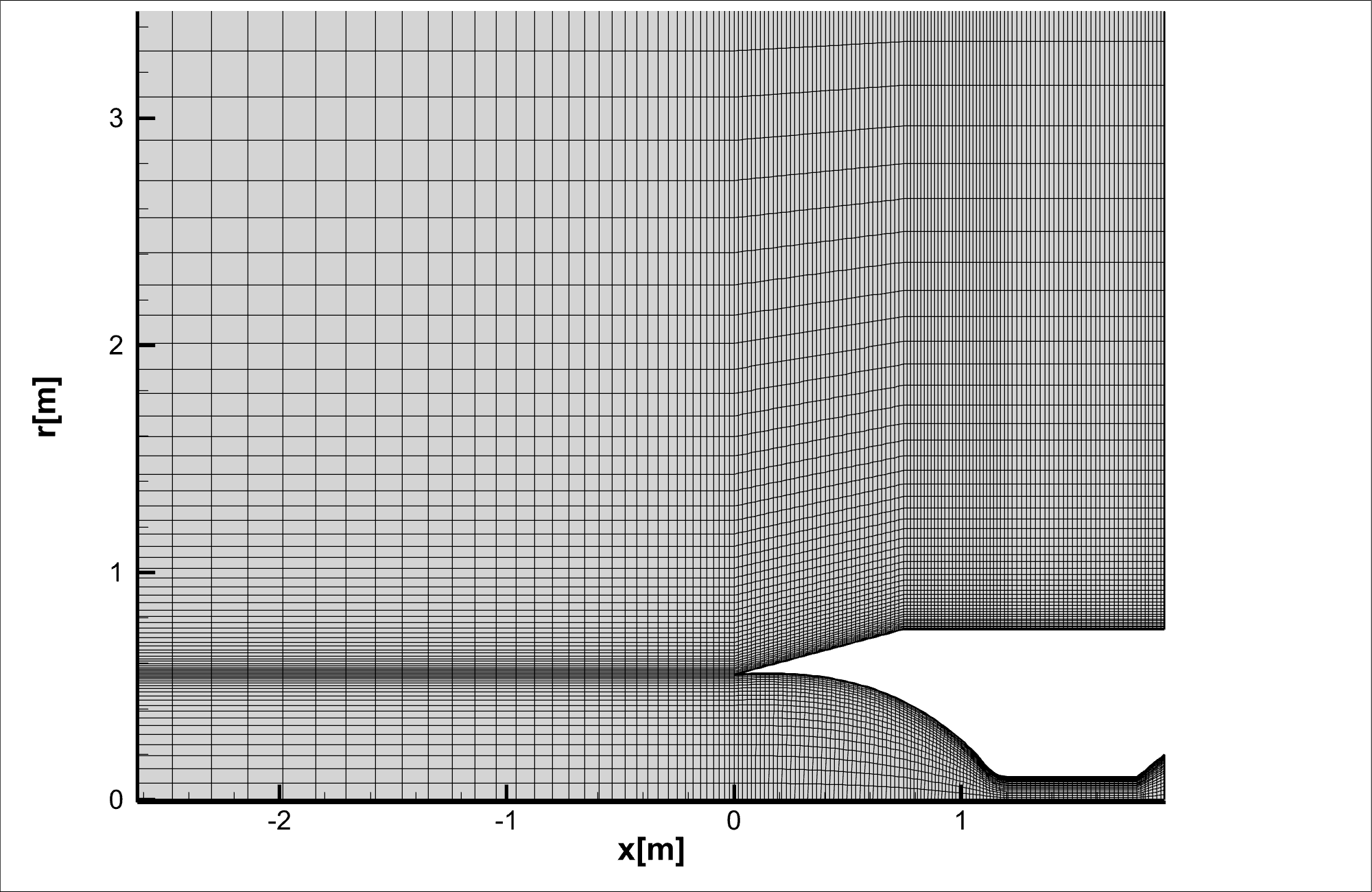}
    }
    \hfill
    \subfloat[Mass density]{
        \includegraphics[width=0.48\textwidth, trim={10 10 10 10}, clip=true]{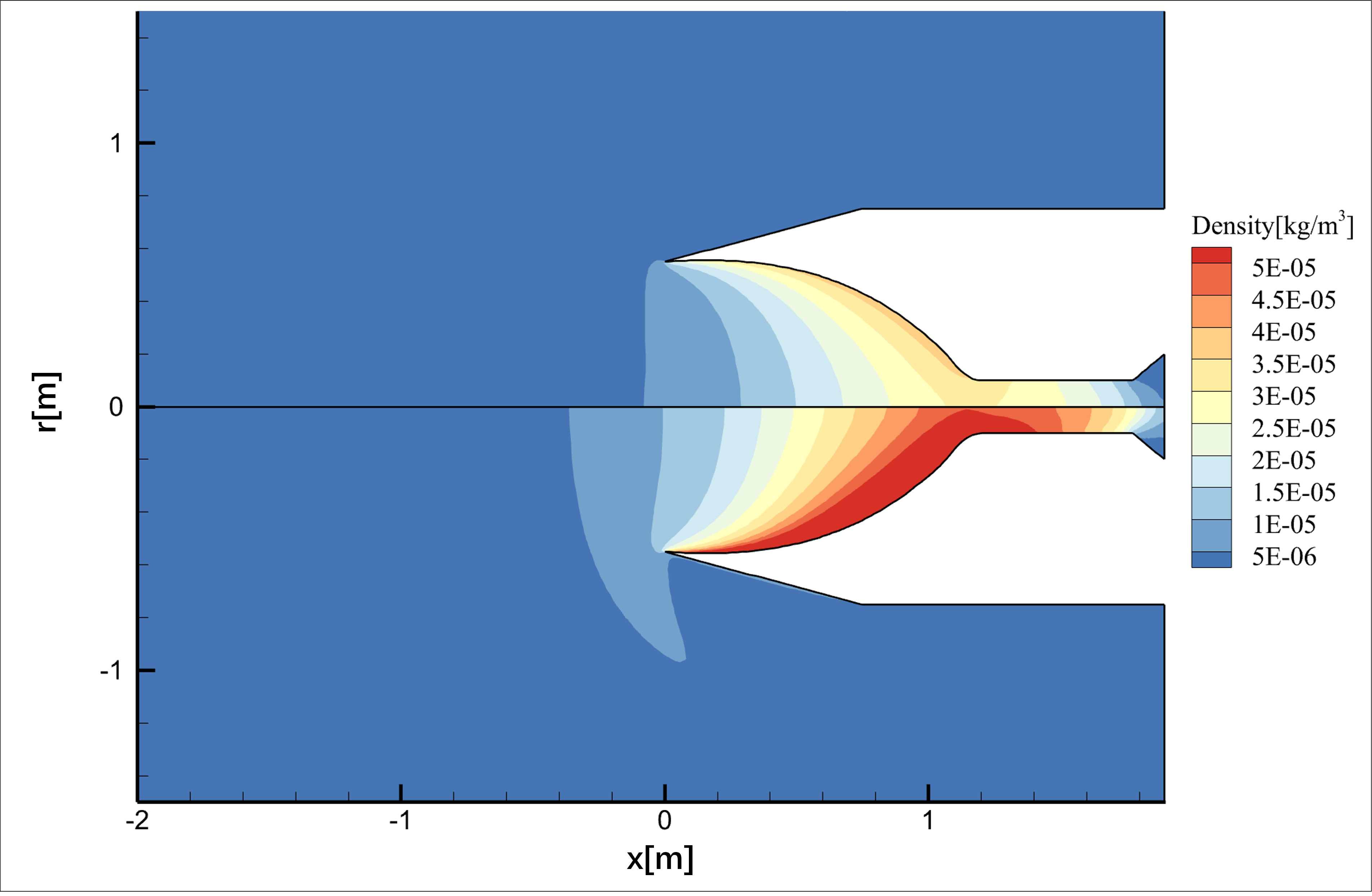}
    }
    \\
    \subfloat[Streamwise velocity]{
        \includegraphics[width=0.48\textwidth, trim={10 10 10 10}, clip=true]{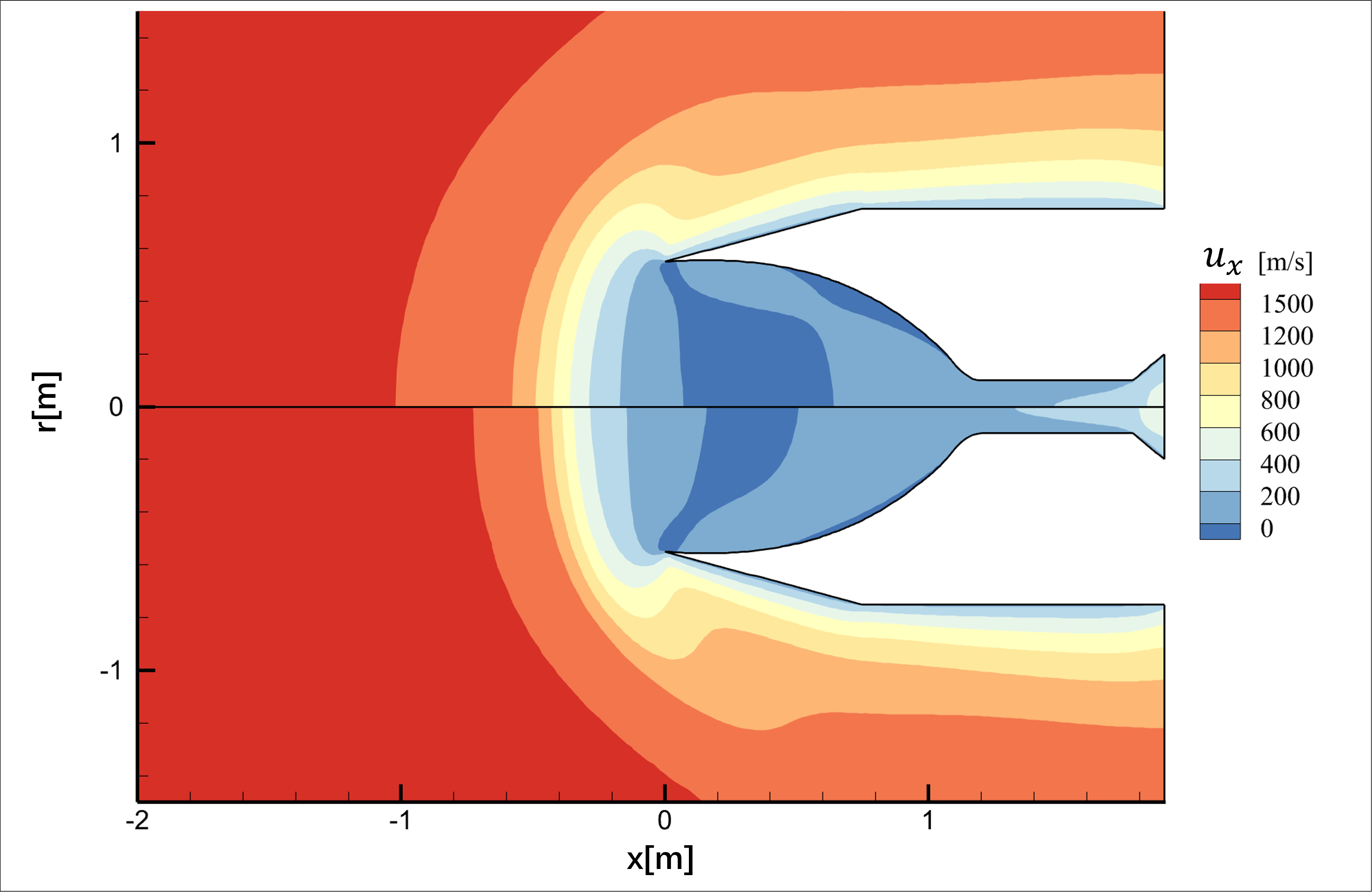}
    }
    \hfill
    \subfloat[Temperature]{
        \includegraphics[width=0.48\textwidth, trim={10 10 10 10}, clip=true]{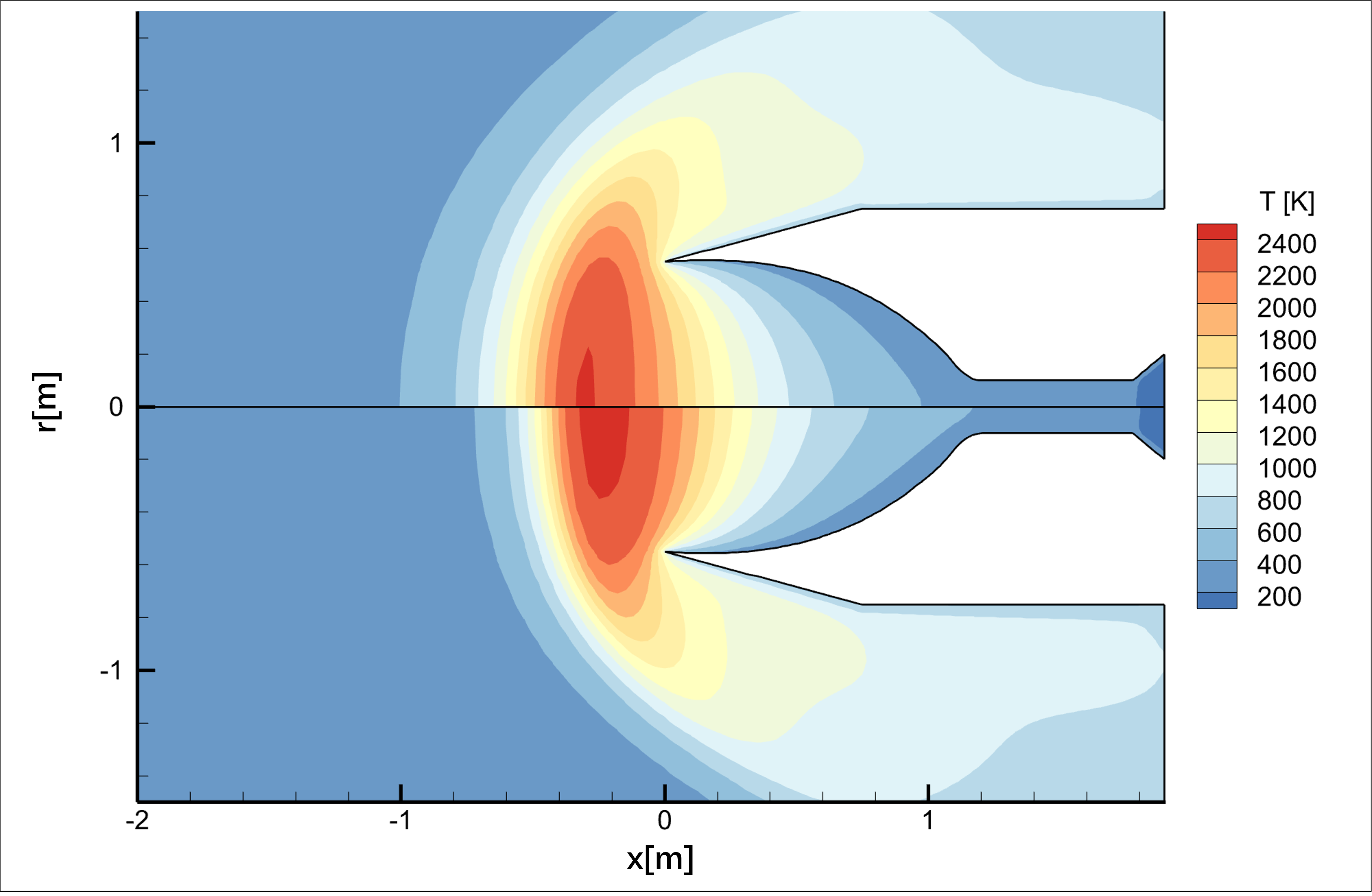}
    }

    \caption{Computational mesh and contour results of AxiGSIS for the neutral nozzle flow.
    (a) Computational mesh, where the nozzle geometry occupies the white region.
    (b) Mass density contour.
    (c) Streamwise velocity contour.
    (d) Temperature contour.
    In panels (b)--(d), the upper and lower half regions show the results for
    \(\mathrm{Kn}=0.057\) and \(\mathrm{Kn}=0.026\), respectively.}
    \label{fig:neutral_nozzle_setup}
\end{figure}

\begin{figure}[p]
    \centering
    \subfloat[Density]{
        \includegraphics[width=0.32\textwidth, clip=true]{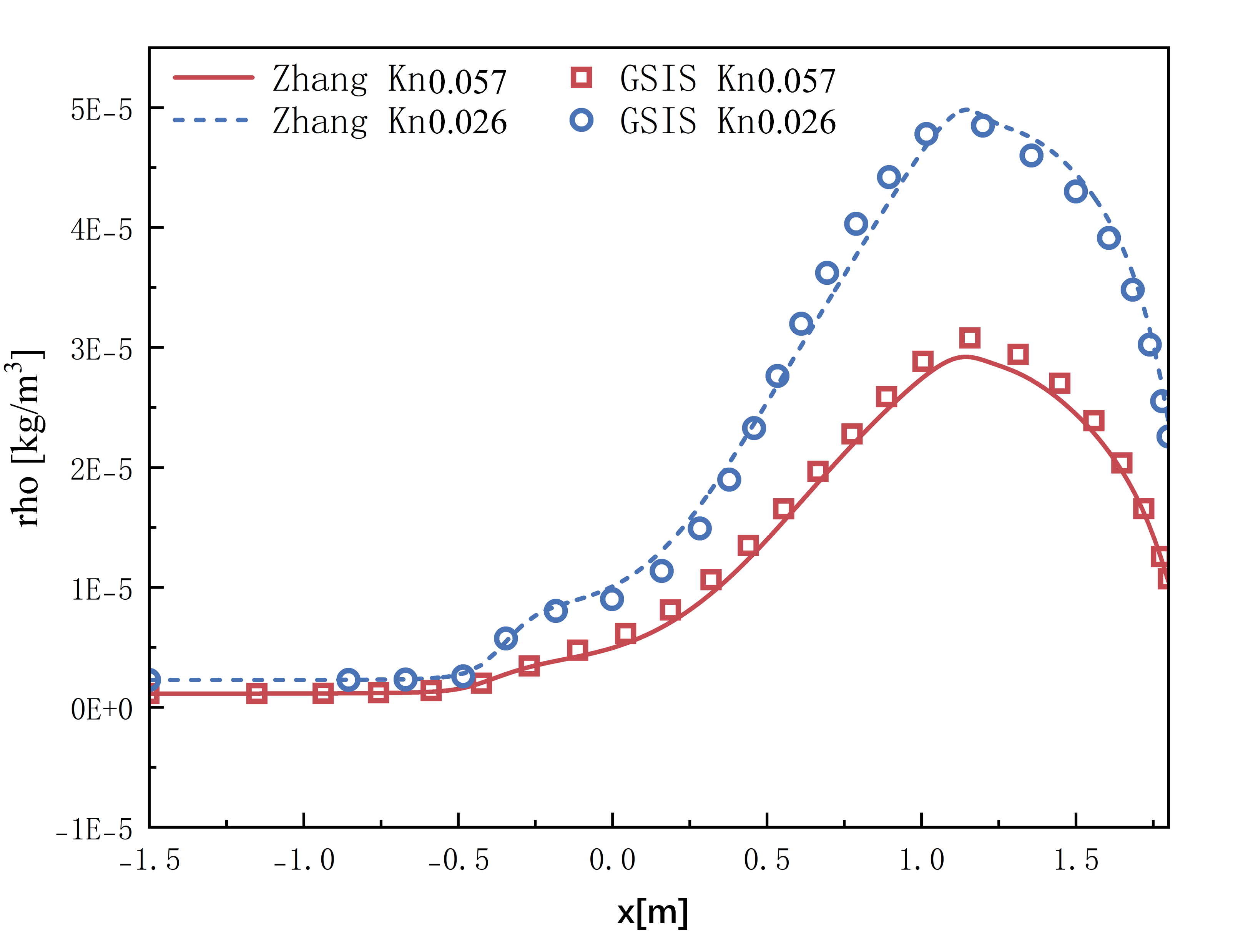}
    }
    \subfloat[Streamwise velocity]{
        \includegraphics[width=0.32\textwidth, clip=true]{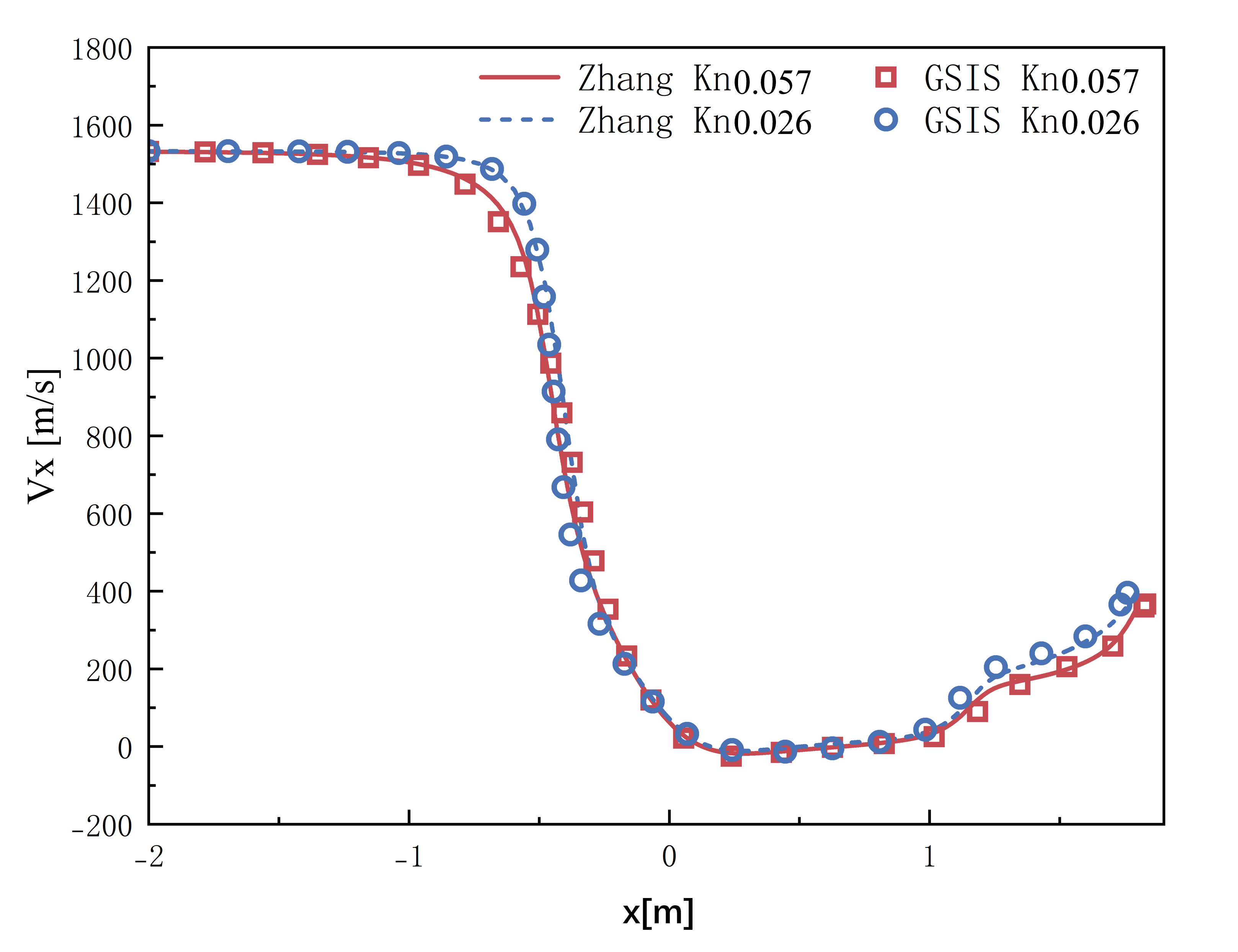}
    }
    \subfloat[Temperature]{
        \includegraphics[width=0.32\textwidth, clip=true]{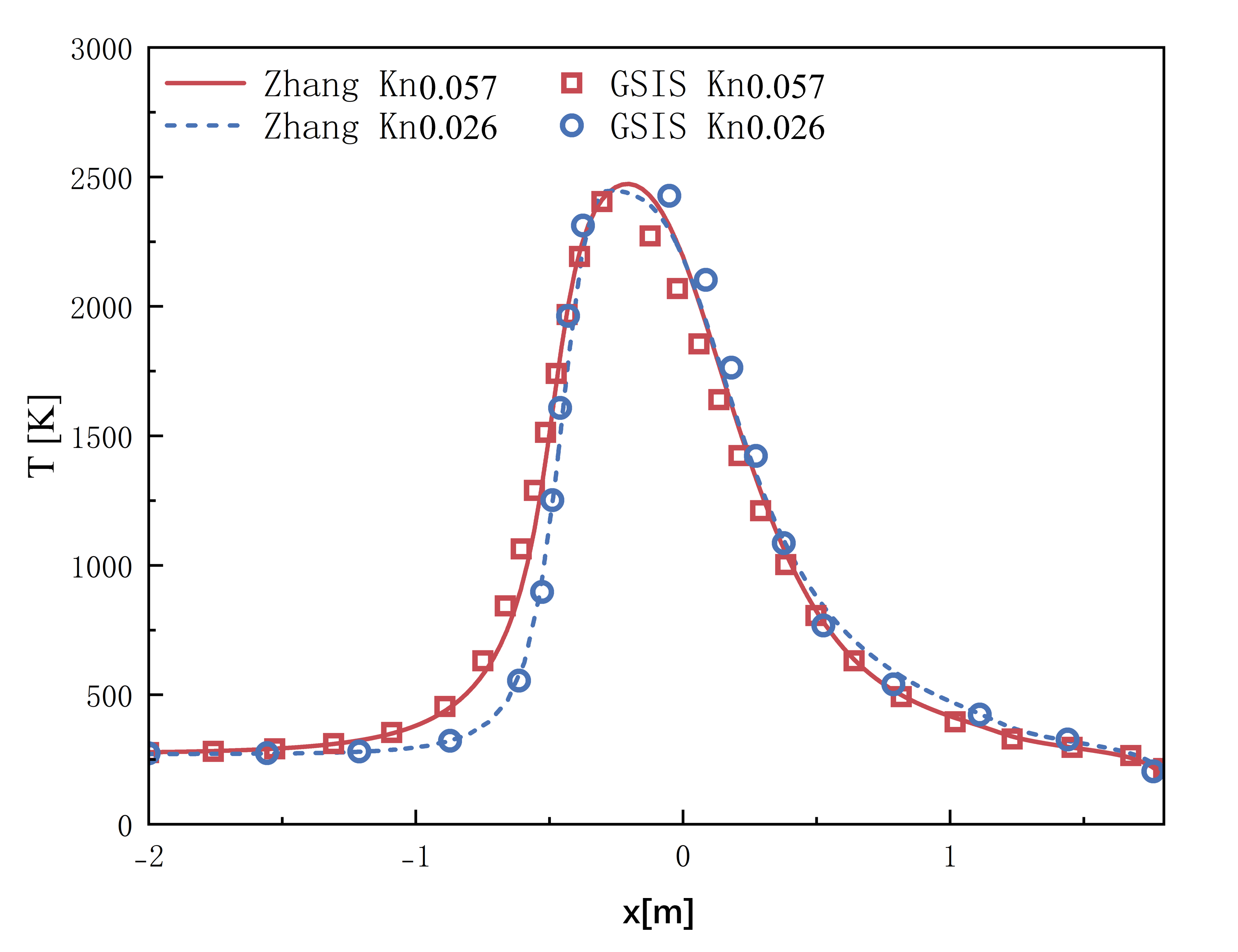}
    }
    \caption{Comparison of the centerline density, streamwise velocity, and temperature obtained by AxiGSIS with the 3D GSIS results of Zhang et al.~\cite{zhang2026efficientheatflux}.}
    \label{fig:neutral_nozzle_profiles}
\end{figure}

We then examine the axisymmetric nozzle flow, which is relevant to the air-breathing electric propulsion \cite{jin2024numerical}. Owing to the axisymmetric nature of the flow, all computations are restricted to the \(x\)-\(r\) meridian plane, as illustrated in Fig.~\ref{fig:neutral_nozzle_setup}(a).  
The outer boundary is assigned far-field conditions with a freestream temperature of \(T_0=273.15~\mathrm{K}\) and an incoming Mach number \(Ma_\infty=5\). The nozzle surface is treated as a diffuse-reflection wall with a wall temperature of \(300~\mathrm{K}\). The reference length is taken as the axial length of the nozzle, namely \(L_{\mathrm{ref}}=1.89~\mathrm{m}\). The global Knudsen number is tuned by varying the inlet number density:   \(n_{\mathrm{in}}=1.678\times10^{19}~\mathrm{m}^{-3}\) for \(\mathrm{Kn}=0.057\), and  \(n_{\mathrm{in}}=3.3559\times10^{19}~\mathrm{m}^{-3}\) for \(\mathrm{Kn}=0.026\). 
The 2D axisymmetric domain is discretized using a structured mesh consisting of \(10,254\) computational cells. The velocity-space domain is set as
$v_x\in[-10,14]\sqrt{k_BT_0/m}$,
$\zeta\in[0,10]\sqrt{k_BT_0/m}$, and $\omega\in[0,2\pi]$.
The number of uniformly distributed discrete velocity points is \(28\times28\times28\).

% The velocity space is truncated on a \((v_x,\zeta,\omega)\) coordinates as 
% $v_x\in[-10,14]\sqrt{k_BT_0/m}$,
% $\zeta\in[0,10]\sqrt{k_BT_0/m}$, and
% $\omega\in[0,2\pi]$,
% which is discretized uniformly with \(28\times28\times28\) discrete nodes. 

Figures~\ref{fig:neutral_nozzle_setup}(b)–(d) present the contours of mass density, axial velocity, and temperature. As the Knudsen number increases, the shock wave becomes substantially thicker. For quantitative validation of the proposed AxiGSIS solver, Fig.~\ref{fig:neutral_nozzle_profiles} compares centerline profiles of density, axial velocity, and temperature against reference data from the 3D GSIS method~\cite{zhang2026efficientheatflux}. For both tested Knudsen numbers, predictions from the AxiGSIS solver exhibit excellent agreement with the 3D GSIS reference data. This demonstrates that the reduced axisymmetric formulation accurately captures the gas expansion, flow acceleration, and  nonequilibrium characteristics of neutral rarefied nozzle flows.

Table~\ref{tab:neutral_nozzle_cost} compares the computational cost of AxiGSIS with the 3D GSIS and 3D CIS methods. For the same mesh, 3D GSIS is much faster than 3D CIS, reducing the cost from \(4819\) to \(97.9\) core-hours at \(\mathrm{Kn}=0.057\), and from \(6470\) to \(86.4\) core-hours at \(\mathrm{Kn}=0.026\). By further exploiting the axisymmetric geometry, the AxiGSIS reduces the number of physical cells from \(335,740\) to \(10,254\), and the corresponding cost decreases to only \(4.9\) and \(4.7\) core-hours, respectively. These results demonstrate that AxiGSIS combines the convergence acceleration of GSIS with the physical-space reduction of the axisymmetric formulation, leading to a substantial reduction in computational cost while maintaining good agreement with the 3D GSIS results.

\begin{table}[h]
    \centering
    % \caption{Computational cost comparison between the proposed AxiGSIS solver and the 3D GSIS scheme~\cite{zhang2026efficientheatflux} for neutral axisymmetric nozzle flows.}
    \caption{Computational cost comparison among AxiGSIS, 3D GSIS, and 3D CIS for neutral axisymmetric nozzle flows. The 3D results are taken from Ref.~\cite{zhang2026efficientheatflux}.}
    \label{tab:neutral_nozzle_cost}
    \begin{tabular}{cccccc}
        \hline
        \(\mathrm{Kn}\)
        & Method
        & Mesh cells
        & Cores
        & \(t_{\mathrm{wall}}~[\mathrm{h}]\)
        & core-hours \\
        \hline
        \multirow{3}{*}{\(0.057\)}
        & AxiGSIS
        & \(10,254\)
        & \(80\)
        & \(0.061\)
        & \(4.9\) \\
        & 3D GSIS
        & \(335,740\)
        & \(192\)
        & \(0.51\)
        & \(97.9\) \\
        & 3D CIS
        & \(335,740\)
        & \(192\)
        & \(25.1\)
        & \(4819\) \\
        \hline
        \multirow{3}{*}{\(0.026\)}
        & AxiGSIS
        & \(10,254\)
        & \(80\)
        & \(0.059\)
        & \(4.7\) \\
        & 3D GSIS
        & \(335,740\)
        & \(192\)
        & \(0.45\)
        & \(86.4\) \\
        & 3D CIS
        & \(335,740\)
        & \(192\)
        & \(33.7\)
        & \(6470\) \\
        \hline
    \end{tabular}
\end{table}

\subsection{Charged-particle flow past an electrostatic sphere}

We then investigate the axisymmetric flow of a single charged-particle species past an electrostatic sphere. Each charged argon particle carries one positive elementary charge.  The freestream Mach number is set to \(Ma_\infty=2\). Both the wall temperature and the freestream temperature are \(T_0=300~\mathrm{K}\). The reference length is chosen as the sphere diameter, namely \(L=1~\mathrm{m}\). The computational mesh contains about \(20,000\) cells. Three Knudsen numbers, $Kn=0.01$, $0.1$, and $1$, are considered, corresponding to freestream number densities of $1.6\times10^{20}$, $1.6\times10^{19}$, and $1.6\times10^{18}\ \mathrm{m}^{-3}$, respectively.
The velocity-space domain is set as
$v_x\in[-6,10]\sqrt{k_BT_0/m}$,
$\zeta\in[0,6]\sqrt{k_BT_0/m}$, and $\omega\in[0,2\pi]$.
The number of uniformly distributed discrete velocity points is \(28\times28\times28\).

\begin{figure}[p]
    \centering
    \subfloat[Computational setup]{
        \includegraphics[width=0.4\textwidth, trim={20 0 40 20},clip=true]{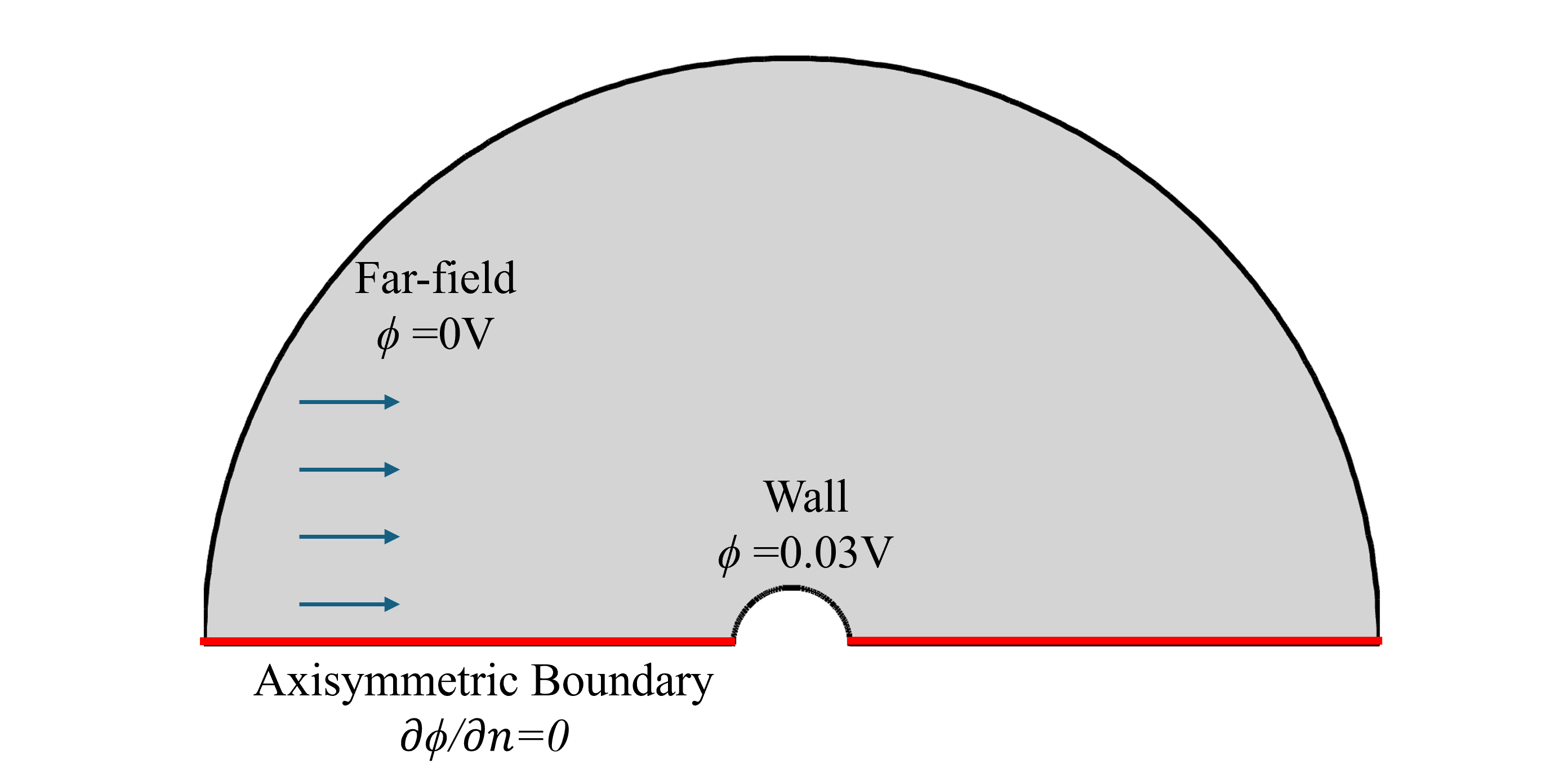}
    }
    \subfloat[Mesh and electrostatic potential]{
        \includegraphics[width=0.4\textwidth, trim={20 0 40 20},clip=true]{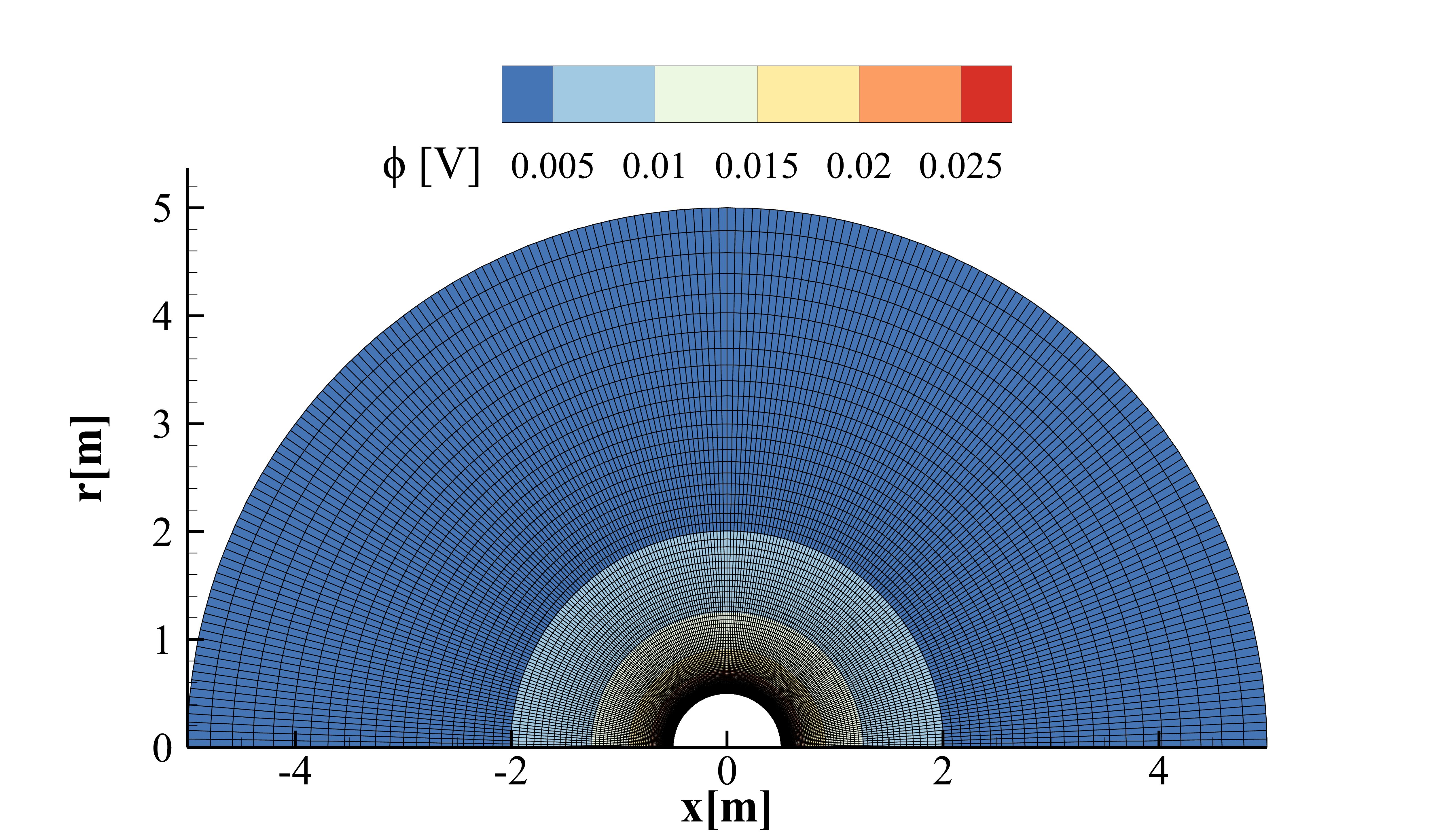}
    }
    \\
    \subfloat[\(T\), \(\mathrm{Kn}=0.01\)]{
        \includegraphics[width=0.4\textwidth, trim={30 15 50 10}, clip=true]{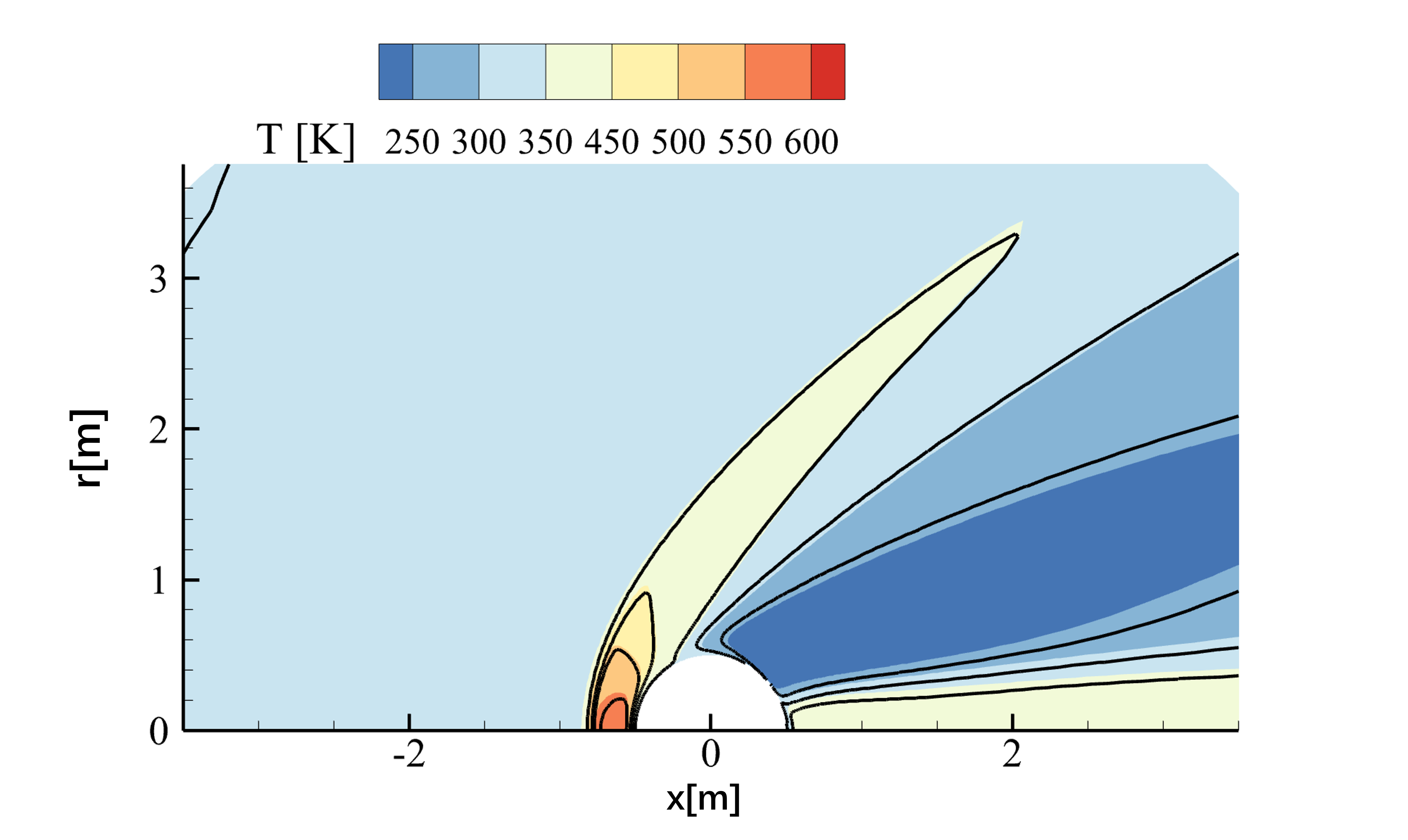}
    }
    \subfloat[\(u_x\), \(\mathrm{Kn}=0.01\)]{
        \includegraphics[width=0.4\textwidth, trim={30 15 50 10}, clip=true]{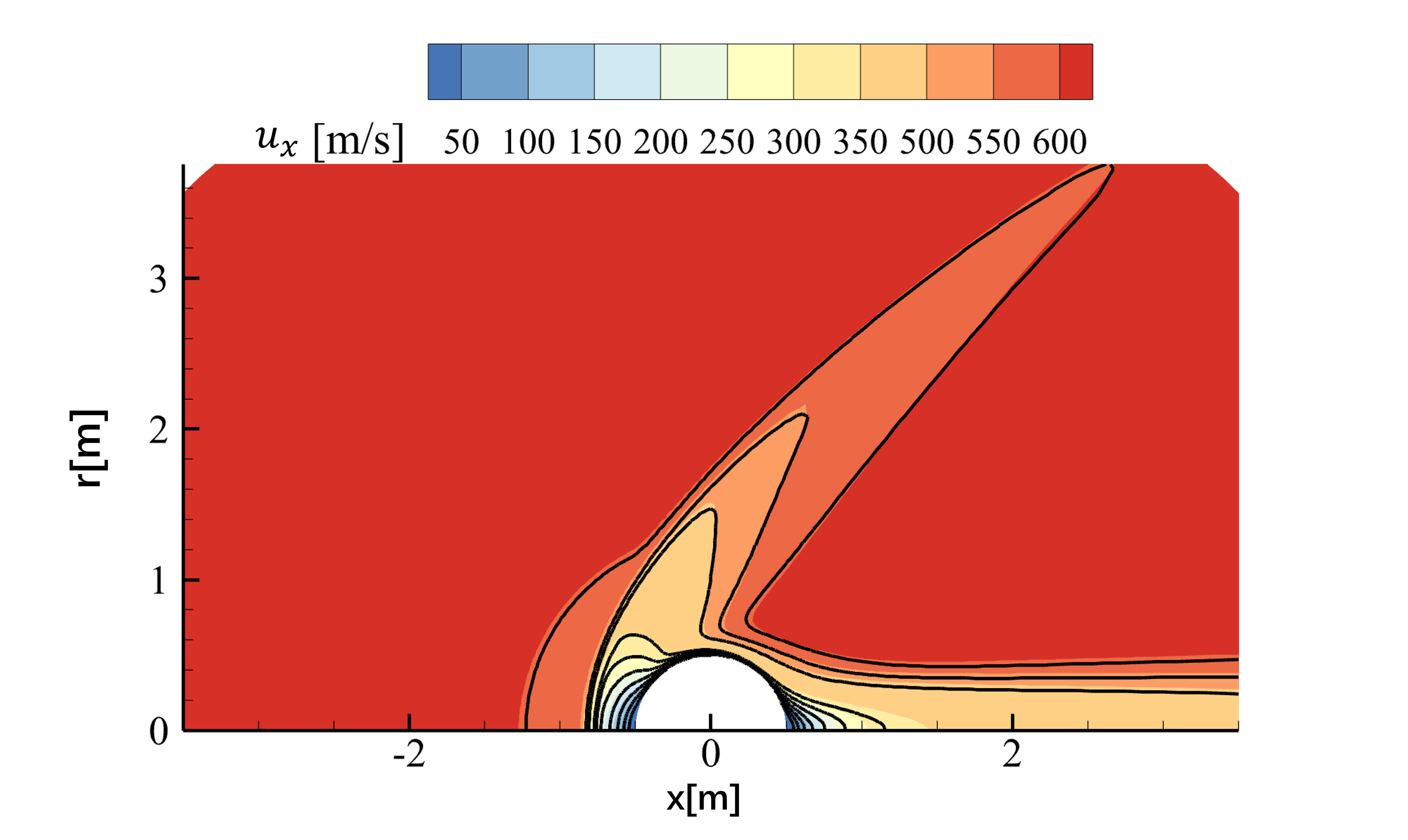}
    }
    \\
    \subfloat[\(T\), \(\mathrm{Kn}=0.1\)]{
        \includegraphics[width=0.4\textwidth, trim={30 15 50 10}, clip=true]{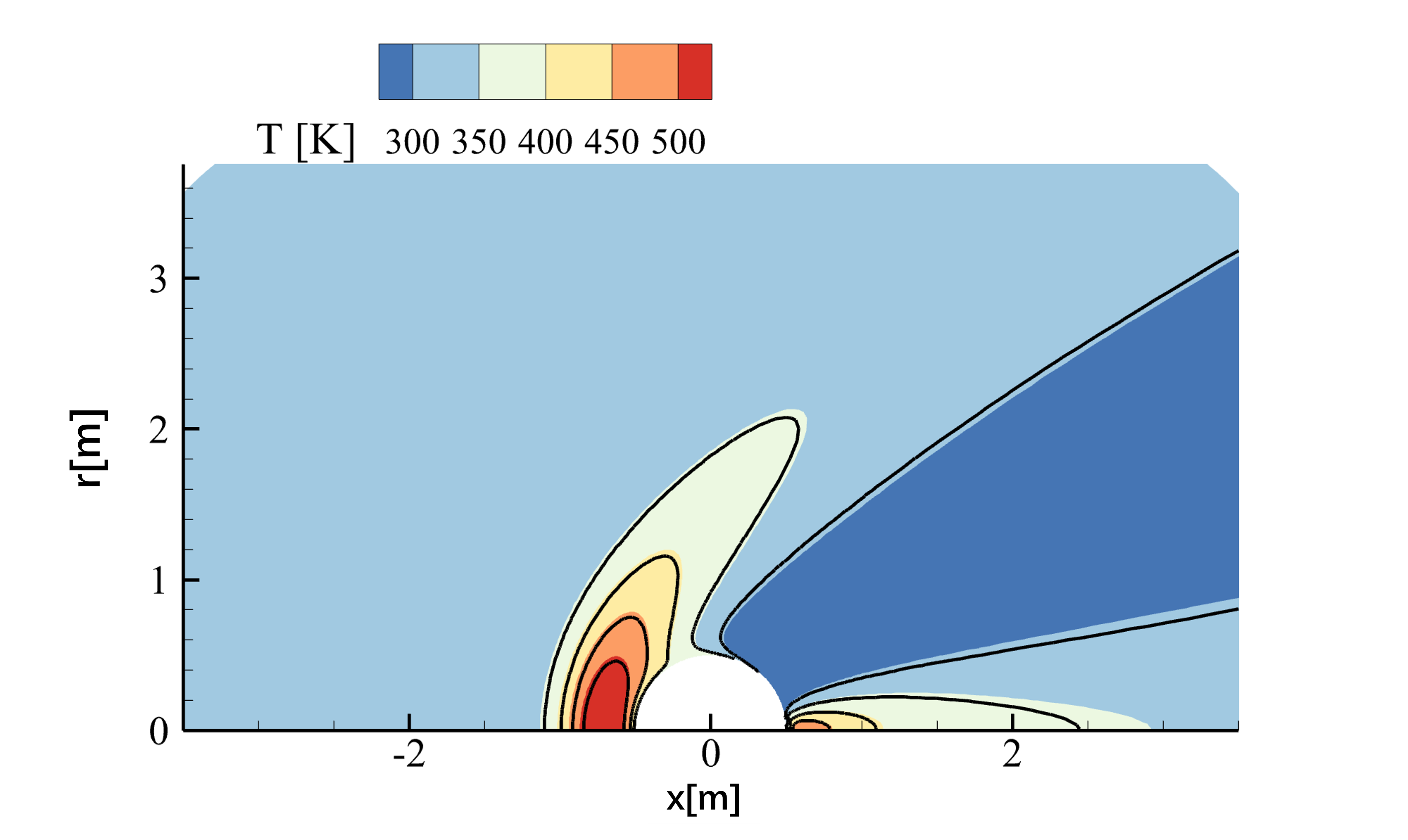}
    }
    \subfloat[\(u_x\), \(\mathrm{Kn}=0.1\)]{
        \includegraphics[width=0.4\textwidth, trim={30 15 50 10}, clip=true]{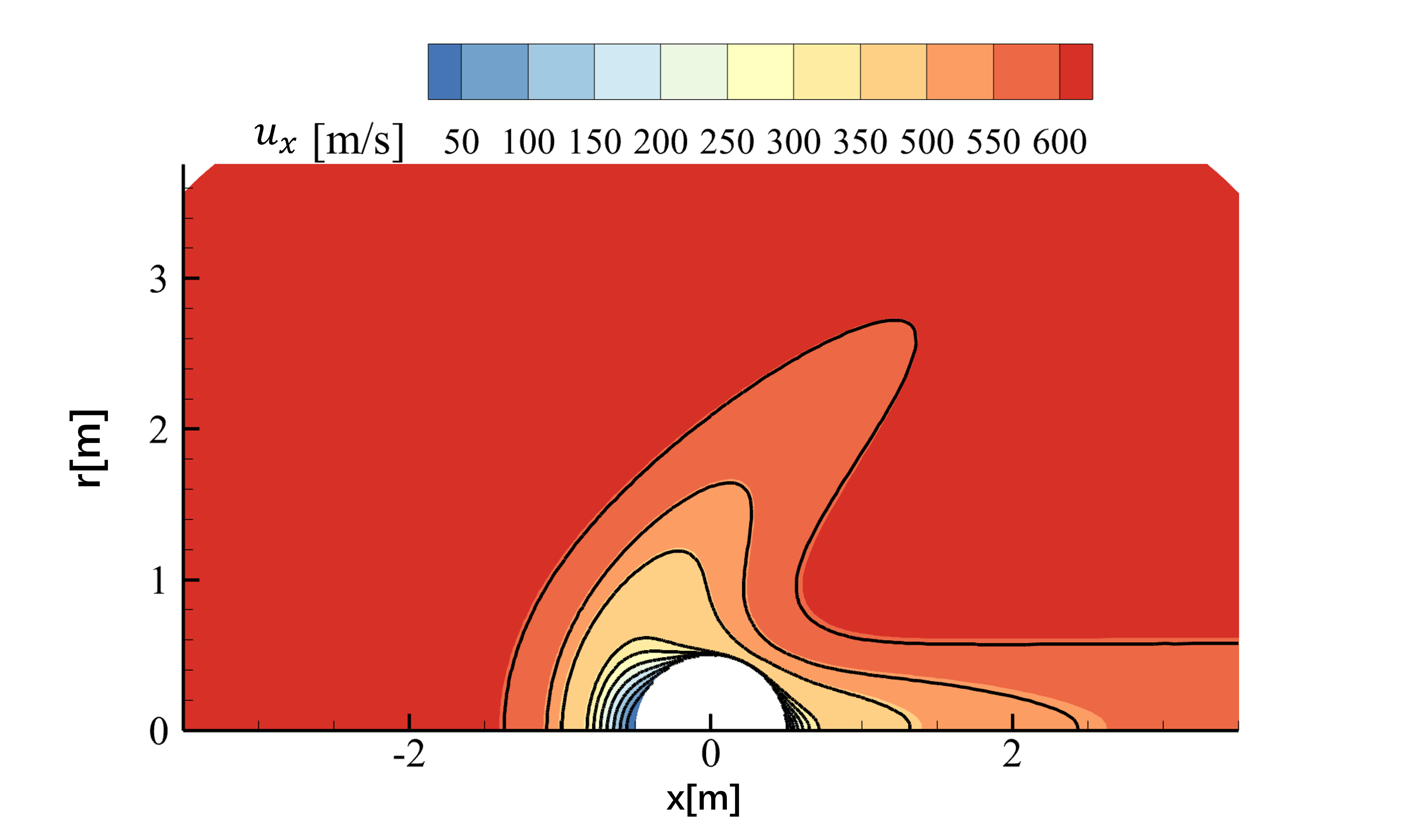}
    }
    \\
    \subfloat[\(T\), \(\mathrm{Kn}=1\)]{
        \includegraphics[width=0.4\textwidth, trim={30 15 50 10}, clip=true]{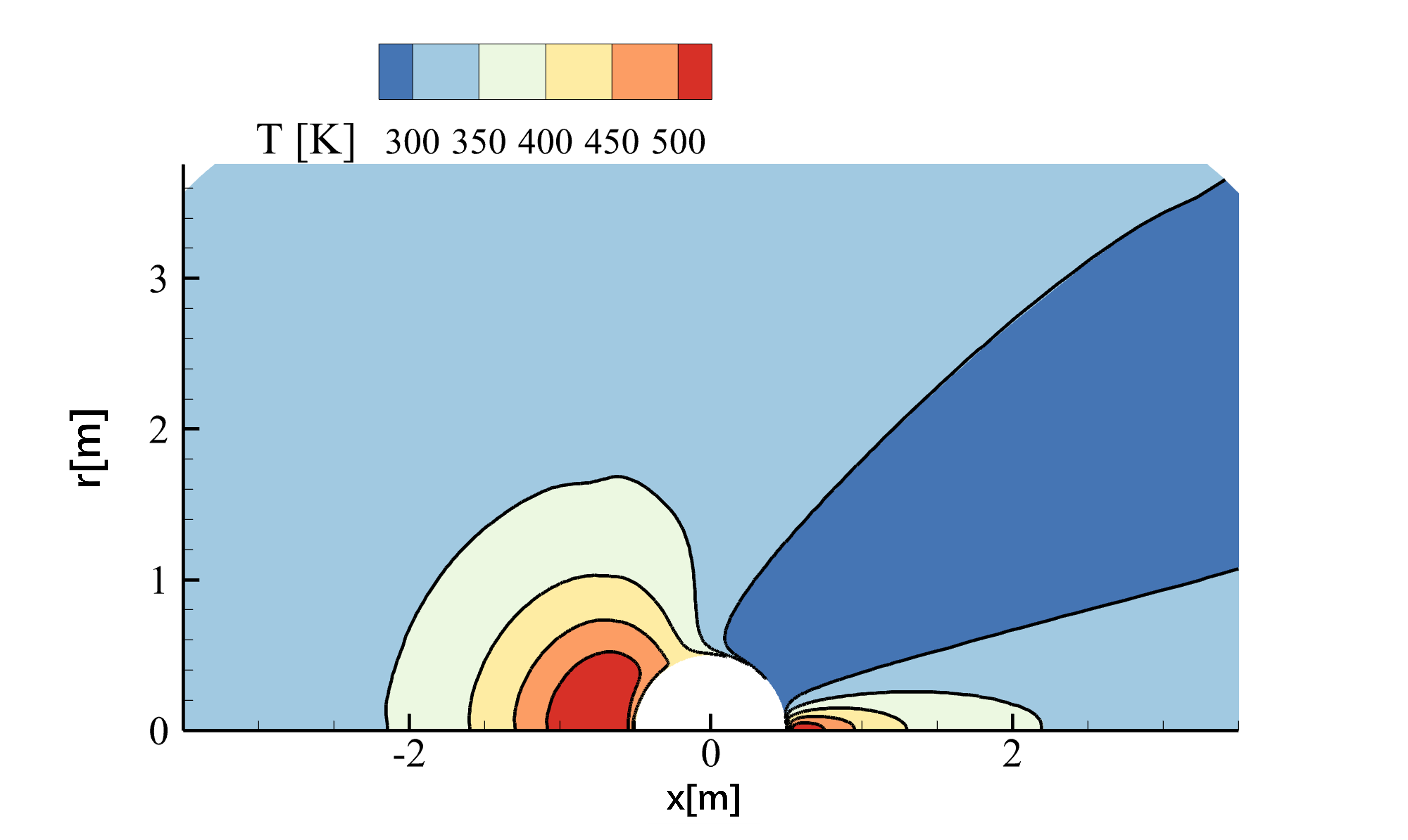}
    }
    \subfloat[\(u_x\), \(\mathrm{Kn}=1\)]{
        \includegraphics[width=0.4\textwidth, trim={30 15 30 10}, clip=true]{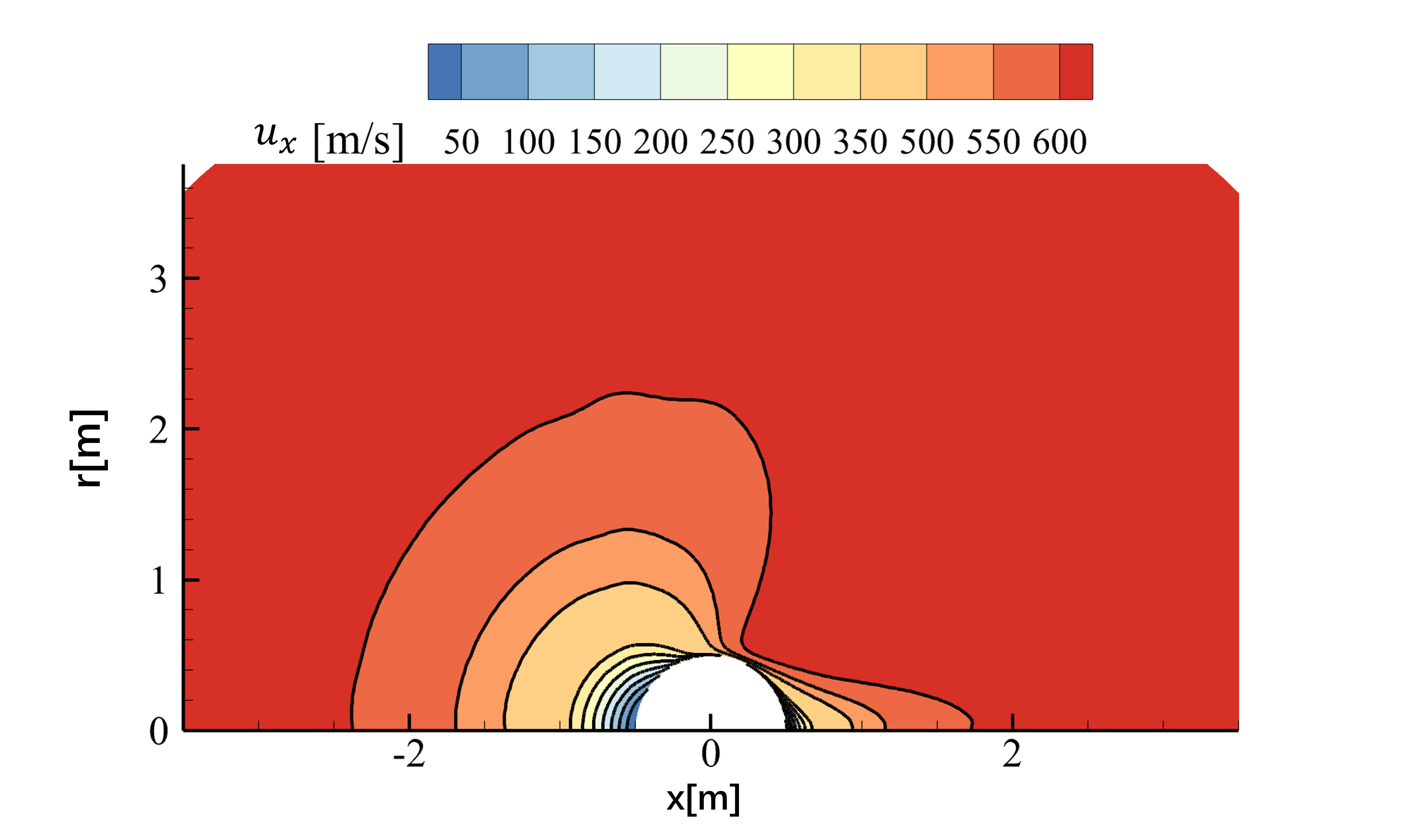}
    }
    \caption{ Charged-particle flow past an electrostatic sphere. (a,b) Computational setup, mesh, and electrostatic potential.   (c-h) Temperature and streamwise velocity contours  at different Knudsen numbers. The filled color contours represent the CIS results, and the black contour lines represent the AxiGSIS results.
    }
    \label{fig:sphere_contours}
\end{figure}

Figure~\ref{fig:sphere_contours}(a,b) shows the computational setup, mesh, and electrostatic potential distribution. A prescribed positive potential of \(0.03~\mathrm{V}\) is imposed on the sphere surface, while the far-field boundary is grounded. The electric field is determined by solving the axisymmetric Laplace equation subject to specified electrostatic boundary conditions, see details in \ref{subsec:poisson_solver}. The resulting electric field is used as an externally imposed field in the kinetic equation. Since the electric field is not updated by the simulated charged-particle density, this case focuses on the kinetic response of positively charged argon particles to a given electrostatic field. The normalized electrostatic acceleration, defined as \(q_pE_{\mathrm{ref}}L/(m v_m^2)\), is about \(4.6\), where \(q_p\) and \(m\) are the charge and mass of one argon ion, \(E_{\mathrm{ref}}\) is the reference electric-field strength, and \(v_m=\sqrt{2k_BT_0/m}\) is the most probable molecular speed.

% \begin{figure}[t!]
%     \centering
%     \subfloat[Velocity]{
%         \includegraphics[width=0.48\textwidth, clip=true]{Sphere_Vx.png}
%     }
%     \hfill
%     \subfloat[Temperature]{
%         \includegraphics[width=0.48\textwidth, clip=true]{Sphere_T.png}
%     }
%     \caption{Streamwise velocity and temperature along the centerline for the charged-particle flow past an electrostatic sphere.
% }
%     \label{fig:sphere_profiles}
% \end{figure}

\begin{figure}[t!]
    \centering
    \subfloat[Velocity]{
        \includegraphics[width=0.45\textwidth, clip=true]{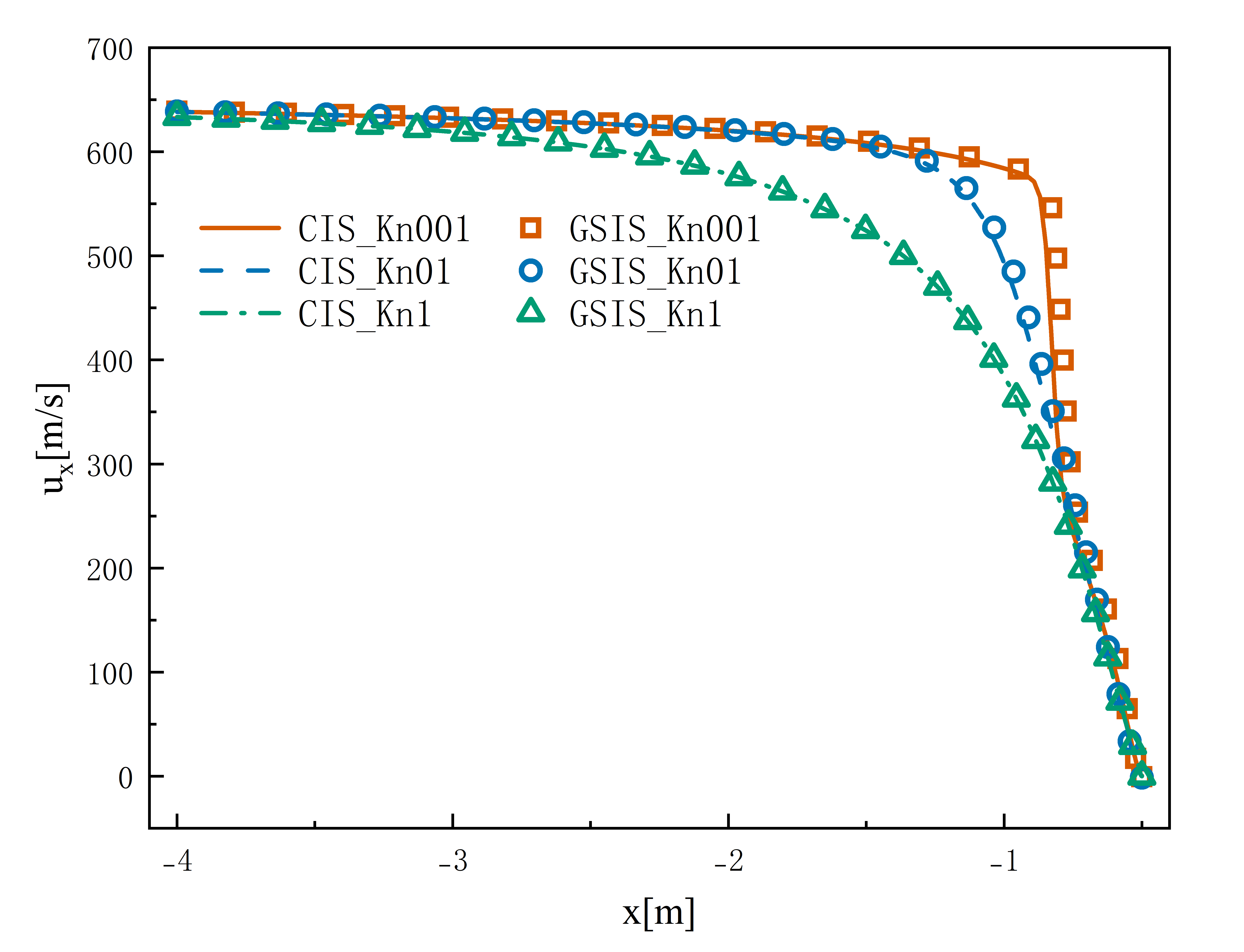}
    }
    \subfloat[Temperature]{
        \includegraphics[width=0.45\textwidth, clip=true]{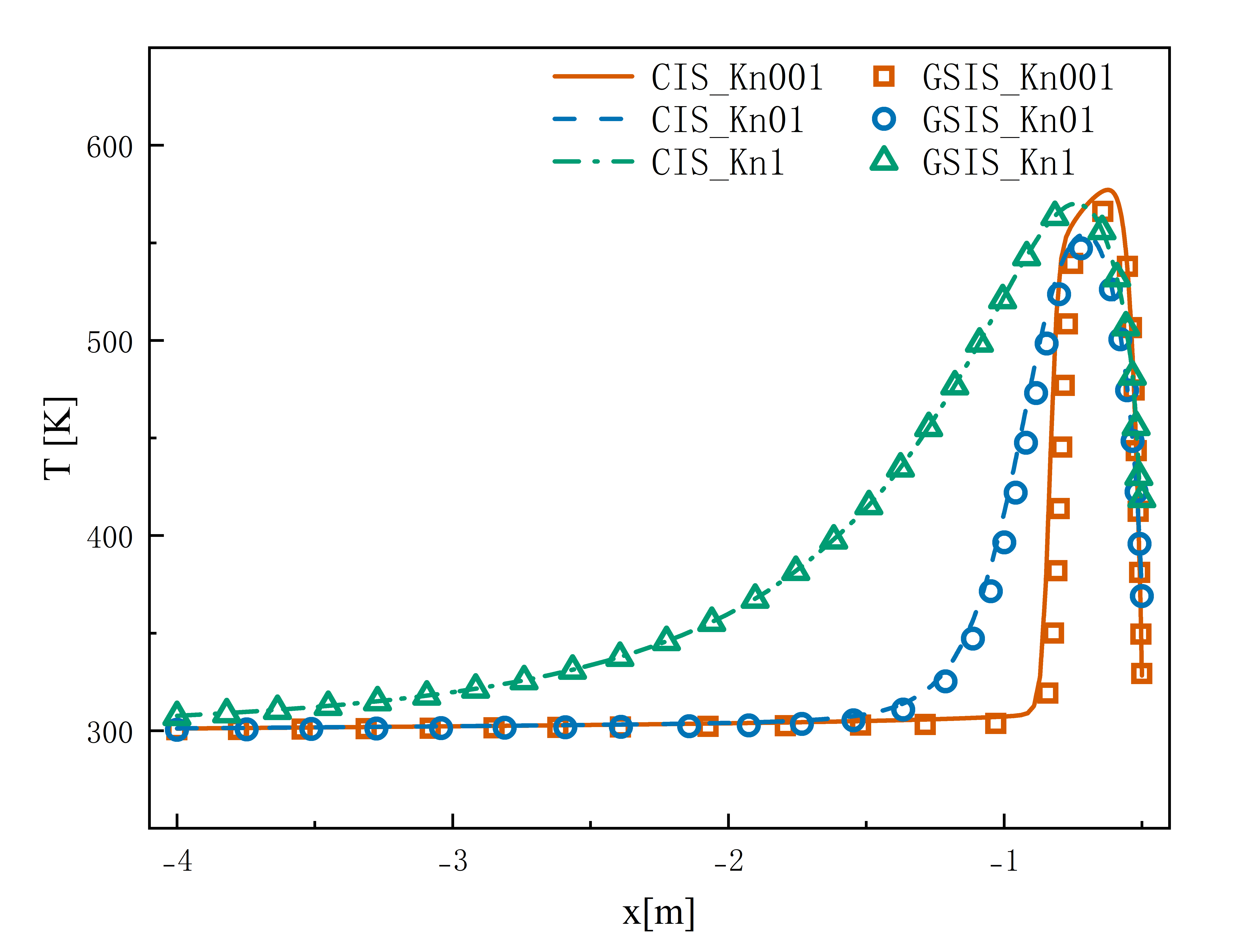}
    }\\
    \subfloat[Pressure]{
        \includegraphics[width=0.45\textwidth, clip=true]{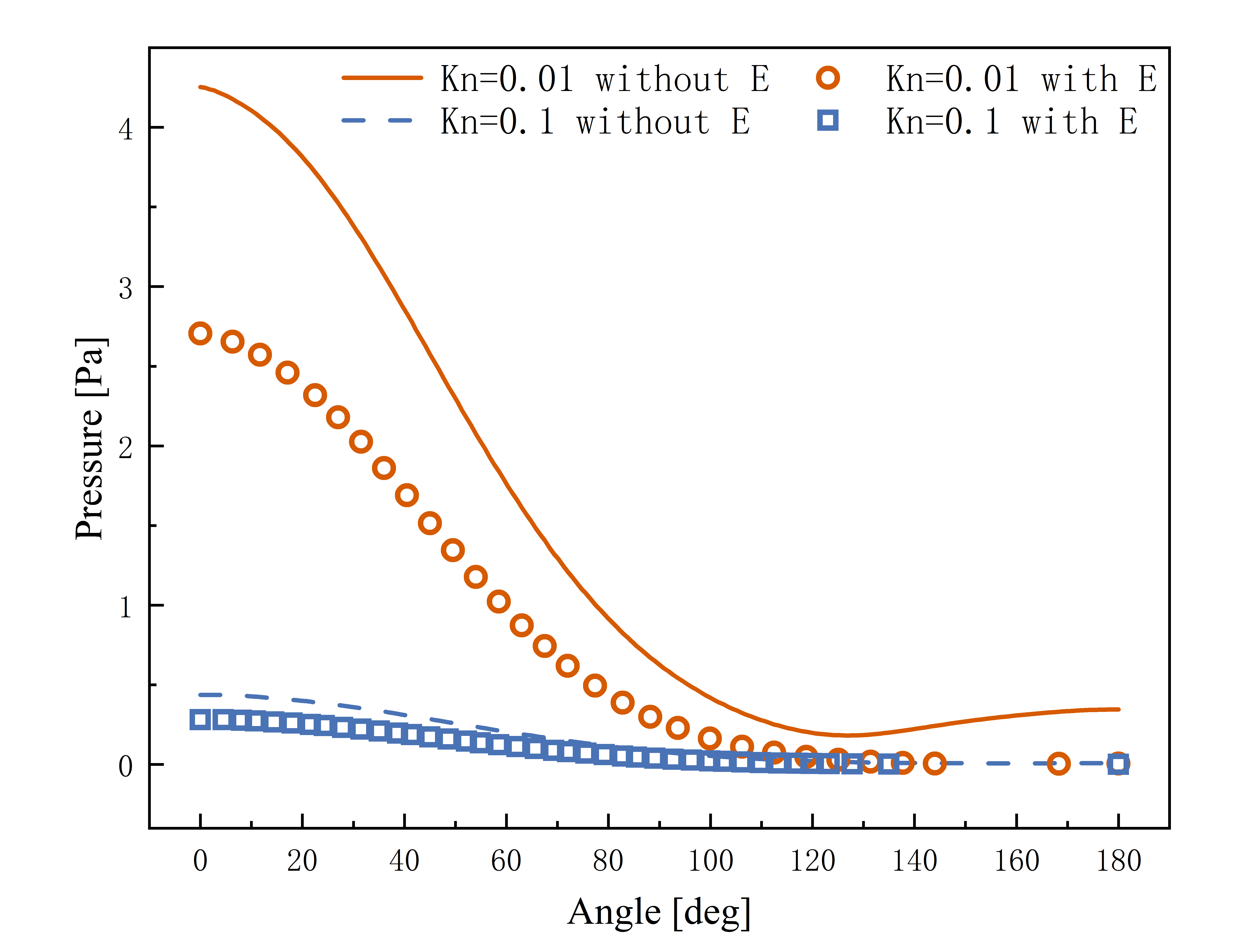}
    }
    \subfloat[Normal heat flux]{
        \includegraphics[width=0.45\textwidth, clip=true]{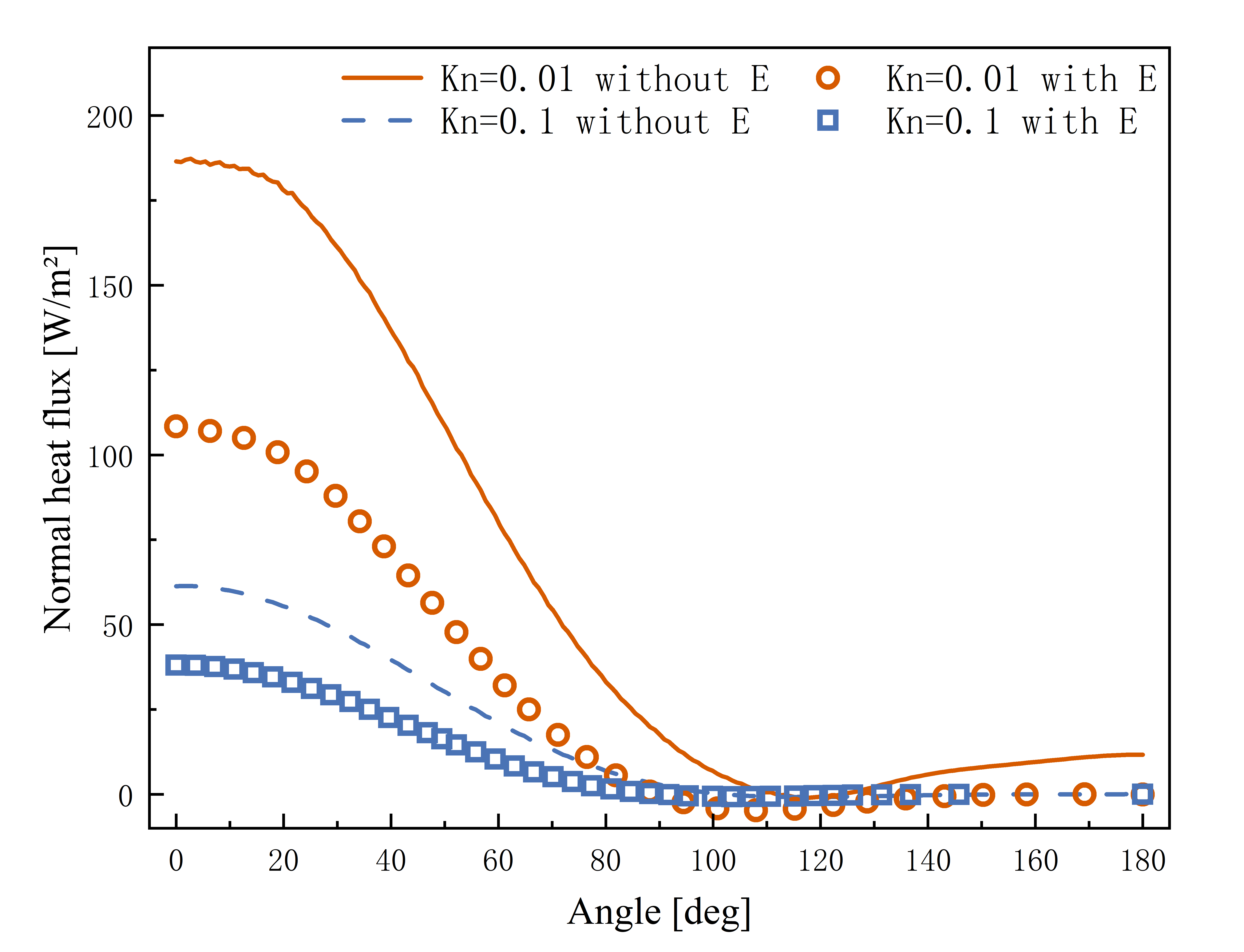}
    }
    % \caption{ Charged-particle flow past an electrostatic sphere. (a,b) Streamwise velocity and temperature along the centerline for the charged-particle flow past an electrostatic sphere. (c,d) Surface pressure and normal heat flux. The symbols and lines show the results with and without the electrostatic field, respectively.
    % The horizontal axis denotes the surface angle measured from the upstream stagnation point, where \(0^\circ\) corresponds to the front stagnation point and \(180^\circ\) corresponds to the rear side of the sphere.
    % }
    \caption{ Charged-particle flow past an electrostatic sphere. (a,b) Streamwise velocity and temperature along the centerline, where the lines and symbols denote the CIS and AxiGSIS results, respectively.
(c,d) Surface pressure and normal heat flux, where the symbols and lines denote the results with and without the electrostatic field, respectively. The surface angle is measured from the upstream stagnation point, with $0^\circ$ and $180^\circ$ corresponding to the front and rear stagnation points, respectively.
    }
    \label{fig:sphere_surface_profiles}
\end{figure}

Figure~\ref{fig:sphere_contours}(c-h) presents the temperature and streamwise velocity contours for \(\mathrm{Kn}=0.01\), \(0.1\), and \(1\). As the Knudsen number rises, rarefaction effects intensify and the shock profiles broaden significantly. Figure~\ref{fig:sphere_surface_profiles}(a,b) compares the streamwise velocity and temperature along the centerline. The AxiGSIS results agree well with the CIS results for all Knudsen numbers.
Figure~\ref{fig:sphere_surface_profiles}(c,d) shows the surface pressure and normal heat flux for the flow past the sphere with and without the electrostatic field. 
The surface pressure generally decreases from the upstream side to the downstream side. For both Knudsen numbers, the pressure obtained with the electrostatic field is lower than that without the electrostatic field. This is because the sphere surface is prescribed with a positive electric potential, while the simulated argon ions are positively charged. The electrostatic repulsion reduces the charged-particle concentration in the near-wall region, thereby lowering the local momentum transfer to the wall. As a result, the surface pressure is reduced when the electrostatic field is included. This reduction is more pronounced in the upstream region, where the impinging particle flux and surface pressure are relatively large.
Similar to the pressure distribution, the normal heat flux is also reduced by the electrostatic field, especially near the upstream side. The positive wall potential repels the positively charged argon particles away from the sphere surface, which decreases the near-wall particle density and weakens the energy exchange between the gas and the wall. Consequently, the heat flux transferred to the wall is reduced. In the downstream region, the normal heat flux approaches zero and may become slightly negative. 

\begin{table}[t!]
    \centering
\caption{Comparison of computational cost between AxiGSIS and CIS for charged-particle flow over an electrostatic sphere. For AxiGSIS, ``Iter'' denotes the number of GSIS outer iterations, and each outer iteration contains three CIS prediction steps. For CIS, ``Iter'' denotes the number of conventional kinetic iterations.}
    \label{tab:sphere_cost}
    \renewcommand{\arraystretch}{1.15}
    \begin{tabular}{cccccc}
        \hline
        \(\mathrm{Kn}\) & Method & Cores & \(t_{\mathrm{wall}}~[\mathrm{h}]\) &  core-hours & Iter \\
        \hline
        \multirow{2}{*}{\(0.01\)}
        & AxiGSIS & \(40\) & \(0.27\) & \(10.8\) & \(28\) \\
        & CIS  & \(200\) & \(1.64\) & \(328\) & \(3060\) \\
        \hline
        \multirow{2}{*}{\(0.1\)}
        & AxiGSIS & \(40\) & \(0.194\) & \(7.8\) & \(28\) \\
        & CIS  & \(200\) & \(0.196\) & \(39.2\) & \(370\) \\
        \hline
        \multirow{2}{*}{\(1\)}
        & AxiGSIS & \(40\) & \(0.29\) & \(11.6\) & \(44\) \\
        & CIS  & \(200\) & \(0.056\) & \(11.2\) & \(111\) \\
        \hline
    \end{tabular}
\end{table}

Table~\ref{tab:sphere_cost} compares the convergence behavior and computational cost of the AxiGSIS and CIS. For \(\mathrm{Kn}=0.01\), the AxiGSIS cuts the iteration steps from 3060 to 28, and the corresponding CPU core-hours are reduced from \(328\) to \(10.8\). For \(\mathrm{Kn}=0.1\), the iteration number is reduced from \(370\) to \(28\), and the CPU core-hours are reduced from \(39.2\) to \(7.8\). At \(\mathrm{Kn}=1\), the computational costs of AxiGSIS and CIS are nearly identical, because the advantage of the macroscopic synthetic iteration becomes less pronounced in the highly rarefied regime. These results demonstrate that the AxiGSIS can significantly accelerate steady-state charged-particle kinetic simulations under a prescribed electrostatic field, especially in the small- and moderate-Knudsen-number regimes.

% \leir{comparison in surface pressure/heat flux with and without electric field? }

\subsection{Charged-particle nozzle flow}

\begin{figure}[t]
    \centering
    \subfloat[Computational setup]{
        \includegraphics[width=0.8\textwidth]{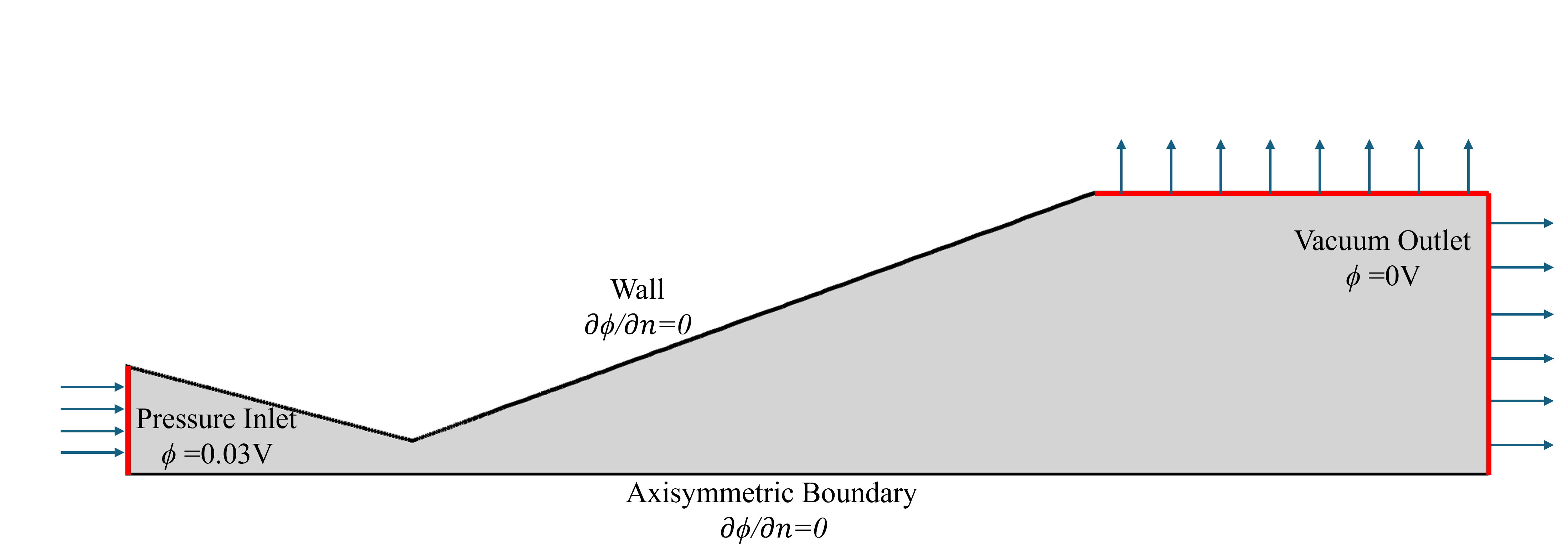}
    }
    \\
    \subfloat[Mesh and electrostatic potential]{
        \includegraphics[width=0.8\textwidth]{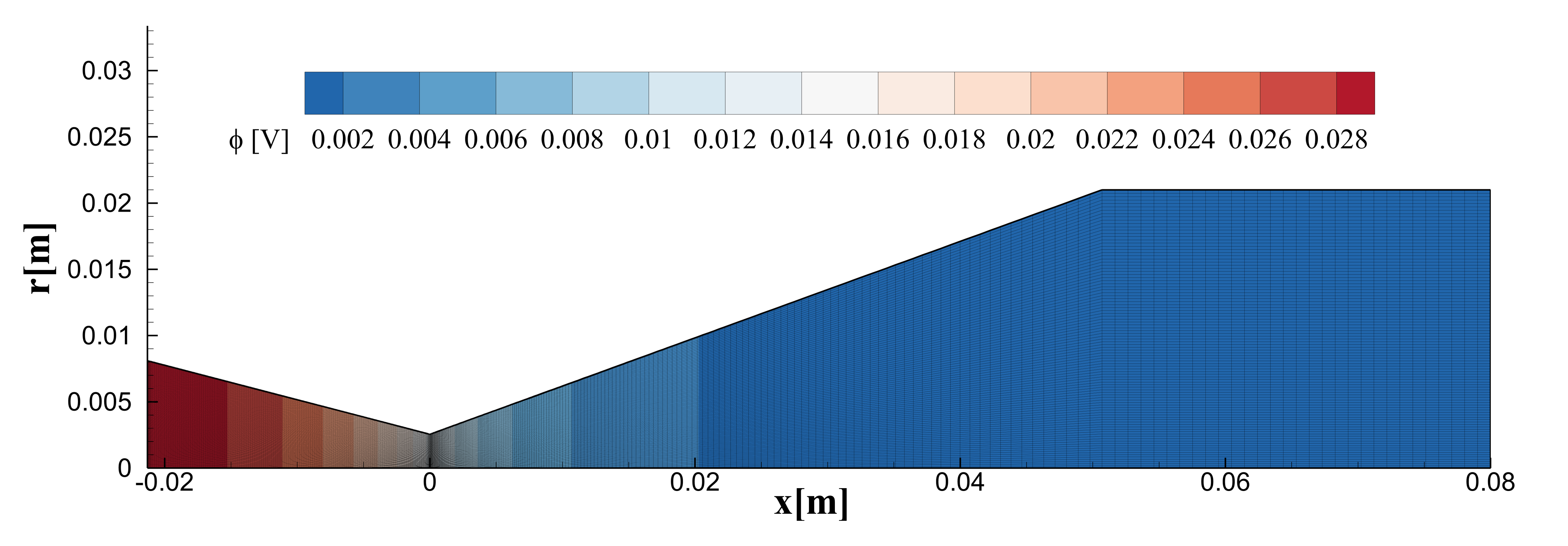}
    }
    \caption{Computational setup, mesh, and electrostatic potential distribution for the charged-particle nozzle flow.
   % (a) Computational configuration and boundary conditions.
    %(b) Computational mesh and electrostatic potential distribution in the nozzle.
    }
    \label{fig:charged_nozzle_setup_phi}
\end{figure}

Lastly, we investigate the flow of charged particles within an axisymmetric nozzle under a predefined electrostatic field. 
%The electric field is determined by solving the axisymmetric Laplace equation subject to specified electrostatic boundary conditions, see details in \ref{subsec:poisson_solver}. 
Specifically, as shown in Fig.~\ref{fig:charged_nozzle_setup_phi}(a), a fixed potential of 0.03 V is imposed at the nozzle inlet, while the outlet potential is set to 0 V. A zero-gradient boundary condition is applied to the nozzle wall. As such, the electrostatic field inside the nozzle is primarily governed by the potential difference between the inlet and outlet, with no fixed potential prescribed on the nozzle wall.
The potential obtained by solving the axisymmetric Laplace equation is shown in Fig.~\ref{fig:charged_nozzle_setup_phi}(b).
The imposed potential decreases from the inlet to the outlet, and the resulting electric field accelerates the positively charged argon particles in the streamwise direction. 
This setup is adopted to characterize the acceleration and transport behaviors of charged argon particles in the nozzle under axial potential differences.

The inlet temperature of the gas and the wall temperature are set to \(T_0=1000~\mathrm{K}\). 
For the flow boundary conditions, a pressure boundary is imposed at the inlet, while a vacuum outlet boundary is applied at the nozzle exit.
Two inlet pressures, \(P_{\mathrm{in}}=100~\mathrm{Pa}\) and \(1000~\mathrm{Pa}\), are considered to examine the influence of pressure boundary conditions on the charged-particle flow in the nozzle. When the reference length is chosen as 0.016 m (inlet diameter), the corresponding Knudsen numbers are 0.03 and 0.003, respectively.
The computational mesh contains about \(13,800\) structured cells, as shown in Fig.~\ref{fig:charged_nozzle_setup_phi}(b). 
The molecular velocity space is truncated as
$v_x\in[-10,14]\sqrt{k_BT_0/m}$,
$\zeta\in[0,10]\sqrt{k_BT_0/m}$, and 
$\omega\in[0,2\pi]$, which is uniformly discretized in the \((v_x,\zeta,\omega)\) coordinates using \(28\times28\times28\) discrete points.

\begin{figure}[p]
    \centering
    \subfloat[Streamwise velocity]{
        \includegraphics[width=0.7\textwidth, trim={0 0 25 0}, clip=true]{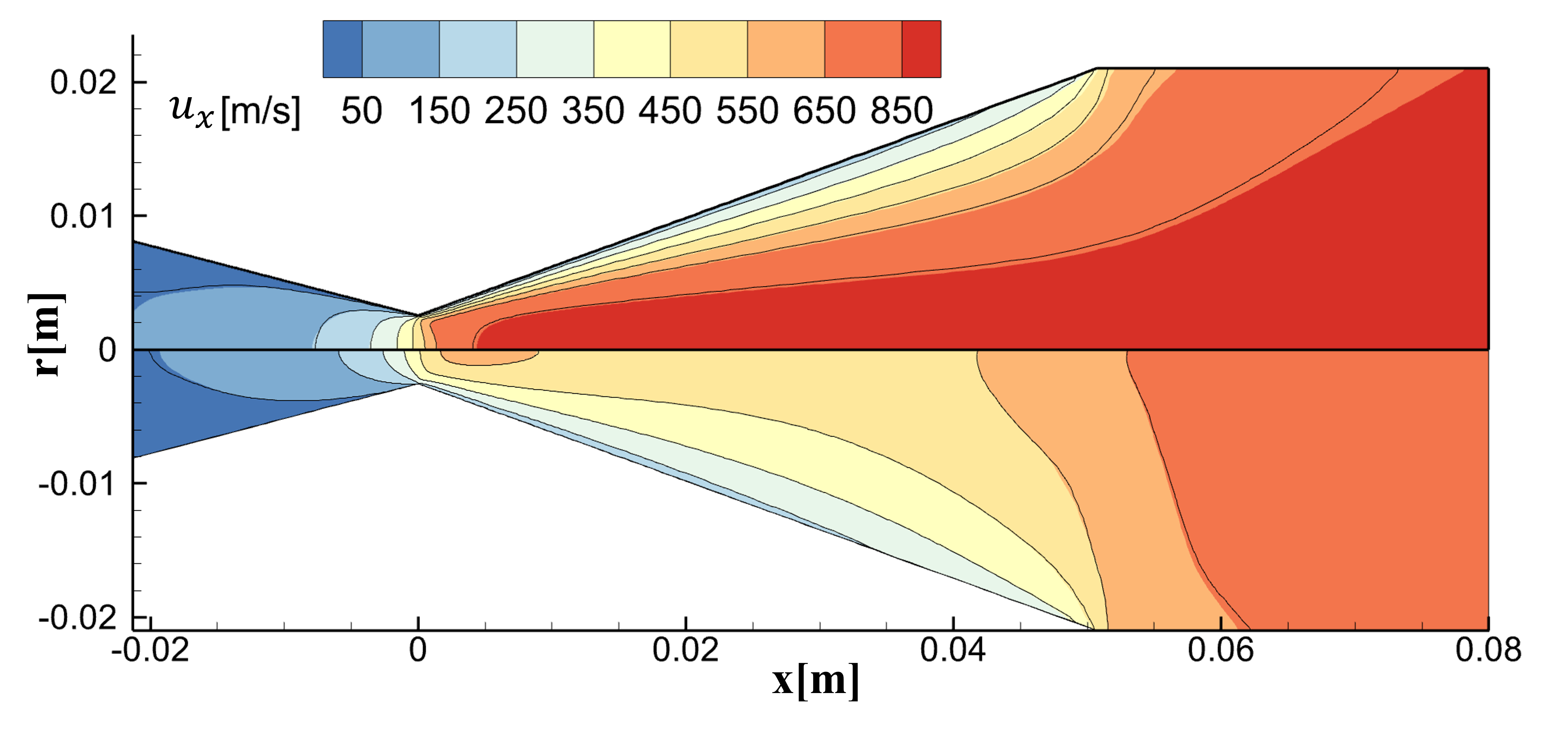}
    }
    \\
    \subfloat[Temperature]{
        \includegraphics[width=0.7\textwidth, trim={0 0 25 0}, clip=true]{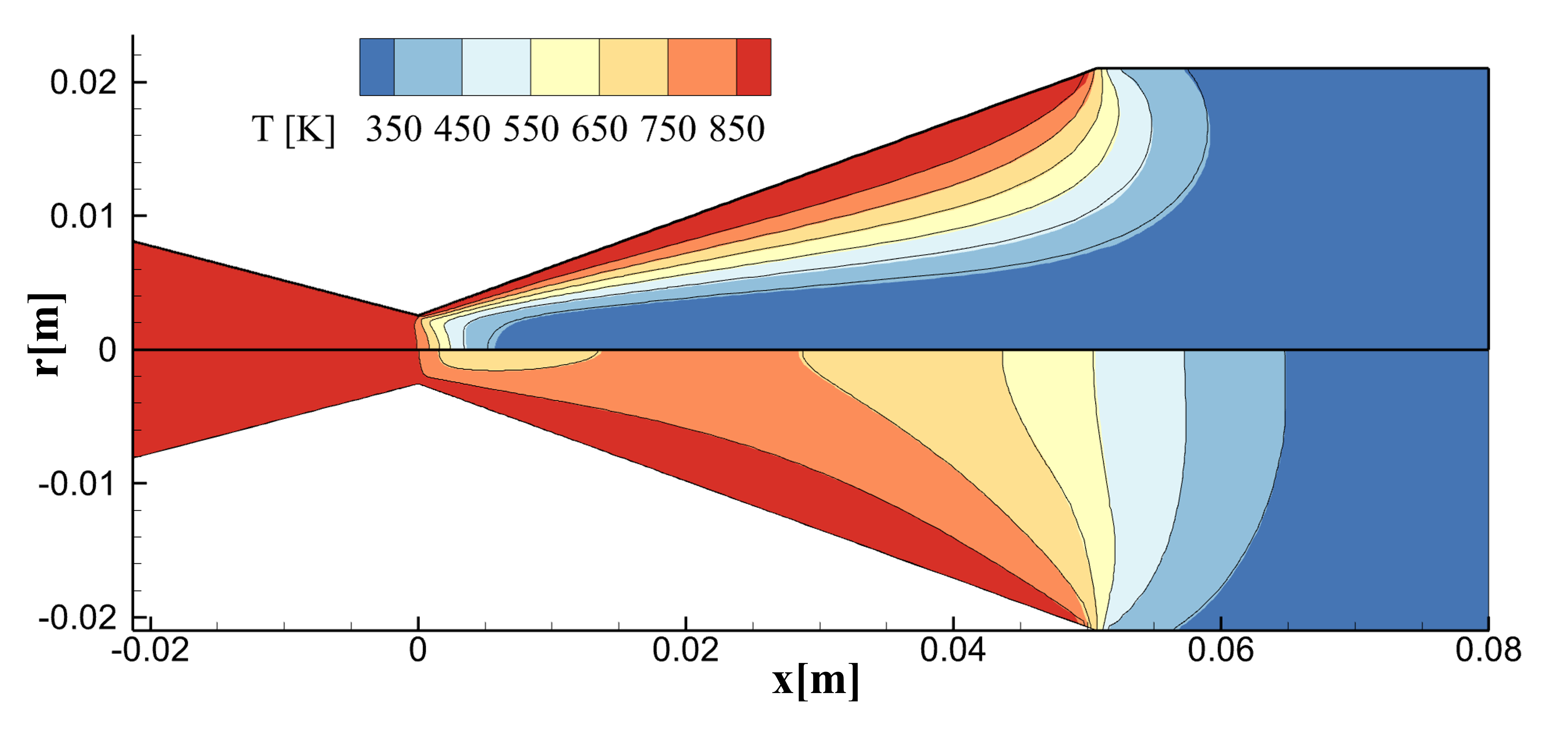}
    }
    \vspace{0.3cm}
    \\
     \subfloat[Streamwise velocity]{
        \includegraphics[width=0.45\textwidth, clip=true]{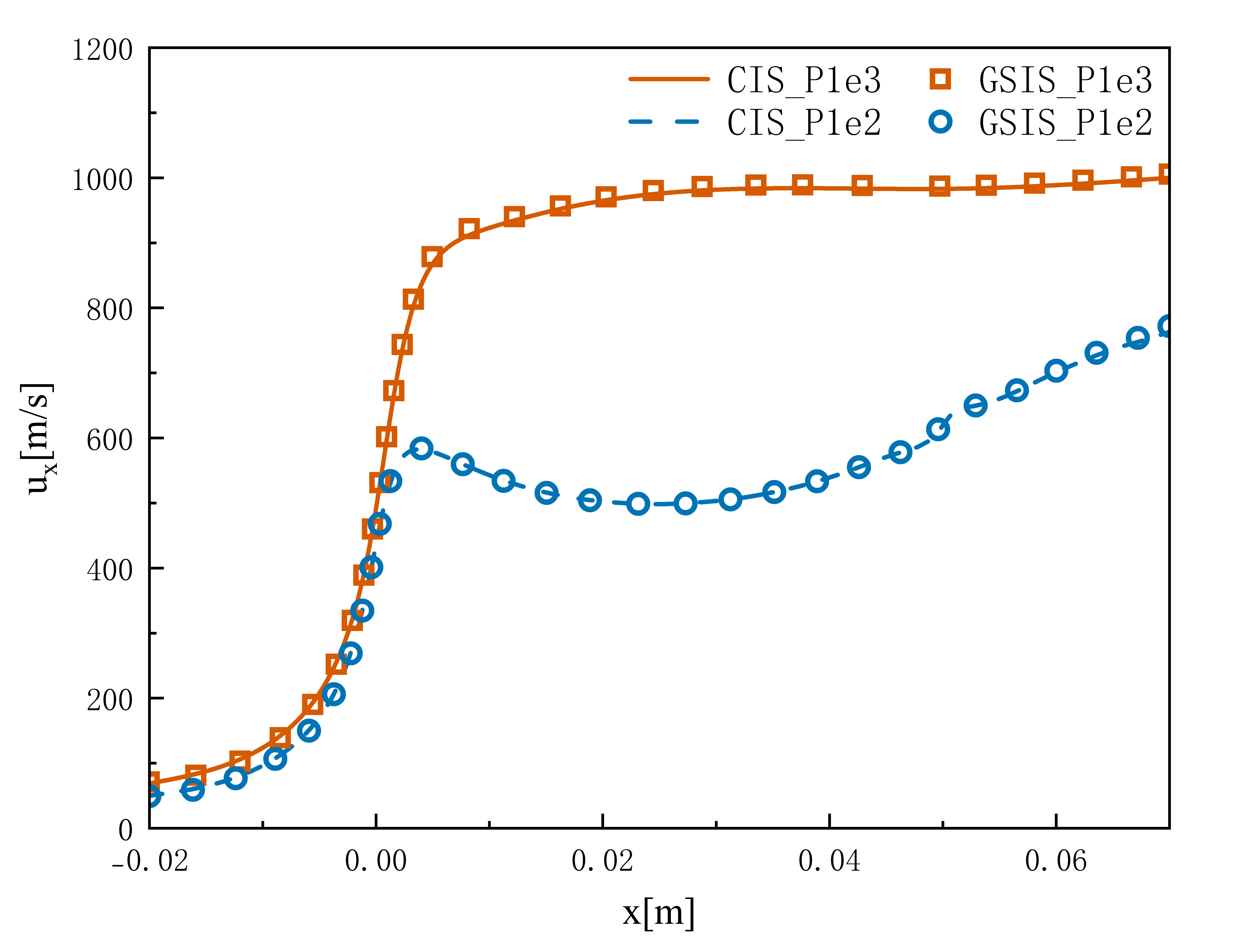}
    }
    \subfloat[Temperature]{
        \includegraphics[width=0.45\textwidth, clip=true]{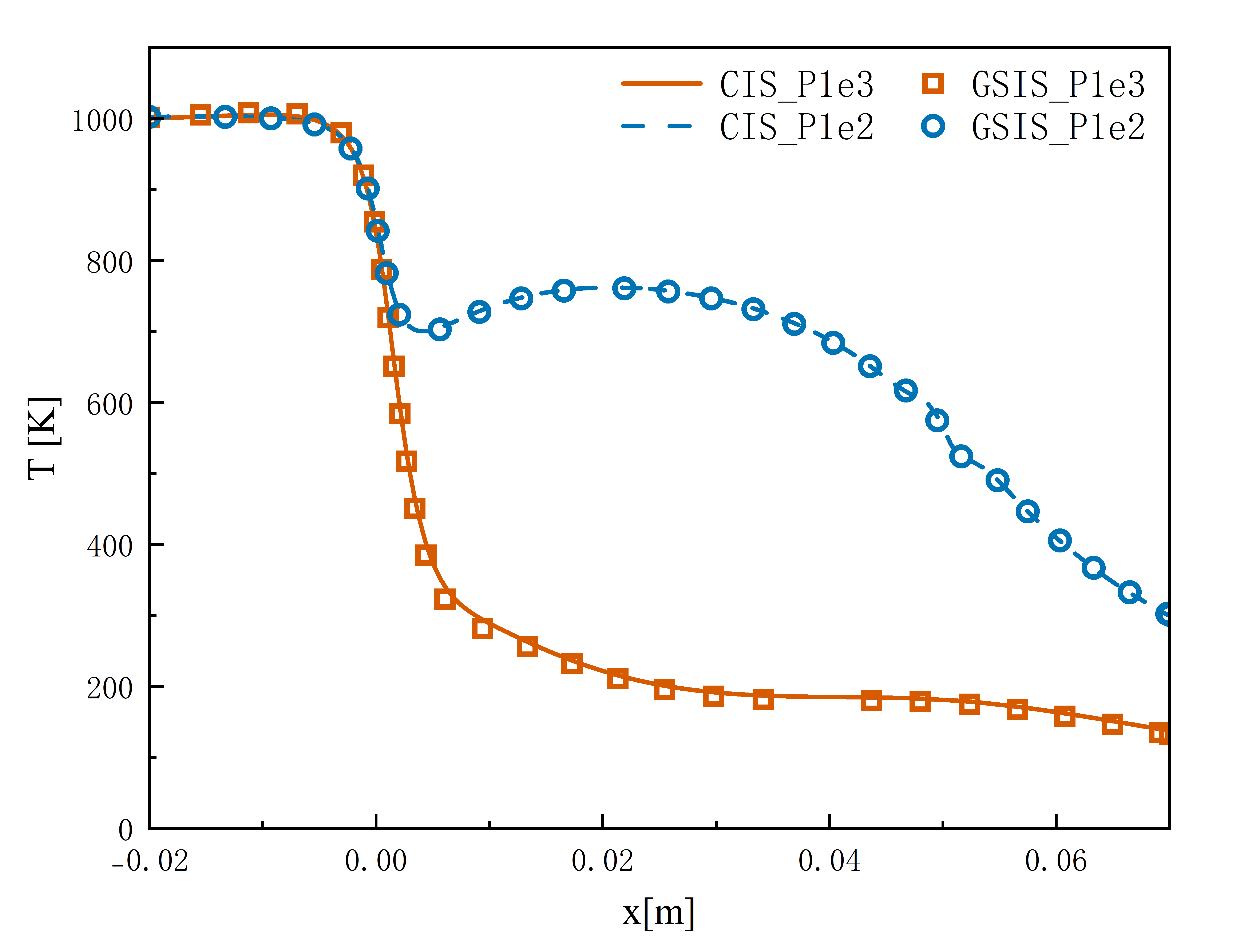}}
    \caption{(a,b) Contours of streamwise velocity and temperature for the charged-particle nozzle flow under different inlet pressures. The CIS and AxiGSIS results are shown by the filled color contours and the black contour lines, respectively. In each subfigure, the upper and lower half regions show the results for
    \(P_{\mathrm{in}}=1000~\mathrm{Pa}\) and \(100~\mathrm{Pa}\), respectively.
    (c,d) Comparison of streamwise velocity and temperature along the nozzle centerline, \(r=0\).
    }
    \label{fig:charged_nozzle_contours}
\end{figure}

Figure~\ref{fig:charged_nozzle_contours} shows the velocity and temperature fields, together with their centerline profiles, for the charged-particle nozzle flow under two inlet pressures \(P_{\mathrm{in}}=10^3~\mathrm{Pa}\) and \(P_{\mathrm{in}}=10^2~\mathrm{Pa}\).
The velocity and temperature contours in Fig.~\ref{fig:charged_nozzle_contours}(a,b) show good agreement between the AxiGSIS and CIS results.

%The pressure inlet and vacuum outlet drive the expansion of particles through the nozzle. Therefore, the streamwise velocity increases rapidly near the throat, as shown in Fig.~\ref{fig:charged_nozzle_contours}(a). The temperature field in Fig.~\ref{fig:charged_nozzle_contours}(b) shows an overall decrease downstream due to expansion cooling and the conversion of random thermal motion into directed kinetic motion. Good agreement between the 

%The centerline velocity profile is affected by three factors: the electrical acceleration and expansion in the divergent nozzle tend to increase the axial velocity, while the 

The effect of inlet pressure on the axial velocity and temperature is clearly observed. At \(P_{\mathrm{in}}=1000~\mathrm{Pa}\), the particle density and collision frequency are higher, so the flow behaves more continuum-like. The diffuse-reflection wall mainly affects the near-wall region, while the centerline flow is dominated by nozzle expansion and electrostatic acceleration. As a result, the centerline velocity in Fig.~\ref{fig:charged_nozzle_contours}(c) increases rapidly after the throat and then approaches a nearly saturated downstream value. The corresponding temperature in Fig.~\ref{fig:charged_nozzle_contours}(d) drops rapidly because of efficient collisional relaxation and expansion cooling.
At \(P_{\mathrm{in}}=100~\mathrm{Pa}\), the gas mean free path increases, giving rise to pronounced rarefaction effects. Particles reflected by the nozzle wall exert influence over an extended spatial domain, and thermalization originating from diffuse reflection propagates toward the nozzle centerline. Although this phenomenon cannot be readily distinguished from contour snapshots, it is clearly manifested in the centerline streamwise velocity shown in Fig.~\ref{fig:charged_nozzle_contours}(c).  Compared with the high-pressure case, the temperature decreases more slowly and remains higher over a longer downstream distance. This indicates that wall thermalization and non-equilibrium transport delay the cooling process in the more rarefied nozzle flow. Likewise, following the initial flow acceleration near the throat, the centerline axial velocity undergoes a mild reduction in the upstream part of the diverging section. Farther downstream, the influence of wall thermalization gradually weakens, while the vacuum outlet and the imposed electric field continue to accelerate the positive ions downstream. Consequently, the centerline velocity increases again.

%After the initial acceleration near the throat, the centerline streamwise velocity decreases slightly in the early divergent section. This is because the wall-induced thermalization redistributes the velocity distribution near the centerline, enhancing random thermal motion and weakening the directed axial motion. 

\begin{table}[t]
    \centering
    \caption{Comparison of computational cost between AxiGSIS and CIS for charged-particle flow through a nozzle. For AxiGSIS, ``Iter'' denotes GSIS outer iterations, each containing three CIS prediction steps; for CIS, it denotes conventional kinetic iterations.}
    \label{tab:charged_nozzle_cost}
    \renewcommand{\arraystretch}{1.15}
    \begin{tabular}{ccccccc} 
        \hline
        \(P_{\mathrm{in}}~[\mathrm{Pa}]\)
        & \(Kn_{\mathrm{in}}\) 
        & Method
        & Cores
        & \(t_{\mathrm{wall}}~[\mathrm{h}]\)
        & core-hours
        & Iter \\
        \hline
        \multirow{2}{*}{\(1000\)}
        & \multirow{2}{*}{\(0.003\)} 
        & AxiGSIS & 40 & 0.15 & 6  & \(17\) \\
        & & CIS  & 200 & 3.76 & 752 & \(5671\) \\
        \hline
        \multirow{2}{*}{\(100\)}
        & \multirow{2}{*}{\(0.03\)} 
        & AxiGSIS &40  &0.45  &18  & \(49\) \\
        & & CIS  &200  &1.20  & 240 & \(1877\) \\
        \hline
    \end{tabular}
\end{table}

Table~\ref{tab:charged_nozzle_cost} compares the convergence behavior and computational cost of AxiGSIS and CIS. For \(P_{\mathrm{in}}=1000~\mathrm{Pa}\), AxiGSIS converges within 17 GSIS outer iterations, while CIS requires 5671 conventional kinetic iterations. The corresponding computational cost is reduced from \(752\) to \(6\) core-hours. For \(P_{\mathrm{in}}=100~\mathrm{Pa}\), AxiGSIS converges within 49 outer iterations, while CIS needs 1877 iterations; AxiGSIS cuts the time cost from \(240\) to \(18\) core-hours. These results demonstrate that the proposed AxiGSIS can significantly accelerate steady-state charged-particle kinetic simulations in nozzle flows with a pressure inlet, a vacuum outlet, and a prescribed electrostatic field.

%\leir{Finally, add a figure to show the asymptotics-preserving property of GSIS}

\section{Conclusions}\label{sec:conclusion}

The AxiGSIS solver has been developed for axisymmetric rarefied gas flows and charged-particle transport under prescribed electrostatic fields.
The axisymmetric kinetic equation is solved by a finite-volume discrete velocity method. The physical-space transport and the velocity-space transport are evaluated with second-order upwind reconstructions in the explicit residual, while the implicit increment equations are constructed with first-order upwind fluxes to improve robustness and reduce the complexity of the linearized iteration.

Four representative test cases have been conducted to assess the accuracy, efficiency, and applicability of the proposed solver. For the Taylor–Couette flow, the method accurately captures the tangential velocity profile and rarefaction-induced slip between two coaxial cylinders with a rotating inner wall and a stationary outer wall. For the neutral nozzle flow, the present AxiGSIS results agree well with the corresponding 3D GSIS results, while the computational cost is substantially reduced due to the elimination of the circumferential physical-space discretization. For the charged-particle flow past an electrostatic sphere, the solver accurately describes the response of positively charged particles to a prescribed electrostatic field, including the reduction of surface pressure and heat flux caused by electrostatic repulsion. For the charged-particle nozzle flow, the method further demonstrates its applicability to internal charged-particle transport with pressure inlet, vacuum outlet, and prescribed electrostatic acceleration.

The convergence comparisons show that AxiGSIS significantly reduces the number of kinetic iterations compared with the conventional iterative scheme. The acceleration is particularly pronounced in the near-continuum and transition regimes, where the conventional kinetic iteration suffers from slow macroscopic information propagation. At elevated Knudsen numbers, the benefits of synthetic iteration diminish, as conventional kinetic iteration already delivers satisfactory computational efficiency. These results confirm that the proposed AxiGSIS provides an accurate and efficient deterministic framework for steady rarefied gas and charged-particle kinetic simulations in axisymmetric configurations.

Future work will encompass several promising extensions, including the integration of a self-consistent Poisson solver, multi-species charged particle transport, and full electromagnetic effect modeling. As observed in Fig.~\ref{fig:charged_nozzle_contours}, rarefied flow conditions can severely degrade nozzle acceleration performance under suboptimal nozzle geometric designs. Furthermore, the present numerical framework can be coupled with adjoint-based optimization techniques for rarefied nozzle flows~\cite{yuan2025adjoint,zhang2026adjoint}, which enables efficient design optimization for axisymmetric kinetic flow devices. Such advancements will further expand the applicability of the AxiGSIS method to plasma apparatuses, charged particle beam systems, rarefied nozzle facilities, and other axisymmetric kinetic configurations characterized by strong field–particle coupling effects.

\section*{Declaration of competing interest}
The authors declare that they have no known competing financial interests or personal relationships that could have appeared to influence the work reported in this paper.

\section*{Acknowledgments}
This work was supported by the National Natural Science Foundation of China (Grant No.~12450002). Special thanks are given to the Center for Computational Science and Engineering at the Southern University of Science and Technology.

\appendix
\appendix
\section{Solution of the electrostatic potential}
\label{subsec:poisson_solver}

% For charged-particle flows, the electrostatic field is obtained from the electrostatic potential $\phi$ as
% $\bm{E}=-\nabla\phi$. In the axisymmetric physical space, the potential satisfies the Poisson equation
% \begin{equation}
% \frac{\partial^2 \phi}{\partial x^2}
% +
% \frac{1}{r}
% \frac{\partial}{\partial r}
% \left(
% r\frac{\partial \phi}{\partial r}
% \right)
% =
% -\frac{\rho_c}{\epsilon_0},
% \label{eq:axisymmetric_poisson}
% \end{equation}
% where $\rho_c$ is the charge density and $\epsilon_0$ is the vacuum permittivity. In the present work, the electrostatic field is prescribed by the boundary potential, and the feedback from charged particles to the electric field is neglected. Therefore, $\rho_c=0$ is used in Eq.~\eqref{eq:axisymmetric_poisson}, and the governing equation reduces to the axisymmetric Laplace equation.

The potential equation is solved on the same finite-volume mesh as the kinetic equation. Integrating Eq.~\eqref{eq:laplace} over the control volume $\Omega_i$ gives
\begin{equation}
\int_{\Omega_i}
\left[
\frac{\partial^2 \phi}{\partial x^2}
+
\frac{1}{r}
\frac{\partial}{\partial r}
\left(
r\frac{\partial \phi}{\partial r}
\right)
\right]
r\,d\Omega
=0.
\end{equation}
Using the divergence theorem, the finite-volume discretization can be written as
\begin{equation}
\sum_{j\in N(i)}
r_{ij}|\Gamma_{ij}|
\left(
\nabla \phi
\right)_{ij}
\cdot
\bm{n}_{ij}
=0,
\label{eq:poisson_fv}
\end{equation}
where $N(i)$ is the set of neighboring cells of cell $i$, $|\Gamma_{ij}|$ is the interface length, $r_{ij}$ is the interface radius, and $\bm{n}_{ij}$ is the outward unit normal vector of cell $i$.

For an interior face $\Gamma_{ij}$, the normal derivative of the potential is approximated by
\begin{equation}
\left(
\nabla \phi
\right)_{ij}
\cdot
\bm{n}_{ij}
=
\frac{\phi_j-\phi_i}{d_{ij}},
\end{equation}
where $d_{ij}$ is the distance between the two cell centers in the normal direction. Substituting this approximation into Eq.~\eqref{eq:poisson_fv}, the discrete equation for an interior cell becomes
\begin{equation}
\sum_{j\in N(i)}
\frac{r_{ij}|\Gamma_{ij}|}{d_{ij}}
\left(
\phi_j-\phi_i
\right)
=0.
\end{equation}
Equivalently, it can be written as
\begin{equation}
A_i\phi_i
=
\sum_{j\in N(i)}
A_{ij}\phi_j
,\qquad 
A_{ij}
=
\frac{r_{ij}|\Gamma_{ij}|}{d_{ij}},
\qquad
A_i
=
\sum_{j\in N(i)}
A_{ij}.
\label{eq:poisson_linear_system}
\end{equation}

The boundary condition is imposed through the boundary-face contribution. For a Dirichlet boundary face, the prescribed potential $\phi_D$ is used to evaluate the normal derivative between the cell center and the boundary face. For a Neumann boundary face, the prescribed normal derivative is directly imposed. On the symmetry axis, the radial derivative of the potential is zero, which corresponds to $E_r=0$.

The resulting linear system is solved iteratively. In the present implementation, a Jacobi-type iteration with under-relaxation is used. At iteration $s$, the potential is updated by
\begin{equation}
\phi_i^{s+1}
=
\left(
1-\eta_\varphi
\right)
\phi_i^s
+
\eta_\varphi
\frac{1}{A_i}
(\sum_{j\in N(i)}
A_{ij}\phi_j^s),
\label{eq:poisson_jacobi}
\end{equation}
where $\eta_\varphi$ is the relaxation factor. The iteration is repeated until the relative change of the potential is smaller than a prescribed tolerance.

\section{Velocity-space integration compensation}
\label{subsec:velocity_space_integration_compensation}

In the discrete velocity method, the continuous velocity-space moments are evaluated by numerical quadrature on a finite discrete velocity grid. For a truncated or non-uniform velocity grid, the discrete quadrature may not exactly reproduce the analytical moments of the Maxwellian distribution. This quadrature error can introduce inconsistencies in macroscopic variables, which may exert a substantial impact in near-continuum flow regimes. To reduce this error, a velocity-space integration compensation is employed in the moment evaluation.

Let \([\cdot]_h\) denote the discrete velocity-space quadrature, and let \(\langle \cdot \rangle\) denote the corresponding analytical velocity-space integral. For a moment function \(\psi(\boldsymbol{v})\), the direct discrete moment is
\begin{equation}
[\psi f]_h
=
\sum_{\alpha}
\psi_\alpha f_\alpha \Delta V_\alpha ,
\end{equation}
where \(\Delta V_\alpha\) is the velocity-space quadrature weight. In the axisymmetric velocity coordinates, this weight contains the cylindrical velocity-space measure.

Following the velocity-space integration compensation idea~\cite{Chen2019Quadrature}, the moment is evaluated by subtracting the discrete moment of a reference Maxwellian and adding back its analytical moment:
\begin{equation}
[\psi f]_c
=
[\psi f]_h
+
\langle \psi g^M(\boldsymbol{W}^*) \rangle
-
[\psi g^M(\boldsymbol{W}^*)]_h .
\label{eq:velocity_integration_compensation}
\end{equation}
Equivalently,
\begin{equation}
[\psi f]_c
=
[\psi (f-g^M(\boldsymbol{W}^*))]_h
+
\langle \psi g^M(\boldsymbol{W}^*) \rangle .
\end{equation}
Here \([\cdot]_c\) denotes the compensated moment, and \(g^M(\boldsymbol{W}^*)\) is the reference Maxwellian constructed from the local macroscopic state \(\boldsymbol{W}^*\). In the present implementation, \(\boldsymbol{W}^*\) is taken as the local macroscopic state before the moment update. This treatment does not modify the VDF or the kinetic transport discretization; it only modifies the moment evaluation.

For the conservative variables, the moment function is
\begin{equation}
\psi_c
=
m
\left(
1,\,
v_x,\,
v_r,\,
v_\theta,\,
\frac{1}{2}|\boldsymbol{v}|^2
\right)^T .
\end{equation}
Therefore, the compensated conservative variables are computed as
\begin{equation}
\boldsymbol{W}
=
[\psi_c f]_h
+
\langle \psi_c g^M(\boldsymbol{W}^*) \rangle
-
[\psi_c g^M(\boldsymbol{W}^*)]_h .
\end{equation}
Since the analytical Maxwellian moments exactly recover the density, momentum and energy associated with \(\boldsymbol{W}^*\), this correction removes the equilibrium quadrature error from the conservative moments.

The same compensation is also applied to the raw moments required for the stress tensor and heat flux. For example, the compensated second-order and third-order raw moments are evaluated as
\begin{equation}
\begin{aligned}
[v_i v_j f]_c
=&
[v_i v_j f]_h
+
\langle v_i v_j g^M(\boldsymbol{W}^*) \rangle
-
[v_i v_j g^M(\boldsymbol{W}^*)]_h ,\\
[v_i |\boldsymbol{v}|^2 f]_c
=&
[v_i |\boldsymbol{v}|^2 f]_h
+
\langle v_i |\boldsymbol{v}|^2 g^M(\boldsymbol{W}^*) \rangle
-
[v_i |\boldsymbol{v}|^2 g^M(\boldsymbol{W}^*)]_h .
\end{aligned}
\end{equation}
For a Maxwellian with number density \(n\), velocity \(\boldsymbol{u}\), and temperature \(T\), the analytical moments are
\begin{equation}
\langle v_i g^M \rangle = n u_i,
\quad
\langle v_i v_j g^M \rangle
=
n u_i u_j
+
n\frac{k_B}{m}T\delta_{ij}, 
\quad 
\langle v_i|\boldsymbol{v}|^2 g^M \rangle
=
n u_i
\left(
|\boldsymbol{u}|^2+5\frac{k_B}{m}T
\right).
\end{equation}

After the compensated raw moments are obtained, the pressure tensor and heat flux are computed in the standard central-moment form:
\begin{equation}
\begin{aligned}
P_{ij}
=&
m
\left(
[v_i v_j f]_c
-
n u_i u_j
\right),
\\
q_i
=&
\frac{m}{2}
\left[
v_i |\boldsymbol{v}|^2 f
\right]_c
-
\frac{m}{2}
\left(
u_i [|\boldsymbol{v}|^2 f]_c
+
2u_j [v_i v_j f]_c
\right)
+
m n u_i
|\boldsymbol{u}|^2.
\end{aligned}
\end{equation}
The stress tensor is then obtained from
\begin{equation}
\Pi_{ij}=P_{ij}-p\delta_{ij},
\qquad
p=nk_BT .
\end{equation}

This compensation can be interpreted as evaluating the non-equilibrium deviation \(f-g^M\) by discrete quadrature while retaining the equilibrium contribution analytically. Hence, when the VDF is close to the Maxwellian, the equilibrium part is exactly integrated at the discrete level, and the remaining quadrature error is associated only with the smaller non-equilibrium part. This improves the consistency between the kinetic moments and the macroscopic synthetic equations, and makes the GSIS correction more robust on finite and truncated velocity grids.

\section{Consistent boundary treatment in the GSIS}
\label{app:gsis_boundary_treatment}

In the GSIS, the macroscopic synthetic equations are used as an essential auxiliary system to accelerate the convergence of the kinetic equation. Therefore, their boundary fluxes should remain consistent with the kinetic boundary conditions specified in Section~\ref{sec:bc}. In the present AxiGSIS, the boundary treatment in Ref.~\cite{zhang2026efficientheatflux} is adapted to the axisymmetric finite-volume formulation.

The boundary flux in the macroscopic synthetic equations is constructed from half-range kinetic moments. 
The incoming half-range VDF, defined for molecules entering the gas computational domain from the boundary, is determined by the prescribed kinetic boundary condition and is not overwritten by the macroscopic synthetic correction. For inflow, far-field, and vacuum-outlet boundaries, the incoming VDF remains fixed during the macroscopic inner iterations. For a diffuse-reflection wall, the incoming VDF keeps the wall-Maxwellian form with prescribed wall velocity and temperature, while the wall density is updated to enforce the zero-normal-mass-flux condition.

The outgoing half-range flux is determined by the gas state inside the computational domain and therefore carries the non-equilibrium information obtained from the kinetic prediction. During the macroscopic inner iterations, the full VDF is not solved again; hence, the non-equilibrium part of the outgoing flux is retained, and only its Maxwellian equilibrium part is updated according to the current macroscopic state. In this way, the boundary flux remains consistent with the updated macroscopic solution without destroying the kinetic boundary condition.

Consider a boundary face \(\Gamma_b\) of the gas computational domain. 
The outward unit normal vector is denoted by \(\boldsymbol{n}_b\). 
At a solid wall, \(\boldsymbol{n}_b\) points from the gas domain toward the solid wall. 
The normal molecular velocity is
$v_n=\boldsymbol{v}\cdot\boldsymbol{n}_b$.
With this convention, molecules with \(v_n>0\) leave the gas domain and move toward the boundary, whereas molecules with \(v_n<0\) enter the gas domain from the boundary.

Let
\begin{equation}
\boldsymbol{\psi}_c
=
m
\left(
1,\,
v_x,\,
v_r,\,
v_\theta,\,
\frac{1}{2}|\boldsymbol{v}|^2
\right)^T
\end{equation}
be the vector of conservative microscopic moments. After the kinetic prediction step in the \(k\)-th GSIS outer iteration, the outgoing kinetic half-range flux is evaluated as
\begin{equation}
\boldsymbol{F}_{\mathrm{out},b}^{K,0}
=
\int_{v_n>0}
v_n\boldsymbol{\psi}_c f^{k+1/2}\,d\boldsymbol{v}.
\end{equation}
Meanwhile, the corresponding outgoing Maxwellian half-range flux is computed from the predicted boundary macroscopic state \(\boldsymbol{W}_b^0\):
\begin{equation}
\boldsymbol{F}_{\mathrm{out},b}^{M,0}
=
\int_{v_n>0}
v_n\boldsymbol{\psi}_c
g^M(\boldsymbol{W}_b^0)
\,d\boldsymbol{v}.
\end{equation}

Following the half-range flux update strategy in Ref.~\cite{zhang2026efficientheatflux}, during the macroscopic inner iterations ($\ell$ is the inner iteration step), the outgoing boundary flux is updated by correcting only the Maxwellian part:
\begin{equation}
\boldsymbol{F}_{\mathrm{out},b}^{\ell}
=
\boldsymbol{F}_{\mathrm{out},b}^{K,0}
+
\left[
\boldsymbol{F}_{\mathrm{out},b}^{M}
(\boldsymbol{W}_b^\ell)
-
\boldsymbol{F}_{\mathrm{out},b}^{M,0}
\right],
\end{equation}
where $
\boldsymbol{F}_{\mathrm{out},b}^{M}
(\boldsymbol{W}_b^\ell)
=
\int_{v_n>0}
v_n\boldsymbol{\psi}_c
g^M(\boldsymbol{W}_b^\ell)
\,d\boldsymbol{v}$.
Thus, the non-equilibrium part extracted from the kinetic prediction is retained at the boundary, while the equilibrium contribution follows the updated macroscopic solution.

The total boundary flux used in the macroscopic synthetic equations is
\begin{equation}
\boldsymbol{F}_{b}^{\ell}
=
\boldsymbol{F}_{\mathrm{out},b}^{\ell}
+
\boldsymbol{F}_{\mathrm{in},b}^{\ell}.
\end{equation}
The incoming flux \(\boldsymbol{F}_{\mathrm{in},b}^{\ell}\) is determined by the corresponding kinetic boundary condition. For inflow or far-field boundaries, the incoming Maxwellian is prescribed and the incoming half-range flux is kept fixed during the macroscopic inner iterations. For pressure-type boundaries, the incoming Maxwellian is constructed from the prescribed pressure and temperature together with the extrapolated boundary velocity. For a vacuum outlet, the incoming VDF is set to zero.

For a diffuse-reflection wall, the incoming VDF is the wall Maxwellian with prescribed wall velocity and temperature. Similar to the isothermal-wall treatment in Ref.~\cite{zhang2026efficientheatflux}, the wall density is updated during the macroscopic inner iterations to satisfy the zero-normal-mass-flux condition:  
\begin{equation}
\rho_w^\ell
=
-
\frac{
F_{\mathrm{out},b,\rho}^{\ell}
}{
\int_{v_n<0}
v_n\
g_w^M(1,\boldsymbol{u}_w,T_w)
\,d\boldsymbol{v}
},
\end{equation}
where \(F_{\mathrm{out},b,\rho}^{\ell}\) is the mass component of
\(\boldsymbol{F}_{\mathrm{out},b}^{\ell}\). The corresponding incoming wall flux is then updated consistently with this wall density.

% If
% \[
% \boldsymbol{U}_w
% =
% \int_{v_n<0}
% v_n\boldsymbol{\psi}_c
% g_w^M(1,\boldsymbol{u}_w,T_w)
% \,d\boldsymbol{v}
% \]
% denotes the incoming wall half-range flux corresponding to unit wall density, then the wall density at the \(\ell\)-th macroscopic inner iteration is obtained from
% \[
% \rho_w^\ell
% =
% -
% \frac{
% F_{\mathrm{out},b,\rho}^{\ell}
% }{
% U_{w,\rho}
% },
% \]
% where \(F_{\mathrm{out},b,\rho}^{\ell}\) and \(U_{w,\rho}\) are the mass components of
% \(\boldsymbol{F}_{\mathrm{out},b}^{\ell}\) and \(\boldsymbol{U}_w\), respectively. The corresponding incoming wall flux is then updated consistently with this wall density.

In the finite-volume residual, the boundary flux is multiplied by the axisymmetric face measure \(r_b|\Gamma_b|\), in the same manner as the interior face fluxes. After the macroscopic synthetic equations are solved, the VDF correction is applied only to the cell-centered VDF. In the next kinetic prediction step, the incoming VDF at each boundary face is reconstructed again from the original kinetic boundary condition. Therefore, the macroscopic synthetic iteration accelerates the propagation of low-order macroscopic information while preserving consistency with the kinetic boundary treatment.

\bibliographystyle{elsarticle-num}
\bibliography{ref}

@book{bird1994molecular,
  title={Molecular gas dynamics and the direct simulation of gas flows},
  author={Bird, Graeme A},
  year={1994},
  publisher={Oxford University Press}
}

@book{birdsall2018plasma,
  title={Plasma physics via computer simulation},
  author={Birdsall, Charles K and Langdon, A Bruce and Langdon, AB},
  year={2018},
  publisher={CRC Press}
}

@article{jin2024numerical,
  author = {Jin, XH and Su, PH and Chen, Z and Cheng, XL and Wang, Q and Wang, B},
  title = {Numerical and experimental investigation of rarefied hypersonic flow in a nozzle},
  journal = {Physics of Fluids},
  year = {2024},
  volume = {36},
  pages = {116131},
}

@article{verboncoeur2005particle,
  title={Particle simulation of plasmas: review and advances},
  author={Verboncoeur, John P},
  journal={Plasma Physics and Controlled Fusion},
  volume={47},
  number={5A},
  pages={A231--A260},
  year={2005}
}

@article{tskhakaya2007particle,
  title={The particle-in-cell method},
  author={Tskhakaya, David and Matyash, Konstantin and Schneider, Ralf and Taccogna, Francesco},
  journal={Contributions to Plasma Physics},
  volume={47},
  number={8-9},
  pages={563--594},
  year={2007},
  publisher={Wiley Online Library}
}

@article{kaganovich2007kinetic,
  title={Kinetic effects in a {H}all thruster discharge},
  author={Kaganovich, ID and Raitses, Yevgeny and Sydorenko, Dmytro and Smolyakov, Andrei},
  journal={Physics of Plasmas},
  volume={14},
  number={5},
  year={2007},
  publisher={AIP Publishing}
}

@article{coche2014two,
  title={A two-dimensional (azimuthal-axial) particle-in-cell model of a {H}all thruster},
  author={Coche, P and Garrigues, L},
  journal={Physics of Plasmas},
  volume={21},
  number={2},
  year={2014},
  publisher={AIP Publishing}
}

@article{davidson2015implementation,
  title={Implementation of a hybrid particle code with a {PIC} description in r--z and a gridless description in $\phi$ into {OSIRIS}},
  author={Davidson, Adam and Tableman, Adam and An, Weiming and Tsung, Frank S and Lu, Wei and Vieira, Jorge and Fonseca, Ricardo A and Silva, Lu{\'\i}s O and Mori, Warren B},
  journal={Journal of Computational Physics},
  volume={281},
  pages={1063--1077},
  year={2015},
  publisher={Elsevier}
}

@article{bruhwiler2001particle,
  title={Particle-in-cell simulations of plasma accelerators and electron-neutral collisions},
  author={Bruhwiler, David L and Giacone, Rodolfo E and Cary, John R and Verboncoeur, John P and Mardahl, Peter and Esarey, Eric and Leemans, WP and Shadwick, BA},
  journal={Physical Review Special Topics-Accelerators and Beams},
  volume={4},
  number={10},
  pages={101302},
  year={2001},
  publisher={APS}
}

@article{pointon2001particle,
  title={Particle-in-cell simulations of electron flow in the post-hole convolute of the Z accelerator},
  author={Pointon, TD and Stygar, WA and Spielman, RB and Ives, HC and Struve, KW},
  journal={Physics of Plasmas},
  volume={8},
  number={10},
  pages={4534--4544},
  year={2001},
  publisher={American Institute of Physics}
}

@article{arber1996hybrid,
  title={Hybrid simulation of the nonlinear evolution of a collisionless, large {Larmor} radius {Z} pinch},
  author={Arber, TD},
  journal={Physical Review Letters},
  volume={77},
  number={9},
  pages={1766},
  year={1996},
  publisher={APS}
}

@article{schmidt2012fully,
  title={Fully kinetic simulations of dense plasma focus {Z}-pinch devices},
  author={Schmidt, A and Tang, V and Welch, D},
  journal={Physical Review Letters},
  volume={109},
  number={20},
  pages={205003},
  year={2012},
  publisher={APS}
}

@article{cheng1976integration,
  title={The integration of the {Vlasov} equation in configuration space},
  author={Cheng, Chio-Zong and Knorr, Georg},
  journal={Journal of Computational Physics},
  volume={22},
  number={3},
  pages={330--351},
  year={1976},
  publisher={Elsevier}
}

@article{sonnendrucker1999semi,
  title={The {semi-Lagrangian method for the numerical resolution of the Vlasov} equation},
  author={Sonnendr{\"u}cker, Eric and Roche, Jean and Bertrand, Pierre and Ghizzo, Alain},
  journal={Journal of Computational Physics},
  volume={149},
  number={2},
  pages={201--220},
  year={1999},
  publisher={Elsevier}
}

@article{filbet2001conservative,
  title={Conservative numerical schemes for the {Vlasov} equation},
  author={Filbet, Francis and Sonnendr{\"u}cker, Eric and Bertrand, Pierre},
  journal={Journal of Computational Physics},
  volume={172},
  number={1},
  pages={166--187},
  year={2001},
  publisher={Elsevier}
}

@article{yuan2025adjoint,
  title={Adjoint shape optimization from the continuum to free-molecular gas flows},
  author={Yuan, Ruifeng and Wu, Lei},
  journal={Journal of Computational Physics},
  volume={537},
  pages={114102},
  year={2025},
  publisher={Elsevier}
}

@article{zhang2026adjoint,
  title={A fast-converging and asymptotic-preserving adjoint shape optimization of rarefied gas flows},
  author={Zhang, Yanbing and Yuan, Ruifeng and Wu, Lei},
  journal={Journal of Computational Physics},
  pages={114960},
  year={2026},
  publisher={Elsevier}
}

@book{CE,
	author = {S. Chapman and T. G. Cowling},
	publisher = {Cambridge University Press},
	title = {{The Mathematical Theory of Non-Uniform Gases}},
	year = {1970}}

@inproceedings{filbet2002direct,
  title={Direct axisymmetric {Vlasov} simulations of space charge dominated beams},
  author={Filbet, Francis and Lemaire, J-L and Sonnendr{\"u}cker, Eric},
  booktitle={International Conference on Computational Science},
  pages={305--314},
  year={2002},
  organization={Springer}
}

@article{shoucri2004study,
  title={Study of the generation of a charge separation and electric field at a plasma edge using {Eulerian Vlasov} codes in cylindrical geometry},
  author={Shoucri, Magdi and Gerhauser, H and Finken, K-H},
  journal={Computer Physics Communications},
  volume={164},
  number={1-3},
  pages={138--149},
  year={2004},
  publisher={Elsevier}
}

@article{valentini2005numerical,
  title={A numerical scheme for the {integration of the Vlasov-Poisson} system of equations, in the magnetized case},
  author={Valentini, Francesco and Veltri, Pierluigi and Mangeney, Andr{\'e}},
  journal={Journal of Computational Physics},
  volume={210},
  number={2},
  pages={730--751},
  year={2005},
  publisher={Elsevier}
}

@article{vogman2018conservative,
  title={Conservative fourth-order finite-volume {Vlasov-Poisson} solver for axisymmetric plasmas in cylindrical ($r$, $v_r$, $v_\theta$) phase space coordinates},
  author={Vogman, GV and Shumlak, Uri and Colella, Phillip},
  journal={Journal of Computational Physics},
  volume={373},
  pages={877--899},
  year={2018},
  publisher={Elsevier}
}

@article{wang2022gas,
  title={A gas-kinetic scheme for collisional {Vlasov-Poisson} equations in cylindrical coordinates},
  author={Wang, Yi and Zhang, Jiexing and Ni, Guoxi},
  journal={Communications in Computational Physics},
  volume={32},
  number={3},
  pages={779--809},
  year={2022}
}

@article{ni2022fourier,
  title={A {Fourier} transformation based {UGKS for Vlasov-Poisson} equations in cylindrical coordinates (r, $\theta$)},
  author={Ni, Anchun and Wang, Yi and Ni, Guoxi and Chen, Yibing},
  journal={Computers \& Fluids},
  volume={245},
  pages={105593},
  year={2022},
  publisher={Elsevier}
}

@article{sone1987steady,
  author  = {Sone, Y. and Aoki, K.},
  title   = {Steady gas flows past bodies at small {Knudsen} numbers---{Boltzmann} and hydrodynamic systems},
  journal = {Transport Theory and Statistical Physics},
  volume  = {16},
  number  = {2--3},
  pages   = {189--199},
  year    = {1987},
}

@article{takata1993sphere,
  author  = {Takata, S. and Sone, Y. and Aoki, K.},
  title   = {Numerical analysis of a uniform flow of a rarefied gas past a sphere on the basis of the {Boltzmann} equation for hard-sphere molecules},
  journal = {Physics of Fluids A},
  volume  = {5},
  number  = {3},
  pages   = {716--737},
  year    = {1993},
}

@article{aoki2003inverted,
  author  = {Aoki, K. and Yoshida, H. and Nakanishi, T. and Garcia, A. L.},
  title   = {Inverted velocity profile in the cylindrical {Couette} flow of a rarefied gas},
  journal = {Physical Review E},
  volume  = {68},
  pages   = {016302},
  year    = {2003},
}

@article{sharipov1996linear,
  author  = {Sharipov, F. M. and Kremer, G. M.},
  title   = {Linear {Couette} flow between two rotating cylinders},
  journal = {European Journal of Mechanics - B/Fluids},
  volume  = {15},
  pages   = {493--505},
  year    = {1996}
}

@article{sharipov1999nonisothermal,
  author  = {Sharipov, F. M. and Kremer, G. M.},
  title   = {Non-isothermal {Couette} flow of a rarefied gas between two rotating cylinders},
  journal = {European Journal of Mechanics - B/Fluids},
  volume  = {18},
  number  = {1},
  pages   = {121--130},
  year    = {1999},
}

@article{li2018ugksas,
  title = {A unified gas-kinetic scheme for axisymmetric flow in all {Knudsen} number regimes},
  author = {Li, Shiyi and Li, Qibing and Fu, Song and Xu, Kun},
  journal = {Journal of Computational Physics},
  volume = {366},
  pages = {144--169},
  year = {2018}
}

@article{luo2024directIntermittentDIG,
  author  = {Luo, L. Y. and Wu, L.},
  title   = {Multiscale simulation of rarefied gas dynamics via direct intermittent {GSIS-DSMC} coupling},
  journal = {Advances in Aerodynamics},
  volume  = {6},
  pages   = {22},
  year    = {2024},

}

@article{zhang2026efficientheatflux,
  title={An efficient treatment of heat-flux boundary conditions in {GSIS} for rarefied gas flows},
  author={Zhang, Yanbing and Yuan, Ruifeng and Luo, Liyan and Wu, Lei},
  journal={Computers \& Fluids},
  volume={315},
  pages={107113},
  year={2026},
}

@book{aristov2001direct,
  author    = {Aristov, V. V.},
  title     = {Direct Methods for Solving the Boltzmann Equation and Study of Nonequilibrium Flows},
  publisher = {Springer},
  address   = {Dordrecht},
  year      = {2001}
}

@article{ wang2018comparative,
Author = {Wang, Peng and Minh Tuan Ho and Wu, Lei and Guo, Zhaoli and Zhang, Yonghao},
Title = {A comparative study of discrete velocity methods for low-speed rarefied gas flows},
Journal = {Computers \& Fluids},
Year = {2018},
Volume = {161},
Pages = {33-46},
}

@article{xu2010ugks,
Author = {Xu, Kun and Huang, Juan-Chen},
Title = {A unified gas-kinetic scheme for continuum and rarefied flows},
Journal = {Journal of Computational Physics},
Year = {2010},
Volume = {229},
Number = {20},
Pages = {7747-7764},
}

@article{guo2013dugks,
Author = {Guo, Zhaoli and Xu, Kun and Wang, Ruijie},
Title = {Discrete unified gas kinetic scheme for all {Knudsen} number flows: Low-speed isothermal case},
Journal = {Physical Review E},
Year = {2013},
Volume = {88},
Number = {3},
}

@article{Su2020GSIS,
author = {W. Su and L. H. Zhu and P. Wang and Y. H. Zhang and L. Wu},
title = {Can we find steady-state solutions to multiscale rarefied gas flows within dozens of iterations?},
journal = {Journal of Computational Physics},
volume = {407},
pages = {109245},
year = {2020},
}

@article{Su2020SIAM,
author = {W. Su and L. H. Zhu and L. Wu},
journal = {SIAM Journal on Scientific Computing},
pages = {B1517--B1540},
title = {Fast convergence and asymptotic preserving of the general synthetic iterative scheme},
volume = {42},
year = {2020}
}

@article{Zhang2024CaF,
author = {Y. B. Zhang and J. N. Zeng and R. F. Yuan and W. Liu and L. Wu},
journal = {Computers \& Fluids},
pages = {106374},
title = {Efficient parallel solver for rarefied gas flow using {GSIS}},
volume = {281},
year = {2024}
}

@article{mieussens2000dvm,
  title = {Discrete-velocity models and numerical schemes for the {Boltzmann-BGK} equation in plane and axisymmetric geometries},
  author = {Mieussens, Luc},
  journal = {Journal of Computational Physics},
  volume = {162},
  Number = {2},
  pages = {429--466},
  year = {2000}
}

@article{tibbs1997,
  title={Anomalous flow profile due to the curvature effect on slip length},
  author={Tibbs, Kevin W and Baras, Florence and Garcia, Alejandro L},
  journal={Physical Review E},
  volume={56},
  number={2},
  pages={2282},
  year={1997},
  publisher={APS}
}

@article{shakhov1968approximate,
  title={Approximate kinetic equations in rarefied gas theory},
  author={Shakhov, EM},
  journal={Fluid Dynamics},
  volume={3},
  number={1},
  pages={112--115},
  year={1968},
  publisher={Springer}
}

@article{RUSANOV1962304,
title = {The calculation of the interaction of non-stationary shock waves and obstacles},
author = {V.V Rusanov},
journal = {USSR Computational Mathematics and Mathematical Physics},
volume = {1},
number = {2},
pages = {304-320},
year = {1962},
}

@inproceedings{venkatakrishnan1993accuracy,
  title={On the accuracy of limiters and convergence to steady state solutions},
  author={Venkatakrishnan, Venkat},
  booktitle={31st Aerospace Sciences Meeting},
  pages={880},
  year={1993}
}

@inproceedings{chen2019quadrature,
  title={Quadrature with equilibrium offset and its application on adaptive velocity grid},
  author={Chen, Songze},
  booktitle={AIP Conference Proceedings},
  volume={2132},
  number={1},
  pages={060003},
  year={2019},
  organization={AIP Publishing LLC}
}
\end{document}